\documentclass[a4paper,11pt]{article}
\pdfoutput=1 

\usepackage{jcappub} 

\usepackage[T1]{fontenc} 
\usepackage{multirow}
\usepackage{tabularx}
\usepackage{subfig}
\usepackage{booktabs}

\usepackage[utf8]{inputenc}
\usepackage{amsmath}
\usepackage{amssymb}
\usepackage{makeidx}
\usepackage{amsfonts}

\title{\boldmath Geodesic structure, shadow and optical appearance of black hole immersed in Chaplygin-like dark fluid}

\author[a,*]{Xiang-Qian Li}
\author[a]{Hao-Peng Yan}
\author[a]{Xiao-Jun Yue}
\author[b]{Shi-Wei Zhou}
\author[a,*]{Qiang Xu}
\note[*]{Corresponding author.}

\affiliation[a]{College of Physics, Taiyuan University of Technology, Taiyuan 030024, China}
\affiliation[b]{Physics and Information Engineering Institute, Shanxi Normal University, Taiyuan 030031, China}

\emailAdd{lixiangqian@tyut.edu.cn}
\emailAdd{yanhaopeng@tyut.edu.cn}
\emailAdd{yuexiaojun@tyut.edu.cn}
\emailAdd{zhousw783@163.com}
\emailAdd{xuqiang01@tyut.edu.cn}

\abstract{In this study, we focus on a black hole immersed in a cosmological Chaplygin-like dark fluid (CDF), characterized by the equation of state $p=-B/\rho$ and an additional parameter $q$ influencing the energy density of the fluid. We investigate the geodesic structure, shadow, and optical appearance of such a black hole. Through analysis on the effective potential and the epicyclic frequencies, it is found that the existence of innermost/outermost stable circular orbits for a timelike particle is governed by the CDF parameters. The behaviors of the orbital conserved quantities and Keplerian frequency are also examined. Due to the existence of pseudo-cosmological horizon, the determination of the shadow radius depends significantly on the position of the observer. By placing the static observer at an approximately flat position between the event and pseudo-cosmological horizons, we constrain the CDF parameters using EHT observations. We investigate the effect of CDF on the shadows and optical images of the black hole, surrounded by various profiles of accretions. For the thin disk accretion, the light trajectories are categorized into direct emission, lensing ring, and photon ring based on impact parameters. Due to the existence of outermost stable circular orbits, outer edges could exist in the direct and lensing ring images. The observed brightness is mainly due to direct emission, with a minor contribution from the lensing ring, while the contribution from the photon ring is negligible due to extreme demagnetization. In the case of spherical accretion, we consider both static and infalling accretion models. The images obtained under infalling accretion are slightly darker than those under static accretion, attributed to the Doppler effect. Throughout the study, we analyze the influence of the parameters $B$ and $q$ on the results.
}

\begin{document}
\maketitle
\flushbottom

\section{Introduction}
\label{sec:intro}
The Event Horizon Telescope (EHT) collaboration has made significant strides in our understanding of supermassive black holes by releasing Very Long Baseline Interferometry (VLBI) observations of the Messier 87 galaxy with a central black hole, know as M87$^{\ast}$~\cite{Narayan:2019imo,EventHorizonTelescope:2019dse,EventHorizonTelescope:2019uob,EventHorizonTelescope:2019jan,EventHorizonTelescope:2019ths,EventHorizonTelescope:2019pgp,EventHorizonTelescope:2019ggy} and the Milky Way galaxy with a central black hole, called Sgr A$^{\ast}$~\cite{EventHorizonTelescope:2022wkp}. These observations, with an angular resolution comparable to that expected of a supermassive black hole, reveal a central dark region known as the black hole shadow, surrounded by a bright ring called the photon ring~\cite{Takahashi:2004xh}. The investigation of light ray deflection around gravitationally intense stars was first explored by Synge~\cite{Synge:1966okc}. Bardeen subsequently extended this work by calculating that the shadow radius of a static Schwarzschild black hole is $r_{\rm sh}=3M$, and also demonstrated that the angular momentum of rotating black holes deforms the shape of the shadow, deviating from a perfect circle observed in static cases~\cite{Bardeen:1972fi}. Furthermore, it is widely recognized that astrophysical black holes are not isolated in empty space but are instead surrounded by luminous accretion flows that significantly impact our observations. The theoretical study of the image of a geometrically thin accretion disk around a Schwarzschild black hole was initially conducted by Luminet in 1979~\cite{Luminet:1979nyg}. Subsequent research investigated the image of a black hole with spherical accretion, affirming the robustness of the shadow's characteristics~\cite{Falcke:1999pj}. Perlick and Tsupko et al. conducted a heuristic study on the shadow of a Schwarzschild black hole in an expanding universe solely driven by a positive cosmological constant, employing analytical methods.~\cite{Perlick:2018iye}. In recent years, extensive research has been conducted on black hole shadows, exploring their properties within various gravitational backgrounds~\cite{Gibbons:2008rj,Werner:2012rc,Jusufi:2018kry,Jusufi:2017mav,Crisnejo:2018uyn,Kumar:2018ple,Guo:2019lur,Gralla:2019drh, Jusufi:2020cpn,Kumar:2020owy,Zeng:2020dco,Gan:2021pwu,Guo:2021bwr,Jusufi:2020zln,Saurabh:2020zqg, Gralla:2019xty,Peng:2020wun,Chakhchi:2022fls,Guo:2021bhr,He:2022yse,Li:2021riw,Zeng:2021dlj,Zeng:2021mok,He:2021htq, Guo:2022rql, Guerrero:2021ues,Yan:2021ygy,Guerrero:2022qkh,Rosa:2022tfv,Atamurotov:2021hck,Kumar:2019ohr,Kumar:2017tdw,Hu:2022lek, Heydari-Fard:2023ent,Pulice:2023dqw,Yang:2022btw,Ma:2022jsy,Wang:2023rjl,Kumaran:2023brp,Huang:2023ilm,Zhang:2022osx,Hu:2023bzy}.

Astronomical observations have revealed that our universe is currently undergoing accelerated expansion, a phenomenon attributed to an unknown component known as dark energy which possesses negative pressure and positive energy density~\cite{SupernovaCosmologyProject:1998vns,SupernovaSearchTeam:1998fmf,SupernovaSearchTeam:1998cav}. One interpretation of the negative pressure was attributed to quintessence dark energy, characterized by the state equation $p=\omega\rho$, with the quintessence state parameter denoted as $\omega$ within the range $-1 < \omega < -1/3$~\cite{Wang:1999fa,Bahcall:1999xn}. The initial static and spherically symmetric black hole solution incorporating quintessence matter was deduced by Kiselev~\cite{Kiselev:2002dx}. Understanding the impact of quintessence dark energy on black hole shadows is a natural progression in this line of inquiry~\cite{Lacroix:2012nz,Haroon:2018ryd,Khan:2020ngg,Zeng:2020vsj,He:2021aeo,Heydari-Fard:2022jdu}. Besides, novel models that combine dark matter and dark energy emerged as potential candidates to explain the dark components of the universe. Among these unified dark fluid models, the Chaplygin gas and its related generalizations have gained significant attention in elucidating the observed accelerated expansion of the universe~\cite{Kamenshchik:2001cp,Bilic:2001cg,Bento:2002ps}. The Chaplygin has been found further applications in addressing the Hubble tension~\cite{Sengupta:2023yxh} and studying the growth of cosmological perturbations~\cite{Abdullah:2021tee}. Although the  Chaplygin gas is usually utilized in cosmological research to depict the evolution of the universe, it is important to note that its equation of state, $p=-\frac{B}{\rho}$, is not merely a phenomenological construct; rather, it can emerges naturally within the context of string theory~\cite{Ogawa:2000gj,Bordemann:1993ep,Jackiw:2000cc}. Recently, we obtained an analytical solution and investigated associated thermodynamic quantities for a charged static spherically-symmetric black hole surrounded by Chaplygin-like dark fluid (CDF) within the framework of Lovelock gravity theory~\cite{Li:2019lhr}. This model has been subsequently extended to incorporate the modified Chaplygin gas (MCG), with equation of state $p=A\rho-\frac{B}{\rho^{\beta}}$, to explore stability aspects of MCG-surrounded black holes in Einstein-Gauss-Bonnet~\cite{Li:2019ndh} and Lovelock~\cite{Li:2022csn} gravity theories. The investigation in Ref.~\cite{Li:2023zfl} focused on the thermodynamical phase transitions and critical behavior of static spherically-symmetric AdS black holes surrounded by CDF in the context of general relativity, exploring the thermodynamics-shadow correspondence. Lovelock black holes sourced by power-Yang-Mills field and CDF were studied in Ref.~\cite{Ali:2020omz}. Ref.~\cite{Ali:2024rrm} considered the CDF as a matter source and worked out the dimensionally continued hairy black hole solutions. In Ref.~\cite{Arora:2023mve}, the thermodynamics of a black hole surrounded by CDF was examined, taking into account the Bekenstein entropy and investigating the Joule-Thomson expansion. Ref.~\cite{Sekhmani:2023plr} provided a detailed study of the critical behavior of the charged AdS black holes with surrounding MCG. Ref.~\cite{Zhang:2024fxj} explored the critical behavior and Joule-Thomson expansion of charged AdS black holes surrounded by MCG.

In this paper, we will investigate the shadows and optical appearances of CDF black hole with different accretions. In particular, we study the geometrically thin and optically thin disk accretion and spherically symmetric accretion.  We expect that the shadow and optical appearance of the CDF black hole will impose constraints on the CDF model in the Universe from the observation of EHT. In many of the disk accretion flows models, the innermost stable circular orbit is of special importance, prompting us to investigate the geodesic structures around CDF black holes. Also we will demonstrate that for CDF black holes, due to the presence of a pseudo-cosmological horizon, the determination of the black hole shadow radius depends on the observer's location. In particular, by considering the asymptotic behavior of CDF black hole spacetime at infinity to be identical to the de Sitter spacetime governed by the cosmological constant responsible for our expanding universe, and assuming an static observer located far away from both event and cosmological horizons, we can use observational data on shadow radius to constrain parameters in the CDF model. When considering optical appearances of the CDF black hole, we assume a static observer located near the pseudo-cosmological horizon, this allows us to obtain a black hole optical image with impact parameter as a coordinate scale.

The structure of this paper is organized as follows: Section \ref{secmetric} provides a concise overview of the CDF black hole solutions. Section \ref{secgeodesic} examines the properties of timelike and lightlike geodesics. Section \ref{disk} presents the shadow images of the black hole with thin disk accretion, utilizing three toy emissivity profile models. In Section \ref{spherical}, we showcase the images of the black hole with spherically symmetric accretions, considering both static and infalling accretion modes, while varying the radial emissivity profile. Finally, Section \ref{conclusion} presents the conclusions and discussions.

\section{Static spherically-symmetric black holes immersed in CDF}
\label{secmetric}
In Ref.~\cite{Li:2023zfl}, we obtained BH solutions with the presence of both the cosmological constant and CDF. When the cosmological constant $\Lambda$ is a negative value less than $-\sqrt{B}$, the BH spacetime is asymptotically anti-de Sitter. In the present work, we suppose the cosmological constant is absent and only asymptotically de Sitter BH solution will be obtained.

For a spacetime that is static and spherically symmetric, we utilize the following metric form
\begin{equation}
ds^2=-f(r)dt^2+\frac{1}{g(r)} dr^2+r^2d\Omega^2,\label{dsf}
\end{equation}
where $f(r)$ and $g(r)$ represent general functions dependent on the radial coordinate $r$, and $d\Omega^2=d\theta^2+{\rm sin}^2\theta d\phi^2$ denotes the standard element on $S^2$.

The stress-energy tensor describing a perfect fluid is given by
\begin{equation}
T_{\mu\nu}= (\rho+p) u_\mu u_\nu +pg_{\mu\nu}, \label{Stress Energy Tensor}
\end{equation}
here $\rho$ and $p$ denote the energy density and isotropic pressure, respectively, as measured by an observer moving with the fluid. The four-velocity of the fluid is represented by $u_\mu$. Semiz~\cite{Semiz:2008ny} extensively studied static spherically-symmetric solutions of Einstein's equations for a perfect fluid source with various equations of state. An interesting scenario arises when considering the pressures of the fluid surrounding a black hole (BH). The cosmological fluid around a BH may exhibit anisotropy due to gravitational attraction near the central body. Kiselev~\cite{Kiselev:2002dx} introduced a BH spacetime model by treating the ambient quintessence matter as an anisotropic fluid. Despite the unclear identification of the generating mechanism for CDF, potential candidates exist in string theory and phenomenological cosmological studies~\cite{Li:2023zfl}.
Considering the presence of kinetic terms in these viable theories and the radial dependence of the essential field of CDF in static spherical symmetry, we propose the CDF to be anisotropic. The stress-energy tensor for CDF, in a covariant form~\cite{Raposo:2018rjn}, can be expressed as
\begin{equation}
T_{\mu\nu}= \rho u_\mu u_\nu +p_r k_\mu k_\nu +p_t \Pi_{\mu\nu}, \label{Stress Energy Tensor}
\end{equation}
where $p_r$ and $p_t$ represent the radial and tangential pressure, respectively, $u_\mu$ is the fluid four-velocity, and $k_\mu$ is a unit spacelike vector orthogonal to $u_\mu$. The vectors $u_\mu$ and $k_\mu$ satisfy $u_\mu u^{\mu}=-1$, $k_\mu k^\mu=1$, and $u^{\mu}k_\mu=0$. The projection tensor is defined as $\Pi_{\mu\nu} = g_{\mu\nu}+u_\mu u_\nu-k_\mu k_\nu$, projecting onto a two-surface orthogonal to $u^{\mu}$ and $k^{\mu}$. In the comoving frame of the fluid, the vectors are given by $u_\mu = (-\sqrt{f},0,0,0)$ and $k_\mu = (0,1/\sqrt{g},0,0)$. Consequently, the stress-energy tensor in Eq.~(\ref{Stress Energy Tensor}) can be reformulated as
\begin{equation}
T_{\mu}{}^{\nu}=-(\rho+p_t)\delta_{\mu}{}^0\delta^{\nu}{}_0 +p_t \delta_{\mu}{}^{\nu} +(p_r-p_t)\delta_{\mu}{}^1\delta^{\nu}{}_1. \label{Stress Energy Tensor2}
\end{equation}
The difference between radial and tangential pressures, $p_r-p_t$, is identified as the anisotropic factor. When $p_r=p_t$, the stress-energy tensor reduces to the standard isotropic form.

Now, we examine the behavior of a matter fluid across an event horizon, as characterized by the stress-energy formulation in Eq.~(\ref{Stress Energy Tensor2}). Within the horizon, where $g_{tt}>0$ and $g_{rr}<0$, the coordinate $r$ serves as time. Consequently, the energy density is given by $-{{T}_r}^r=-p_r$, and the pressure along the spatial $t$ direction is ${{T}_t}^t=-\rho$. To maintain continuity across the horizon, it is necessary for $p_r=-\rho$. If $p_r\neq-\rho$ and $\rho(r_h)\neq0$, the pressure must exhibit discontinuity at the horizon $r_h$, thus making the solution dynamical. In this study, the condition $p_r=-\rho$ is imposed to ensure the CDF remains static, and the energy density remains continuous across the horizon, placing a constraint on the solution.

For a cosmological fluid with an equation of state in the form $p=p(\rho)$, even when exhibiting anisotropy in the gravitational field generated by a BH, the equation of state should appear as $p=p(\rho)$ at a cosmological scale. This allows the tangential pressure $p_t$ to be constrained by taking an isotropic average over angles and requiring $\langle {T}_i{}^j\rangle=p(\rho)\delta_i{}^j$, expressed as
\begin{equation}
p_t+\frac{1}{3}(p_r-p_t)=p(\rho),\label{angles_average}
\end{equation}
utilizing the relation $\langle \delta_i{}^1\delta^j{}_1\rangle=\frac{1}{3}$. For quintessence matter with an equation of state $p =\omega\rho$ ($-1<\omega<-1/3$), the tangential pressure is deduced from Eq.~(\ref{angles_average}) as $p_t=\frac{1}{2}(1+3\omega)\rho$, compatible with this is radial pressure $p_r=-\rho$, in accordance with the results obtained by Kiselev~\cite{Kiselev:2002dx}.

In our given context, the CDF is characterized by a non-linear equation of state, expressed as $p=-\frac{B}{\rho}$, where $B$ is a positive constant. Ensuring $p_r=-\rho$, the tangential pressure is derived as $p_t=\frac{1}{2}\rho-\frac{3B}{2\rho}$. Consequently, the stress-energy tensor of the CDF is formulated as
\begin{equation}\label{Texpression}
{{T}_t}^t={{T}_r}^r=-\rho,\hspace{1cm} {{T}_{\theta}}^{\theta}={{T}_{\phi}}^{\phi}=\frac{1}{2}\rho-\frac{3B}{2\rho}.
\end{equation}
As we will demonstrate later, the anisotropy of the CDF diminishes, and the equation of state yields $p=-B/\rho$ at a cosmological scale.

Since we employ ${{T}_t}^t={{T}_r}^r$, without any loss of generality, the relationship between the metric components $g(r)=f(r)$ can be achieved through an appropriate rescaling of time. Subsequently, the components of the Einstein tensor are given by
\begin{equation}\label{Gexpression}
{{G}_t}^t={{G}_r}^r=\frac{1}{r^2}(f+rf'-1),\hspace{1cm} {{G}_{\theta}}^{\theta}={{G}_{\phi}}^{\phi}=\frac{1}{2r}(2f'+rf'').
\end{equation}

Combining Eqs.~(\ref{Texpression}) and~(\ref{Gexpression}), gravitational equations are obtained as
\begin{equation}\label{graviequations}
\frac{1}{r^2}(f+rf'-1)=-\rho,\hspace{1cm} \frac{1}{2r}(2f'+rf'')=\frac{1}{2}\rho-\frac{3B}{2\rho}.
\end{equation}
Thus, there are two unknown functions, $f(r)$ and $\rho(r)$, which can be analytically determined by the aforementioned differential equations. Solving the set of differential equations (\ref{graviequations}), one can first obtain the solution for the energy density of CDF
\begin{equation}
\rho(r)=\sqrt{B+\frac{q^2}{r^6}}, \label{CDFenergydensity}
\end{equation}
where $q>0$ is a normalization factor representing the intensity of the CDF. Additionally, Eq.~(\ref{CDFenergydensity}) is a direct consequence of the conservation law for the stress-energy tensor $\nabla_{\nu}T^{\mu\nu}=0$. Notably, for small radial coordinates (i.e., $r^6\ll q^2/B$), the CDF energy density is approximated by
\begin{equation}
\rho(r)\approx\frac{q}{r^3}, \label{CDFenergydensitysmallr}
\end{equation}
indicating that the CDF behaves like a matter content with an energy density varying as $r^{-3}$. For large radial coordinates (i.e., $r^6\gg q^2/B$), it follows that
\begin{equation}
\rho(r)\approx \sqrt{B}, \label{CDFenergydensitylarger}
\end{equation}
suggesting that the CDF acts as a positive cosmological constant at a large-scale regime. It is also observed that $p_r\rightarrow-\sqrt{B}$ and $p_{\theta,\phi}\rightarrow-\sqrt{B}$ as $r\rightarrow\infty$, indicating that the CDF appears isotropic, and its equation of state reverts to $p=-B/\rho$ at a cosmological scale. It is noteworthy that for a cosmological fluid with a general equation of state $p=p(\rho)$, whose radial pressure satisfies $p_r=-\rho$ when surrounding a central BH, it tends to be isotropic at a cosmological scale.
\begin{figure}[htbp]
  \centering
  \begin{minipage}{0.45\textwidth}
    \includegraphics[width=\linewidth]{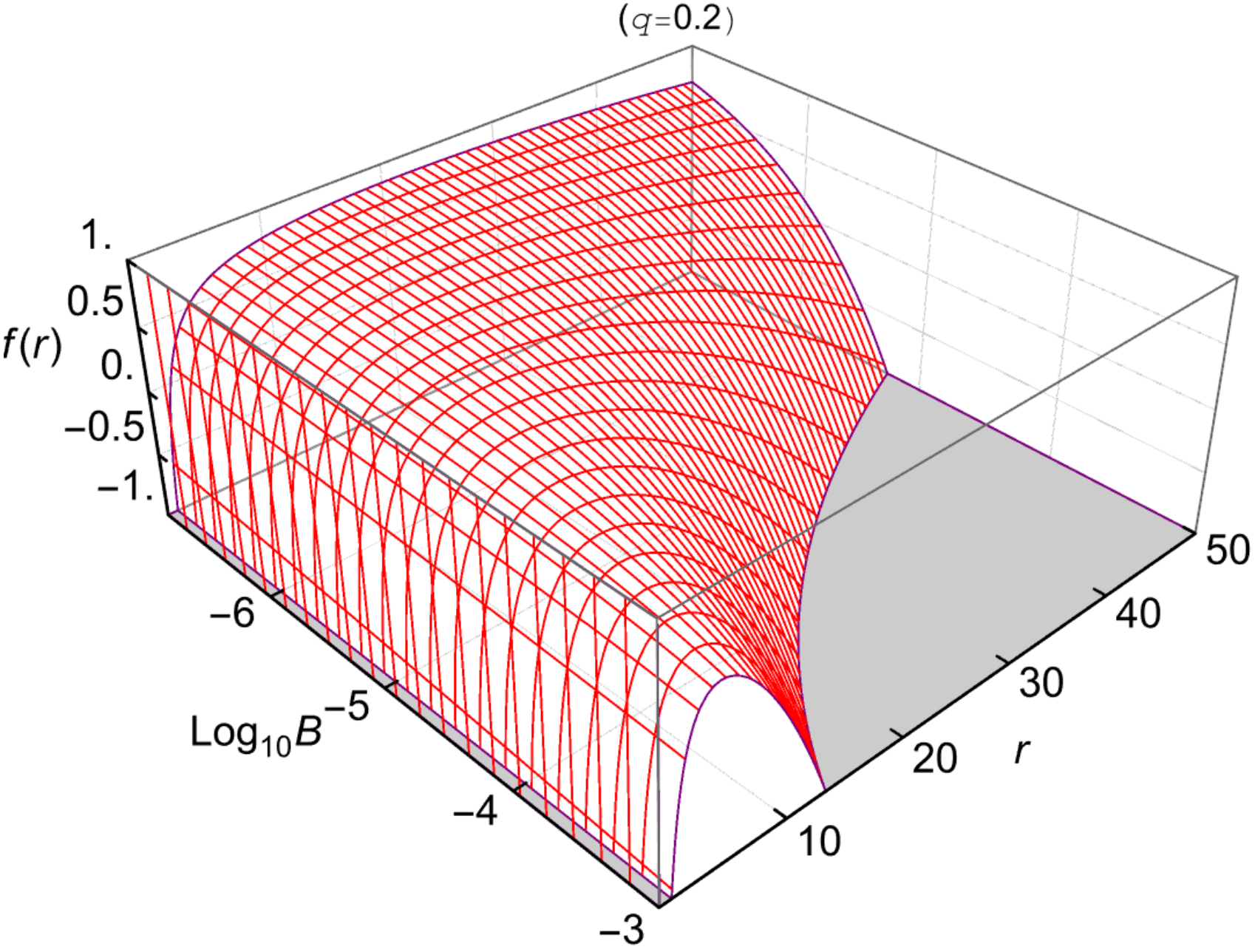}
  \end{minipage}
  \begin{minipage}{0.45\textwidth}
    \includegraphics[width=\linewidth]{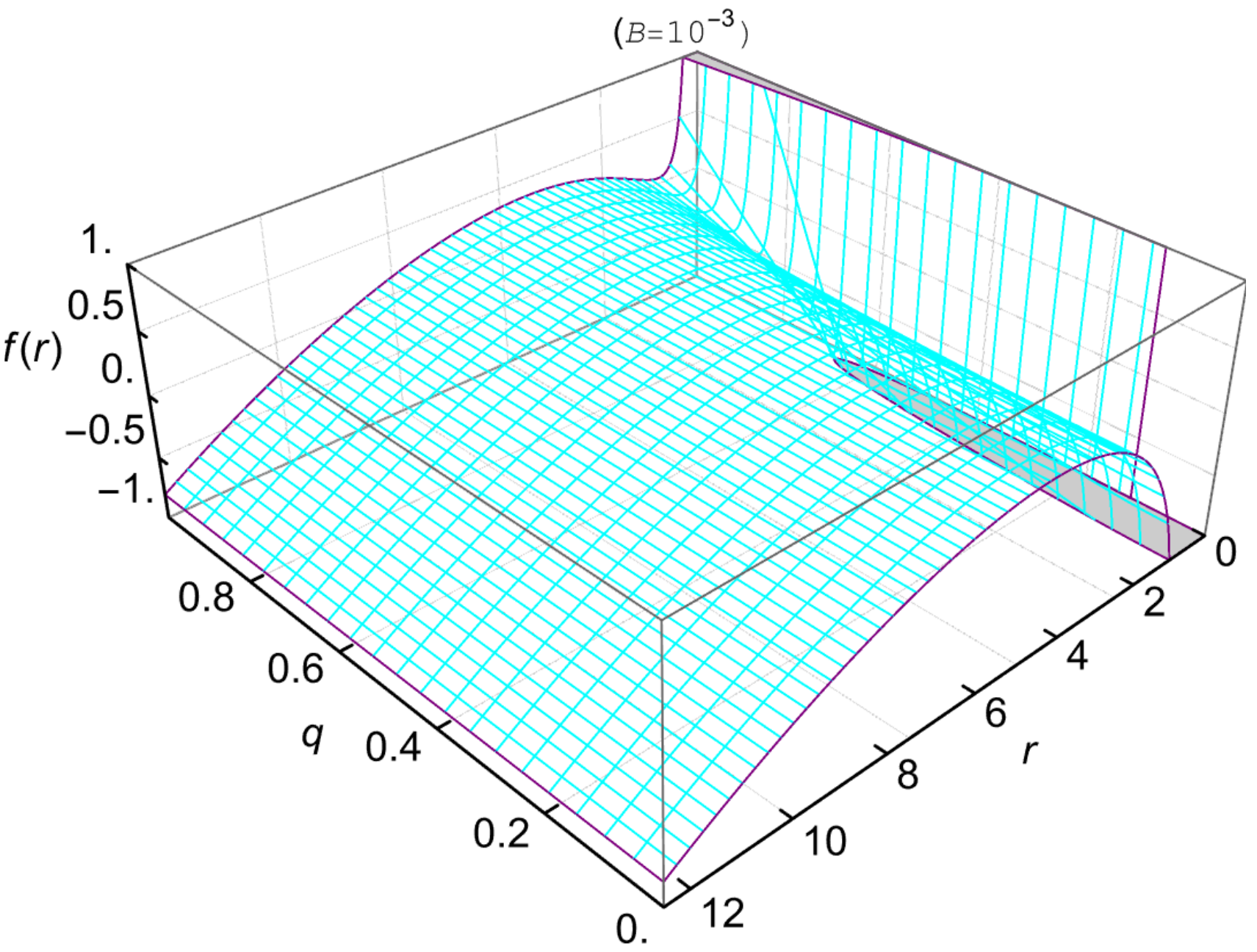}
  \end{minipage}
  \caption{\label{figfr} \textbf{Left panel:} The $f(r)$ function with varying $B$ for $q = 0.2$. \textbf{Right panel:} The $f(r)$ function with varying $q$ for $B = 10^{-3}$. In both panels, the point mass of the BH has been set as $M=1$.}
\end{figure}

Substituting Eq.~(\ref{CDFenergydensity}) into Eq.~(\ref{graviequations}), we derive the analytical solution for $f(r)$
\begin{equation}
f(r)=1-\frac{2M}{r}-\frac{r^2}{3}\sqrt{B+\frac{q^2}{r^6}}+\frac{q}{3r}{\rm ArcSinh}\frac{q}{\sqrt{B}r^3}, \label{ffr}
\end{equation}
where $M$ denotes the mass of the BH, here we consider the BH as a point mass BH, thus $M$ arises as a constant. Hereafter, the BH represented by Eq.~(\ref{ffr}) will be referred to as CDF-BH. To study the asymptotic behavior of $f(r)$, we take $r\rightarrow\infty$ and find that
\begin{equation}
f(r)\rightarrow 1-\frac{r^2}{3}\sqrt{B},
\label{frasymp}
\end{equation}
which reveals that, the spacetime described by the function $f(r)$ in Eq.~(\ref{ffr}) exhibits asymptotic de Sitter-like characteristics, thereby endowing the BH with both an event horizon $r_{h}$ and a pseudo-cosmological horizon $r_{c}$. The effects of the parameters $B$ and $q$ on $f(r)$ are shown in Fig.~\ref{figfr}. By examining Fig.~\ref{figfr}, the following conclusions can be drawn: the existence and location of the event horizon are primarily governed by the parameter $q$, while the position of the pseudo-cosmological horizon is mainly determined by the parameter $B$. The region between the two horizons is called the domain of outer communication \cite{Friedman:1993ty,Chrusciel:1994tr}, since any two observers in this region may communicate with each other without being hindered by a horizon.

\section{Geodesic structure around a static spherically-symmetric CDF-BH}
\label{secgeodesic}
In this section, we explore the presence and stability of both timelike and lightlike circular orbits for a CDF-BH within Einstein gravity. We employ the Euler-Lagrange equation given by
\begin{equation}
\frac{d}{ds}\left(\frac{\partial \mathcal{L}}{\partial \dot{x}^{\mu}}\right)=\frac{\partial \mathcal{L}}{\partial x^{\mu}},
\label{eleq}
\end{equation}
where $s$ is the affine parameter of the light trajectory, $\dot{}$ denotes the derivative with respect to $s$, $\dot{x}^{\mu}$ represents the four-velocity of the light ray, and $\mathcal{L}$ is the Lagrangian given by
\begin{equation}
 \mathcal{L}=\frac{1}{2}g_{\mu\nu}\dot{x}^{\mu}\dot{x}^{\nu}=\frac{1}{2}\left( - f(r)\dot{t}^2+\frac{\dot{r}^2}{f(r)}+r^2\left(\dot{\theta}^2+\sin^2{\theta}~ \dot{\phi}^2 \right)\right).
\label{laeq}
 \end{equation}
We also impose initial conditions $\theta = \pi/2$, $\dot{\theta} = 0$, indicating that the particle always moves in the equatorial plane. Additionally, as the metric coefficients do not explicitly depend on time $t$ and azimuthal angle $\phi$, there are two corresponding conserved quantities.

Combining Eqs.~(\ref{ffr}), (\ref{eleq}), and (\ref{laeq}), the time, azimuthal, and radial components of the four-velocity satisfy the following equations of motion:
\begin{eqnarray}
  &&\dot{t}=\frac{E}{1-\frac{2M}{r}-\frac{r^2}{3}\sqrt{B+\frac{q^2}{r^6}}+\frac{q}{3r}{\rm ArcSinh}\frac{q}{\sqrt{B}r^3}},  \label{time} \\
  &&\dot{\phi}=\frac{L}{r^2}, \label{psi} \\
  &&\dot{r}^2+\left(\delta+\frac{L^2}{r^2}\right) \left(1-\frac{2M}{r}-\frac{r^2}{3}\sqrt{B+\frac{q^2}{r^6}}+\frac{q}{3r}{\rm ArcSinh}\frac{q}{\sqrt{B}r^3}\right)=E^2,
 \label{radial}
\end{eqnarray}
where $E$ and $L$ are the conserved quantities corresponding to time and azimuthal direction, respectively, defined as
\begin{equation}\label{ELexpression}
E=-\frac{\partial \mathcal{L}}{\partial \dot{t}},\quad L=-\frac{\partial \mathcal{L}}{\partial \dot{\phi}},
\end{equation}
and
\begin{equation}
\delta=\left\{
               \begin{array}{ll}
                 1, & \hbox{for the timelike particle;} \\
                 0, & \hbox{for the lighlike particle.}
               \end{array}
             \right.
\end{equation}

Furthermore, Eq.~(\ref{radial}) can be expressed as
\begin{equation}
 \dot{r}^2+V_{eff}(r)=E^2,  \label{vbr}
\end{equation}
where
\begin{equation} \label{epotential}
V_{\rm eff}(r)=\left(\delta+\frac{L^2}{r^2}\right)\left(1-\frac{2M}{r}-\frac{r^2}{3}\sqrt{B+\frac{q^2}{r^6}}+\frac{q}{3r}{\rm ArcSinh}\frac{q}{\sqrt{B}r^3}\right),
\end{equation}
represents an effective potential.

\subsection{The structure of timelike geodesics}
\begin{figure}[h]
\centering 
\includegraphics[width=.48\textwidth]{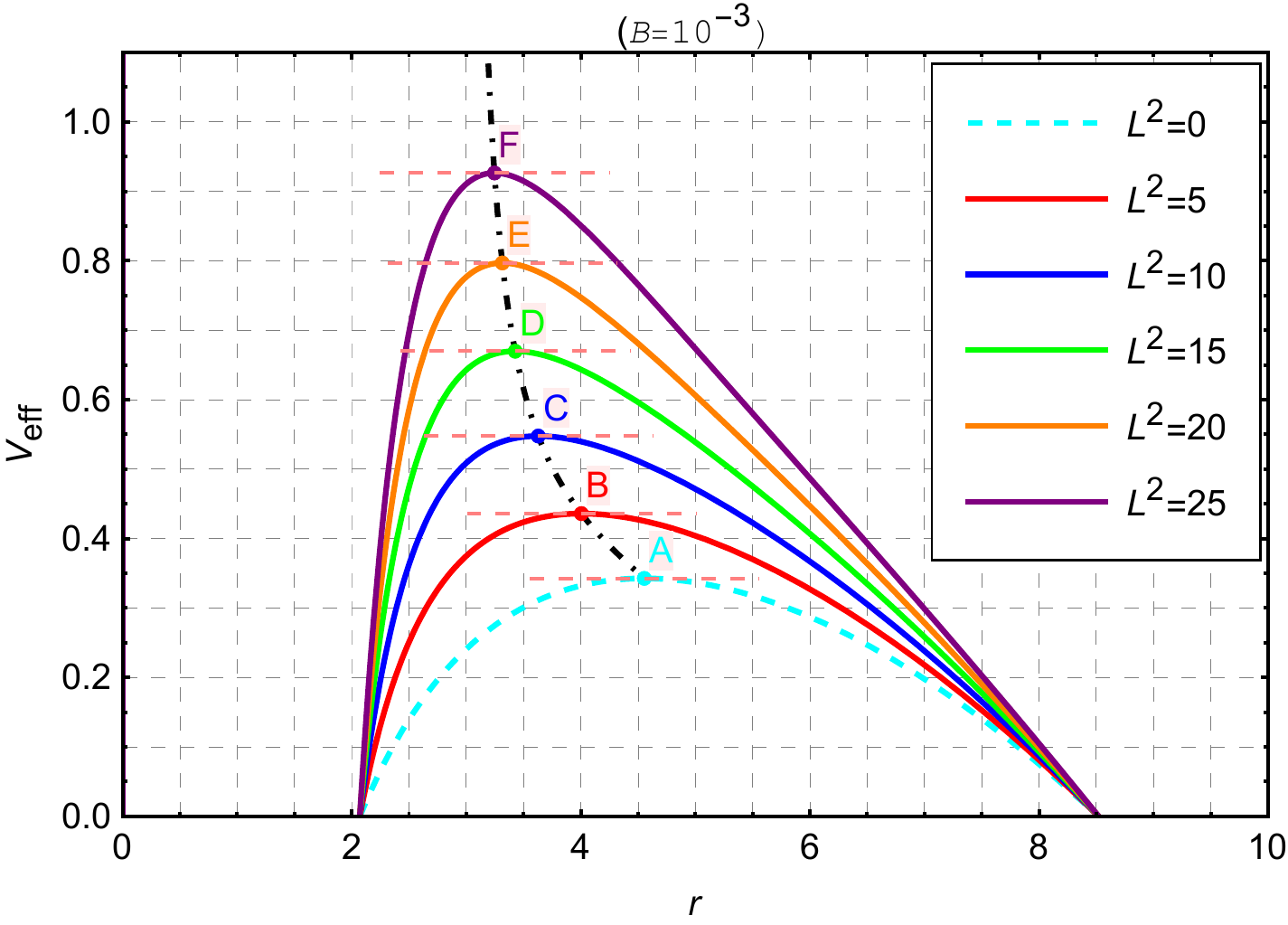}
\includegraphics[width=.49\textwidth]{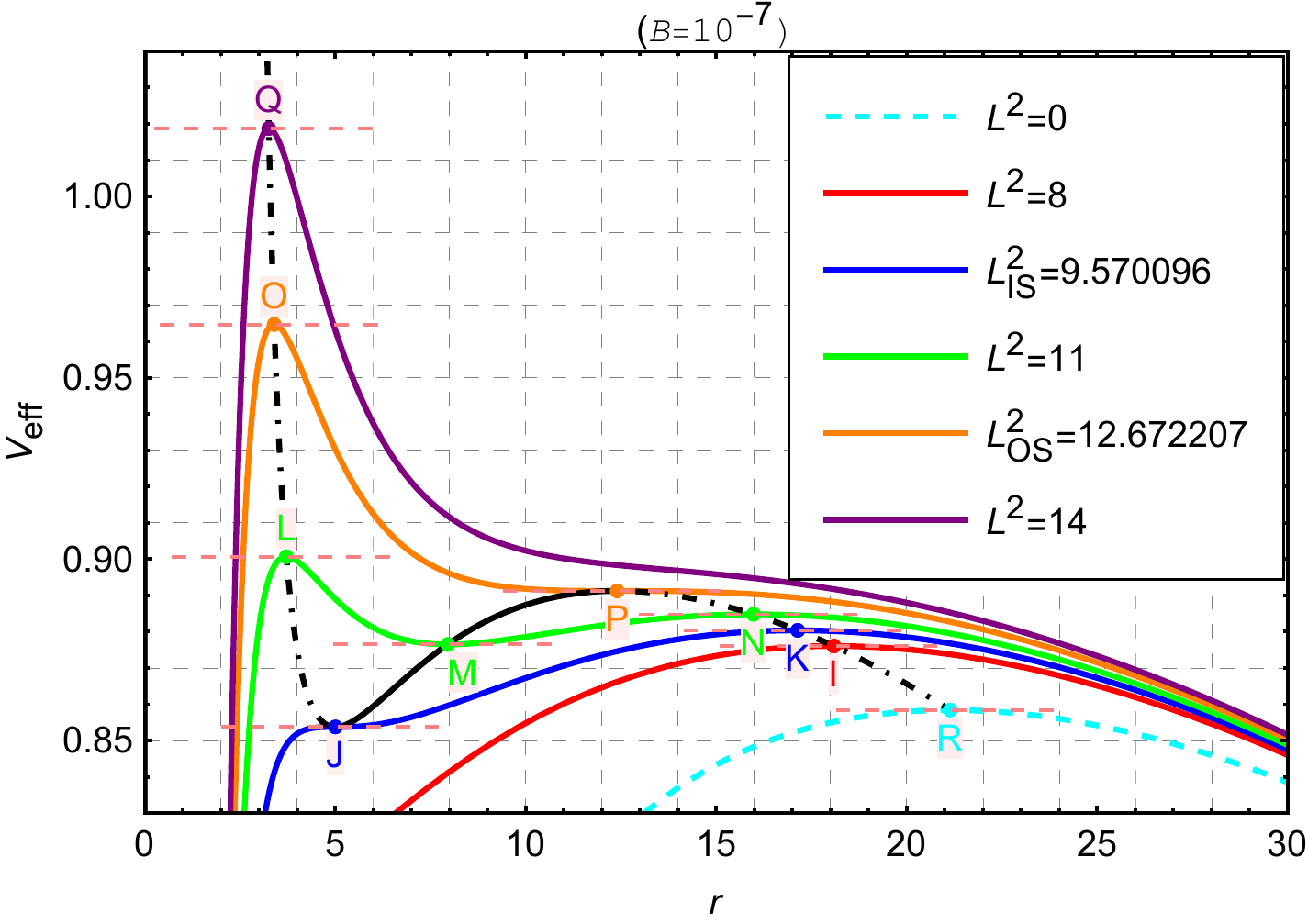}
\caption{\label{FigUeff} The profile of the effective potential curves of timelike particles for $B=10^{-3}$ (left) and $B=10^{-7}$ (right) with $M=1$, $q=0.2$. With varying $L^2$, the dot-dashed and solid black curves represents the unstable and stable circular orbit points, respectively. The cyan dashed curves correspond to $L^2=0$, which are unrealistic. The radial coordinates of points $A$, $J$, $P$ and $R$ are $r_A=4.554003$, $r_J=5.004919$, $r_P=12.399015$ and $r_R=21.137809$.}
\end{figure}
Examining the motion of timelike particles is crucial, since it may significantly impacts the profiles of the accretion matters. According to Eq.~(\ref{epotential}), for timelike particles, $\delta=1$, the effective potential is written as
\begin{equation}\label{EquUeff}
V_{\rm eff}(r)=\left(1+\frac{L^2}{r^2}\right)\left(1-\frac{2M}{r}-\frac{r^2}{3}\sqrt{B+\frac{q^2}{r^6}}+\frac{q}{3r}{\rm ArcSinh}\frac{q}{\sqrt{B}r^3}\right).
\end{equation}
The effective potential depends on the mass $M$, the angular momentum $L$, and the CDF parameters $B$ and $q$. By studying the potential energy functions corresponding to different parameter values, one finds that, from the perspective of timelike orbits, CDF-BHs can be divided into two types: those with stable orbits and those without. We will illustrate this with examples below. Fixing $M=1$ and $q=0.2$, the effective potential curves for various values of $L^2$ with $B=10^{-3}$ (left panel) and $B=10^{-7}$ (right panel) are shown in Fig.~\ref{FigUeff}. For $B=10^{-3}$, the effective potential curves corresponding to different values of $L^2$ all exhibit only one maximum point. This indicates that particles for $B=10^{-3}$ have only one unstable circular orbit (when $E^2$ equals the maximum value of the potential energy curve). For $B=10^{-7}$, the situation with the effective potential energy curves becomes slightly more complex, therefore we categorize the angular momentums as follows
\begin{itemize}
  \item For $0 < L^2 < L^2_{\rm IS}$, the effective potential curves have only one maximum point, indicating that particles in this range have only one unstable circular orbit (e.g. point I).
  \item For $L^2 =L^2_{\rm IS}$, particles have an unstable orbit (e.g. point K) and the Innermost Stable Circular Orbit (ISCO) (e.g. point J).
  \item For $L^2_{\rm IS}< L^2< L^2_{\rm OS} $, the curves exhibit one minimum point and two maximum points, indicating that particles with $L^2$ in this range have one Stable Circular Orbit (SCO) (e.g. point M) and two unstable circular orbits (e.g. points L and N).
  \item For $L^2 = L^2_{\rm OS}$, particles have an unstable orbit (e.g. point O) and the Outermost Stable Circular Orbit (OSCO) (e.g. point P).
  \item For $L^2 >L^2_{\rm OS}$, particles have only an unstable orbit (e.g. point Q).
\end{itemize}
\begin{figure}[h]
\centering 
\includegraphics[width=.485\textwidth]{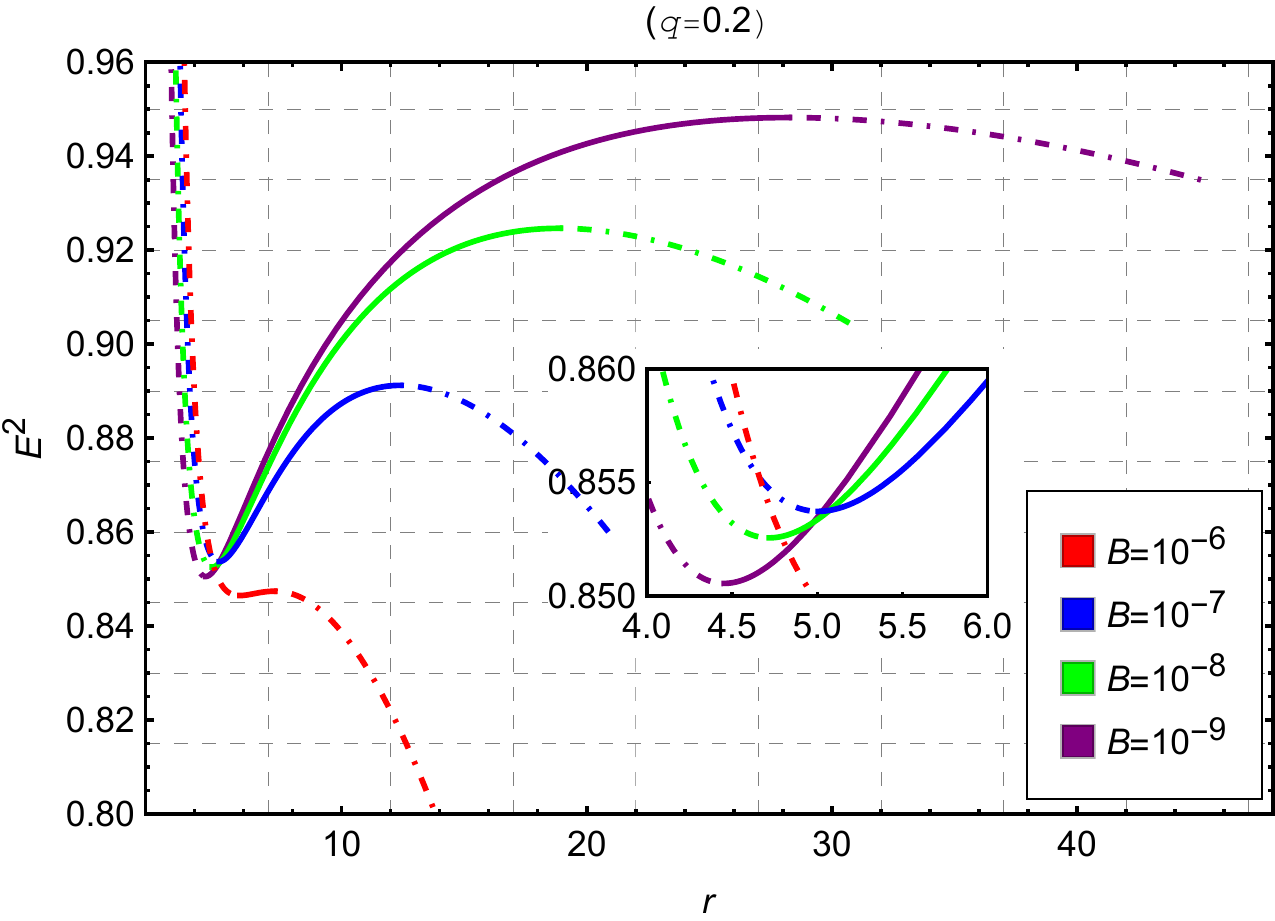}
\includegraphics[width=.485\textwidth]{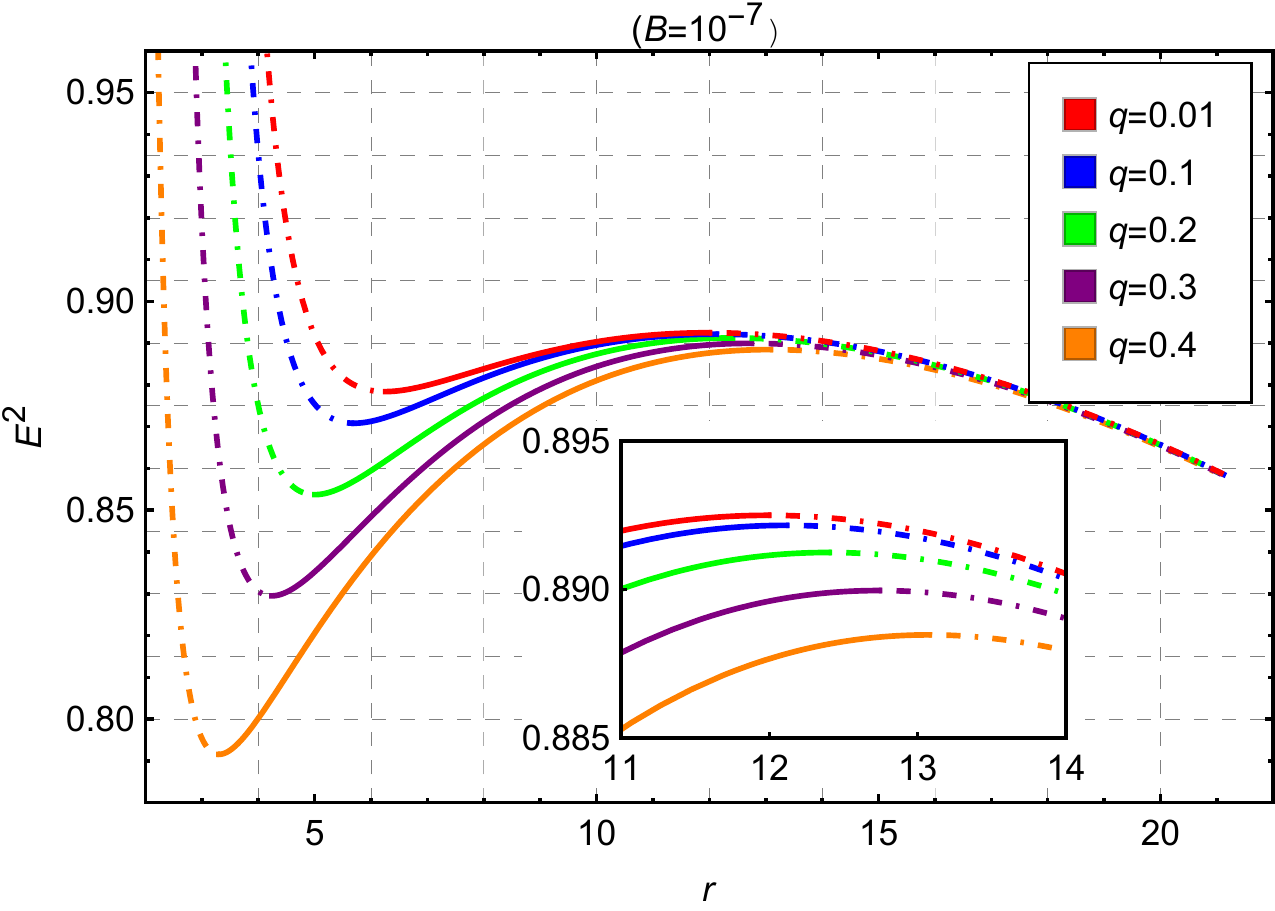}\\
\includegraphics[width=.485\textwidth]{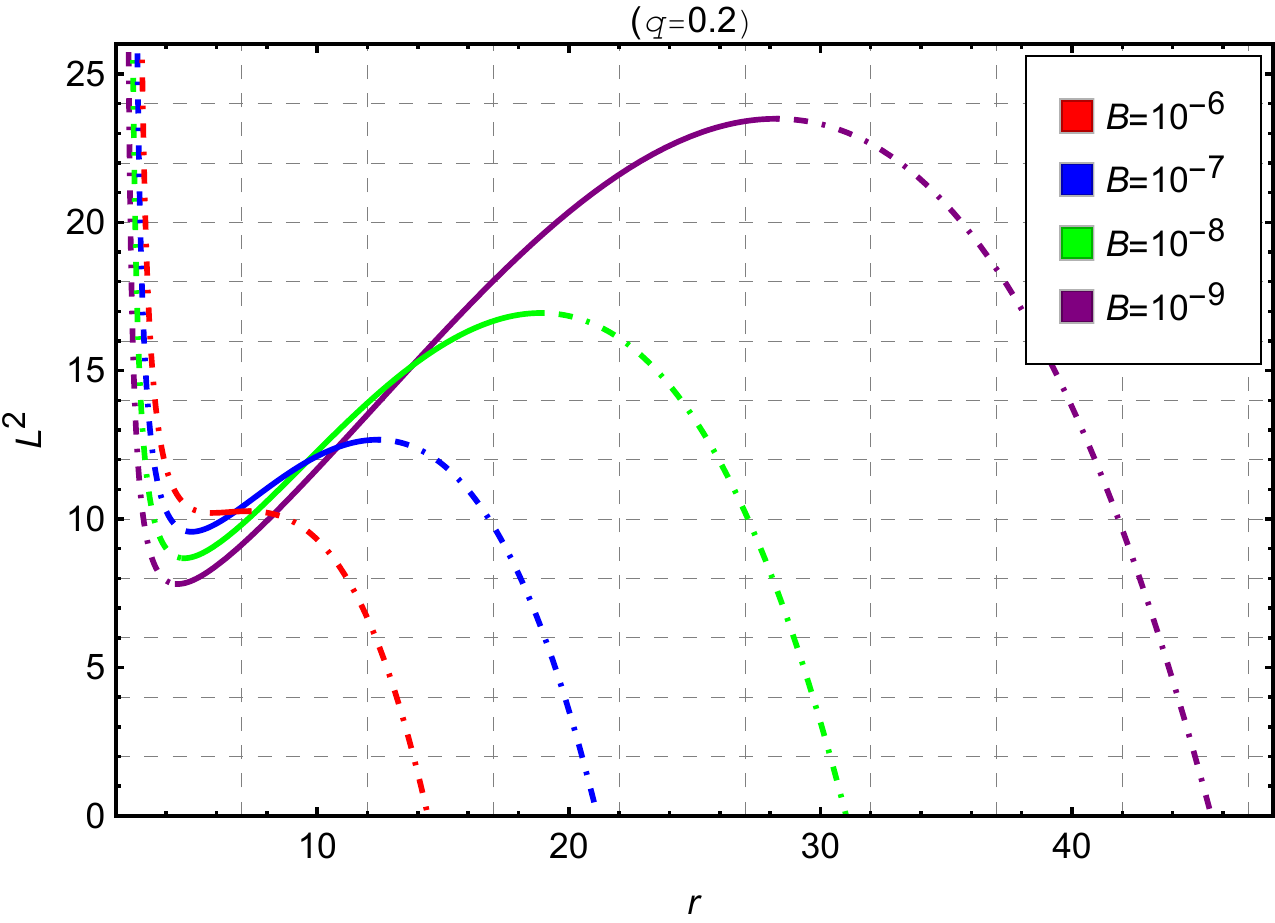}
\includegraphics[width=.485\textwidth]{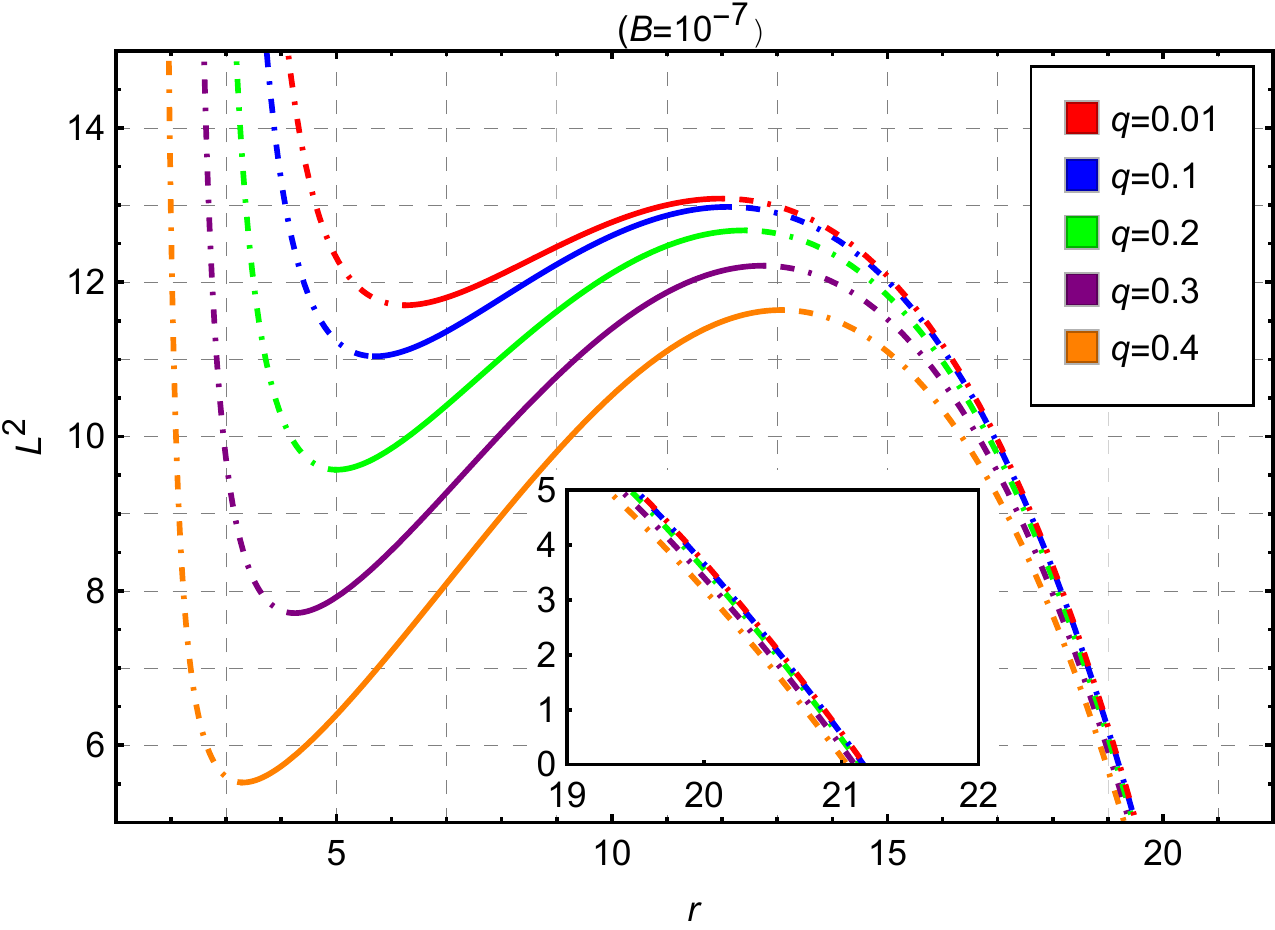}
\caption{\label{FigEELLvaryBq}The radial profiles of $E^2$ and $L^2$ of the circular orbits with varying values of $B$ (fixing $q=0.2$, left panels) and $q$ (fixing $B=10^{-7}$, right panels). The dotted-dashed and solid segments of a curve depict the positions of unstable and stable circular orbits, respectively. Here we have set the point mass of the CDF-BH as $M=1$. }
\end{figure}
\begin{table}[htbp]
\caption{The radii of the ISCO and OSCO with various CDF parameters, the dash `-' indicates the absence of stable circular orbits.}
\label{tabsco}
\vspace{2mm}
\centering
\scriptsize
\renewcommand{\arraystretch}{1.5}
\resizebox{\textwidth}{!}{
\begin{tabular}{c|c|cccc|c|ccc|c|c}
\toprule[\lightrulewidth]
\hline
$B$&$10^{-3}$ &\multicolumn{4}{c|}{$10^{-5}$} & \multicolumn{1}{c|}{$10^{-6}$}& \multicolumn{3}{c|}{$10^{-7}$}& \multicolumn{1}{c|}{$10^{-8}$}& \multicolumn{1}{c}{$10^{-9}$} \\
\hline
$q$ & $(0,0.9649)$ & 0.2& 0.4 & 0.5& 0.6 & 0.2 & 0.01 &0.2 &0.4 &0.2 &0.2 \\
\hline
$r_{\rm ISCO}$ & - & -& - & 4.50184& 3.14823& 5.84933 & 6.25026 &5.00492 &3.30513 & 4.71761 &4.45334 \\
\hline
$r_{\rm OSCO}$ & - & -& - & 4.83642& 5.67846& 7.25026 & 11.99247 &12.39901 &13.07427 & 18.8983 &28.2441\\
\hline
\bottomrule[\lightrulewidth]
\end{tabular}
}
\end{table}

The requirement $V_{\rm{eff}}(r)=V'_{\rm{eff}}(r)=0$ for circular orbits yields
\begin{equation}\label{ELexpressions}
E^2=\frac{2f(r)^2}{2f(r)-rf'(r)},~~\quad~~L^2=\frac{r^3f'(r)}{2f(r)-rf'(r)}.
\end{equation}
We show, respectively, the radial profiles of $E^2$ and $L^2$ of the circular orbits with varying values of $B$ and $q$ in Fig.~\ref{FigEELLvaryBq}. When the circular orbit radius approaches the photon sphere radius $r_{\rm ph}$, as defined later in Eq.~(\ref{Eqrph}), both $E^2$ and $L^2$ tend to infinity. Therefore, $r_{\rm ph}$ can be considered as the minimum cutoff radius for timelike circular orbits. As the circular orbit radius increases continuously, $L^2$ approaches 0, indicating the maximum cutoff radius for circular orbits. The SCOs can be obtained by solving the equations $V'_{\rm{eff}}(r)=0$  and $V''_{\rm{eff}}(r)=0$. Formally, the radii of SCOs satisfy
\begin{equation}\label{RISCO}
r_{\rm{SCO}}=-\frac{3f(r_{\rm{SCO}})f'(r_{\rm{SCO}})}{f(r_{\rm{SCO}})f''(r_{\rm{SCO}})-2f'(r_{\rm{SCO}})^2}.
\end{equation}
It can be observed from Fig.~\ref{FigUeff} that, $r_{\rm SCO}$ is a monotonic function of $E^2$ (as well as $L^2$, as shown in Fig.~\ref{FigEELLvaryBq}), thus the minimum and maximum values of $E^2$ for bound orbits correspond to the ISCO and OSCO, respectively. If Eq.~(\ref{RISCO}) has no solution, it indicates the absence of SCO, which is denoted by `-' in Table~\ref{tabsco}. From Table~\ref{tabsco}, we observe that, for a fixed $q$ value, as $B$ increases, the radius of the ISCO increases, while the radius of the OSCO decreases, indicating a reduction in the radial span of stable circular orbits. For a fixed $B$ value, as $q$ increases, the radius of the ISCO decreases, and the radius of the OSCO increases, signifying an expansion in the radial span of stable circular orbits.

\begin{figure}[h]
\centering
\includegraphics[width=.49\textwidth]{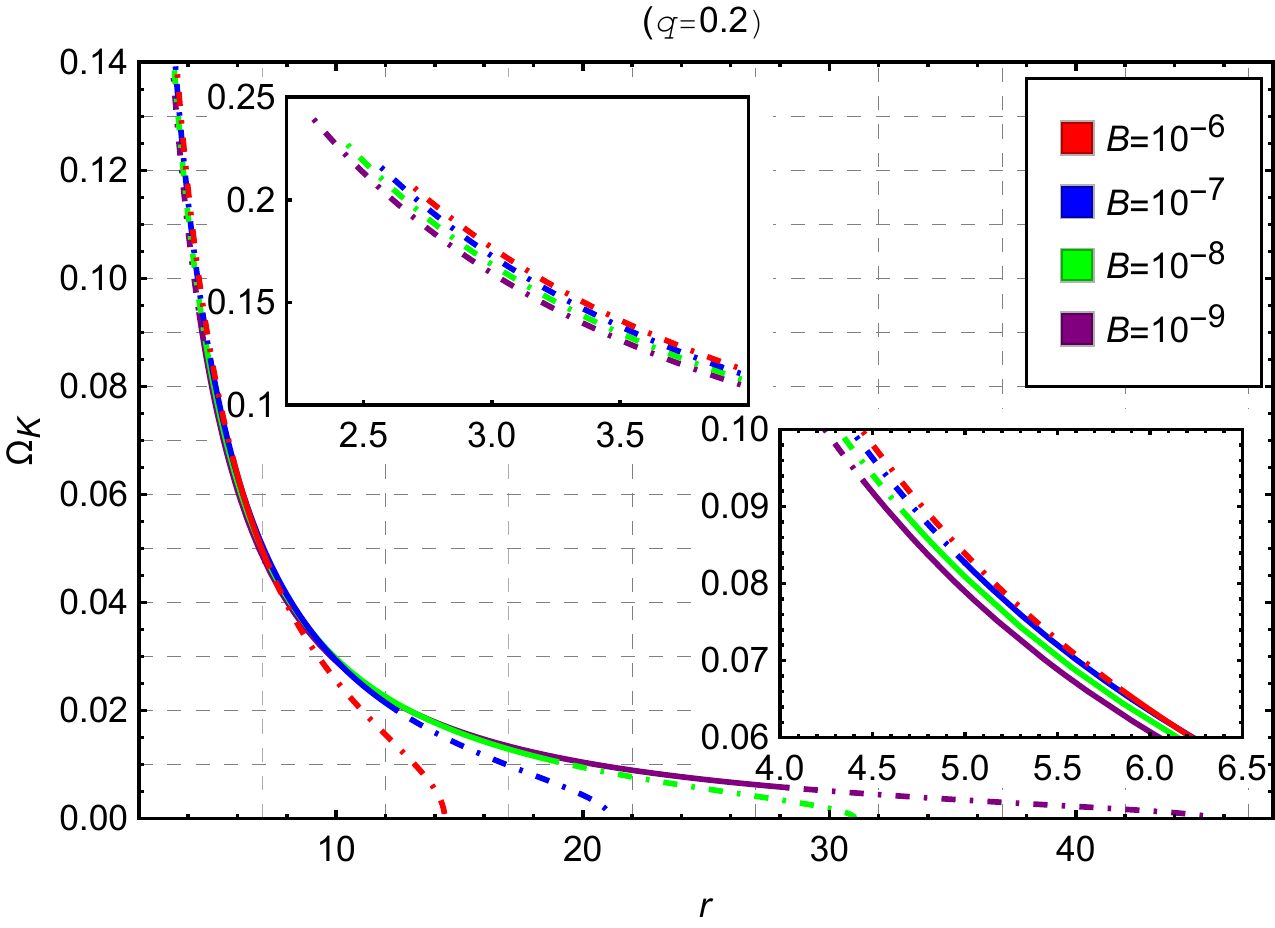}
\includegraphics[width=.49\textwidth]{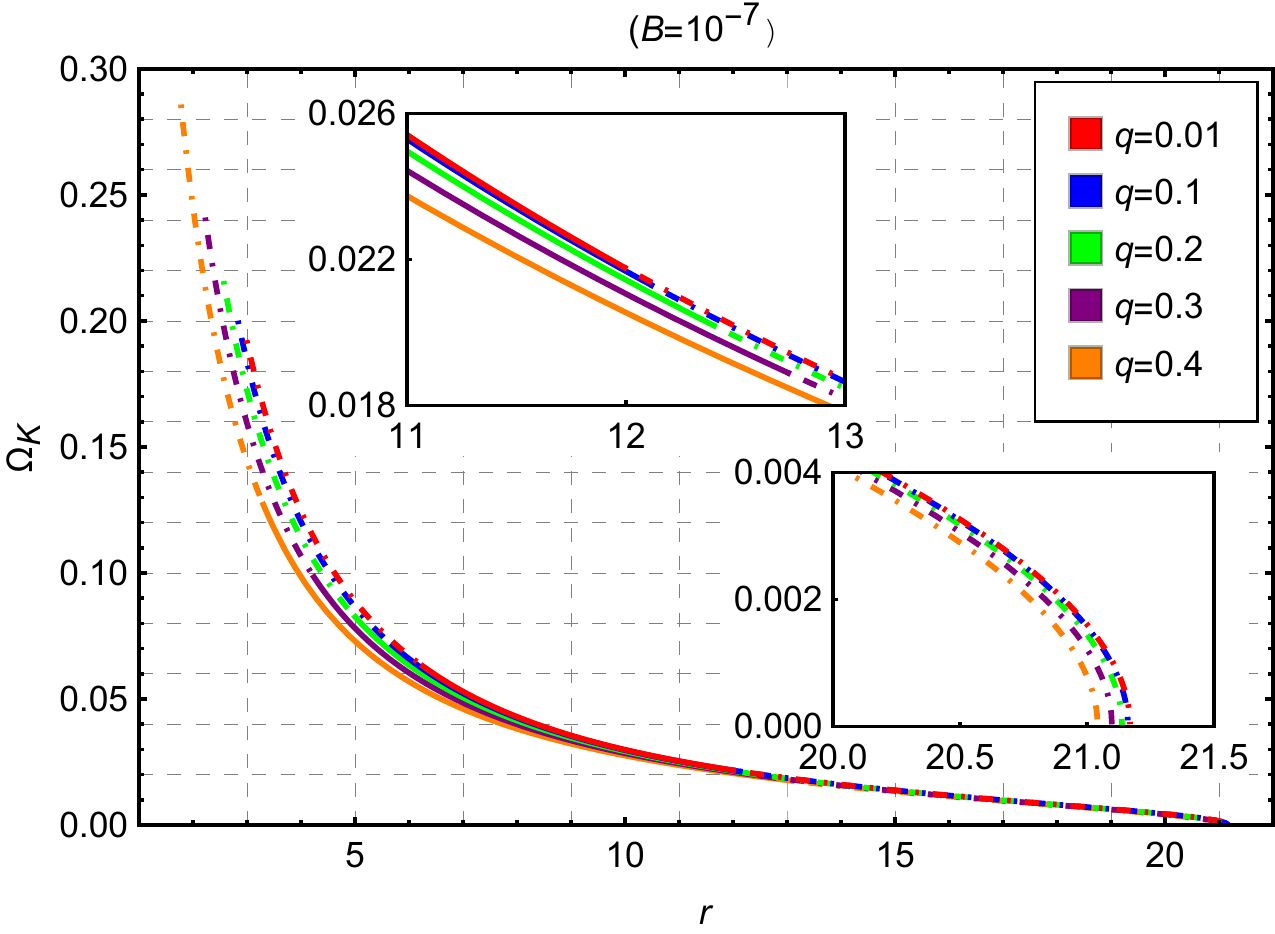}
\caption{\label{FigOKwithBq}Radial dependence of Keplerian frequencies of test particles around a CDF-BH for different values of $B$ (fixing $q=0.2$, left panel) and $q$ (fixing $B=10^{-7}$, right panel). The dotted-dashed and solid segments of a curve depict the positions of unstable and stable circular orbits, respectively. The point mass of the BH has been set as $M=1$.}
\end{figure}

The angular velocity of a particle orbiting a BH measured by an observer located at infinity, known as the Keplerian frequency, is defined by
\begin{equation}\label{Omigaexpression}
\Omega_{\rm{K}}=\frac{d\phi}{dt}\equiv\frac{\dot{\phi}}{\dot{t}}\Rightarrow\Omega_{\rm{K}}^2=\frac{f'(r)}{2r}.
\end{equation}
The radial dependence of the Keplerian frequencies for test particles around a CDF-BH is shown in Fig.~\ref{FigOKwithBq}. From Fig.~\ref{FigOKwithBq}, one can observe that $\Omega_{\rm K}(r)$ is a monotonically decreasing function of $r$. As $q$ increases, circular orbits at the same radius exhibit smaller Keplerian frequencies. The influence of $B$ on $\Omega_{\rm K}(r)$ is segmented: in the proximity of $r_{\rm ph}$, larger $B$ leads to a higher orbital frequency; at the outer regions, larger $B$ results in a lower orbital frequency.

The orbit properties of timelike particles can also be studied by examining the oscillatory motions and the epicyclic frequencies. In order to dertermine the radial locations where circular equatorial motion is either stable or unstable in the radial or vertical directions, one can calculate the radial and vertical epicyclic frequencies $\Omega_r$ and $\Omega_{\theta}$ as follows. According to Eqs.~(\ref{time}) and (\ref{radial}), radial and vertical motions around a circular equatorial orbit are governed by the equations
\begin{eqnarray}
  &&\frac{1}{2}\left(\frac{dr}{dt}\right)^2=-\frac{1}{2}\frac{f(r)^3}{E^2}\left[1-\frac{E^2}{f(r)}+\frac{L^2}{r^2\sin^2{\theta}}\right]\equiv V_{\rm{eff}}^{(r)},  \label{radialeom} \\
  &&\frac{1}{2}\left(\frac{d\theta}{dt}\right)^2=-\frac{1}{2}\frac{f(r)^2}{r^2E^2}\left[1-\frac{E^2}{f(r)}+\frac{L^2}{r^2\sin^2{\theta}}\right]\equiv V_{\rm{eff}}^{(\theta)},
 \label{verticaleom}
\end{eqnarray}
where the factor $\sin^2{\theta}$ is recovered to study the orbital perturbations. We then introduce small perturbations $\delta r$ and $\delta \theta$, and take the coordinate time derivative of Eqs.~(\ref{radialeom}) and (\ref{verticaleom}), which yields
\begin{equation}\label{Eqperturbation}
\frac{d^2(\delta r)}{dt^2}=\frac{d^2V_{\rm{eff}}^{(r)}}{dr^2}\delta r,~~\quad~~\frac{d^2(\delta \theta)}{dt^2}=\frac{d^2V_{\rm{eff}}^{(\theta)}}{d\theta^2}\delta \theta.
\end{equation}
The radial and vertical epicyclic frequencies can be derived as
\begin{equation}\label{epifrequncies}
\Omega_r^2=\frac{d^2V_{\rm{eff}}^{(r)}}{dr^2},~~\quad~~\Omega_\theta^2=\frac{d^2V_{\rm{eff}}^{(\theta)}}{d\theta^2}.
\end{equation}
By combining Eqs.~(\ref{ELexpressions}), (\ref{radialeom}) and (\ref{verticaleom}), one can obtain the explicit expressions for epicyclic frequencies
\begin{equation}\label{Omigaexpressions}
\Omega_r^2=-f'(r)^2+\frac{f(r)}{2r}\left[3f'(r)+rf''(r)\right],~~\quad~~\Omega_\theta^2=\frac{f'(r)}{2r}.
\end{equation}
\begin{figure}[tbp]
\centering 
\includegraphics[width=.48\textwidth]{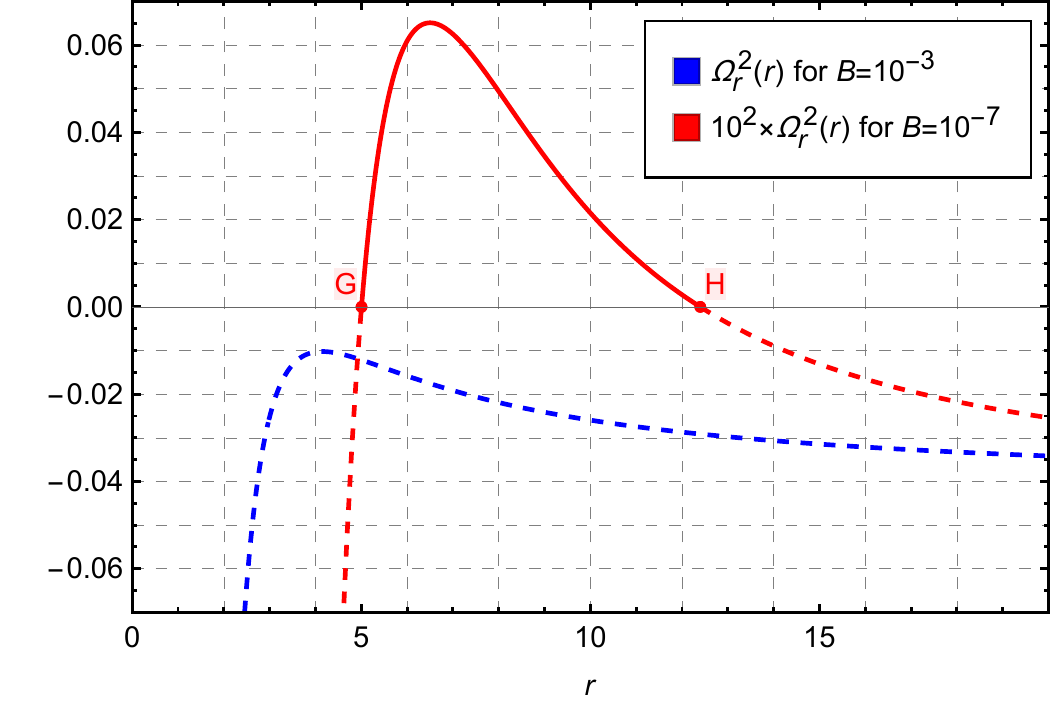}
\includegraphics[width=.49\textwidth]{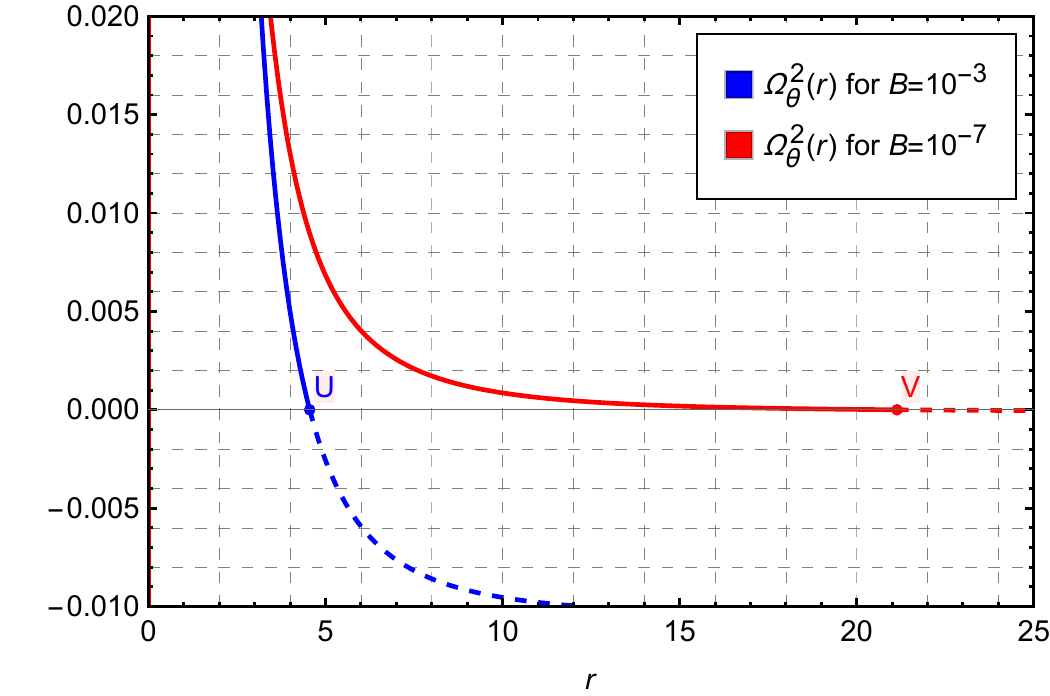}
\caption{\label{Figomiga2}  Radial (left) and vertical (right) epicyclic frequencies for $B=10^{-3}$ and $B=10^{-7}$ with $M=1$, $q=0.2$. The radial coordinates of points $\mathrm{G}$, $\mathrm{H}$, $\mathrm{U}$ and $\mathrm{V}$ are $r_\mathrm{G}=5.004919$, $r_\mathrm{H}=12.399015$, $r_\mathrm{U}=4.554003$ and $r_\mathrm{V}=21.137809$.}
\end{figure}Fig.~\ref{Figomiga2} reflects the radial dependence of radial epicyclic frequencies on the left panel and vertical epicyclic frequencies on the right panel, of test particles around the CDF-BH for $B=10^{-3}$ and $B=10^{-7}$. From Fig.~\ref{Figomiga2}, it can be seen that $\Omega_r^2(r)$ keeps negative as the radial coordinate varies for $B=10^{-3}$, indicating that the particles have no stable circular orbit. This is consistent with the conclusion we obtained from the left panel of Fig.~\ref{FigUeff}. $\Omega_{\theta}^2(r)$ reaches to zero at $r_\mathrm{G}=5.004919$, which is exactly the radius $r_\mathrm{A}$ in the left panel of Fig.~\ref{FigUeff} at which the angular momentum $L$ vanishes. For $B=10^{-7}$, $\Omega_r^2(r)$ is positive at the interval $r_\mathrm{G}<r<r_\mathrm{H}$, indicating the radial stability of the circular equatorial motion at this interval. The right panel of Fig.~\ref{Figomiga2} shows that the circular equatorial orbits at the interval $r_\mathrm{G}<r<r_\mathrm{H}$ are also vertically stable, observing that $r_\mathrm{H}<r_\mathrm{V}$ with $r_\mathrm{V}$ the point at which $\Omega_{\theta}^2(r)$ reaches to zero (corresponding to the radius $r_\mathrm{R}$ in the right panel of Fig.~\ref{FigUeff} at which the angular momentum $L$ vanishes). One can also observe that, $r_\mathrm{G}=r_\mathrm{J}$ and $r_\mathrm{H}=r_\mathrm{P}$, thus the points $\mathrm{G}$ and $\mathrm{H}$ represents the ISCO and OSCO, respectively.

\subsection{The structure of lightlike geodesics}
According to Eq.~(\ref{epotential}), for lightlike particles, $\delta=0$, the effective potential is written as
\begin{equation}\label{EquUefflight}
V_{\rm eff}(r)=\frac{L^2}{r^2}\left(1-\frac{2M}{r}-\frac{r^2}{3}\sqrt{B+\frac{q^2}{r^6}}+\frac{q}{3r}{\rm ArcSinh}\frac{q}{\sqrt{B}r^3}\right).
\end{equation}
At the photon sphere, the motion of the light ray satisfies $\dot{r}=0$ and $\ddot{r}=0$, implying
\begin{eqnarray}
V_{\rm{eff}}(r)=E_{\rm{ph}}^2, ~~~ V_{\rm{eff}}^{'}(r) = 0,
\label{condition1}
\end{eqnarray}
where the prime $'$ denotes the derivative with respect to the radial coordinate $r$. Considering Eq.~(\ref{EquUeff}), the radial coordinate of the photon sphere is determined by
\begin{equation}\label{Eqrph}
f(r_{\rm{ph}})-\frac{1}{2}r_{\rm{ph}}f'(r_{\rm{ph}})=0.
\end{equation}
Based on Eq.~(\ref{Eqrph}) and with the help of Eq.~(\ref{ELexpressions}), we can obtain the radius $r_{\rm{ph}}$ and impact parameter $b_{\rm{ph}}={|L_{\rm{ph}}|}/{E_{\rm{ph}}}$ of the photon sphere for different values of $B$ and $q$. Generally, obtaining analytic results for the radius and impact parameter is challenging, so we resort to numerical methods for their determination. The numerical results of radii and impact parameters of the photon sphere, as well as the event horizons and pseudo-cosmological horizons, are listed in Table \ref{tab:physical_quantities}. From this table, it can be observed that as the value of $B$ increases, the event horizon $r_{\rm{h}}$, the radius $r_{\rm{ph}}$, and impact parameter $b_{\rm{ph}}$ of the photon sphere also increase, while the pseudo-cosmological horizon $r_{\rm{c}}$ decreases. The parameter $q$ affects the associated physical quantities in the opposite direction to that of $B$. As the value of $q$ increases, the values of $r_{\rm{h}}$, $r_{\rm{ph}}$, and $b_{\rm{ph}}$ decrease, while the value of $r_{\rm{c}}$ increases. Consequently, an increase in the parameter $B$ results in a narrower outer communication region, while an increase in $q$ expands the outer communication region.

\begin{table}[htbp]
\caption{The values of radius $r_{\rm{ph}}$ and impact parameter $b_{\rm{ph}}$ of the photon sphere, the event horizon $r_{\rm{h}}$ and pseudo-cosmological horizon $r_\mathrm{c}$, as well as the rays classification parameters $b_m^{\pm}$ with varying $B$ and $q$ for $M=1.0$.}
\label{tab:physical_quantities}
\vspace{2mm}
\centering
\scriptsize 
\renewcommand{\arraystretch}{1.5} 
\begin{tabularx}{\textwidth}{c|c|*{10}{X}}
\toprule[\lightrulewidth]
\hline
$B$ & $q$ & \multicolumn{1}{c}{$r_{\mathrm{h}}$} & \multicolumn{1}{c}{$r_{\mathrm{c}}$} & \multicolumn{1}{c}{$r_{\mathrm{ph}}$} & \multicolumn{1}{c}{$b_{\mathrm{ph}}$} & \multicolumn{1}{c}{$b_1^-$} & \multicolumn{1}{c}{$b_2^-$} & \multicolumn{1}{c}{$b_2^+$} & \multicolumn{1}{c}{$b_3^-$} & \multicolumn{1}{c}{$b_3^+$} \\
\hline
\multirow{4}{*}{$\begin{array}{c} 10^{-3}\end{array}$} & $10^{-3}$ & 2.09723 & 8.52058 & 3.00000 & 6.14340 & 3.77645 & 6.00058 & 6.48810 & 6.13704 & 6.15765 \\
& $10^{-2}$ & 2.09717 & 8.52058 & 2.99994 & 6.14332 &  3.77632 & 6.00048 & 6.48803 & 6.13696 & 6.15757 \\
& 0.2 & 2.07073 & 8.52084 & 2.97623 & 6.10951 & 3.72433 & 5.95982 & 6.46324 & 6.10259 & 6.12466 \\
& 0.4 & 1.99026 & 8.52163 & 2.90048 & 6.00309 & 3.56588 & 5.83074 & 6.38686 & 5.99411 & 6.02145 \\
\hline
\multirow{3}{*}{$\begin{array}{c}10^{-4}\\10^{-5}\\10^{-6}\end{array}$} & \multirow{3}{*}{$\begin{array}{c} 0.2\end{array}$} & 1.95737 & 16.21745 & 2.92692 & 5.37247 & 3.09929 & 5.20594 & 5.88608 & 5.36413 & 5.39462 \\
&  & 1.86928 & 29.74732 & 2.82444 & 5.08701 & 2.81652 & 4.90003 & 5.79038 & 5.07696 & 5.11743 \\
&  & 1.78290 & 53.74347 & 2.69841 & 4.85819 & 2.62476 & 4.66285 & 5.71446 & 4.84741 & 4.89368 \\
\hline
\multirow{4}{*}{$\begin{array}{c} 10^{-7} \end{array}$} & $10^{-3}$ & 2.00078 & 96.38454 & 2.99994 & 5.20351 & 2.89373 & 5.03064 & 6.03561 & 5.19558 & 5.23211 \\
& $10^{-2}$ & 1.99648 & 96.38454 & 2.99499 & 5.19888 & 2.88805 & 5.02500 & 6.03334 & 5.19085 & 5.22773 \\
& 0.2 & 1.69616 & 96.38455 & 2.56873 & 4.63326 & 2.46830 & 4.43750 & 5.57041 & 4.62230 & 4.67088 \\
& 0.4 & 1.14147 & 96.38459 & 1.77817 & 3.49840 & 1.68256 & 3.25723 & 4.60648 & 3.47784 & 3.55956 \\
\hline
\bottomrule[\lightrulewidth] 
\end{tabularx}
\end{table}

\begin{figure}[tbp]
\centering 
\includegraphics[width=.32\textwidth]{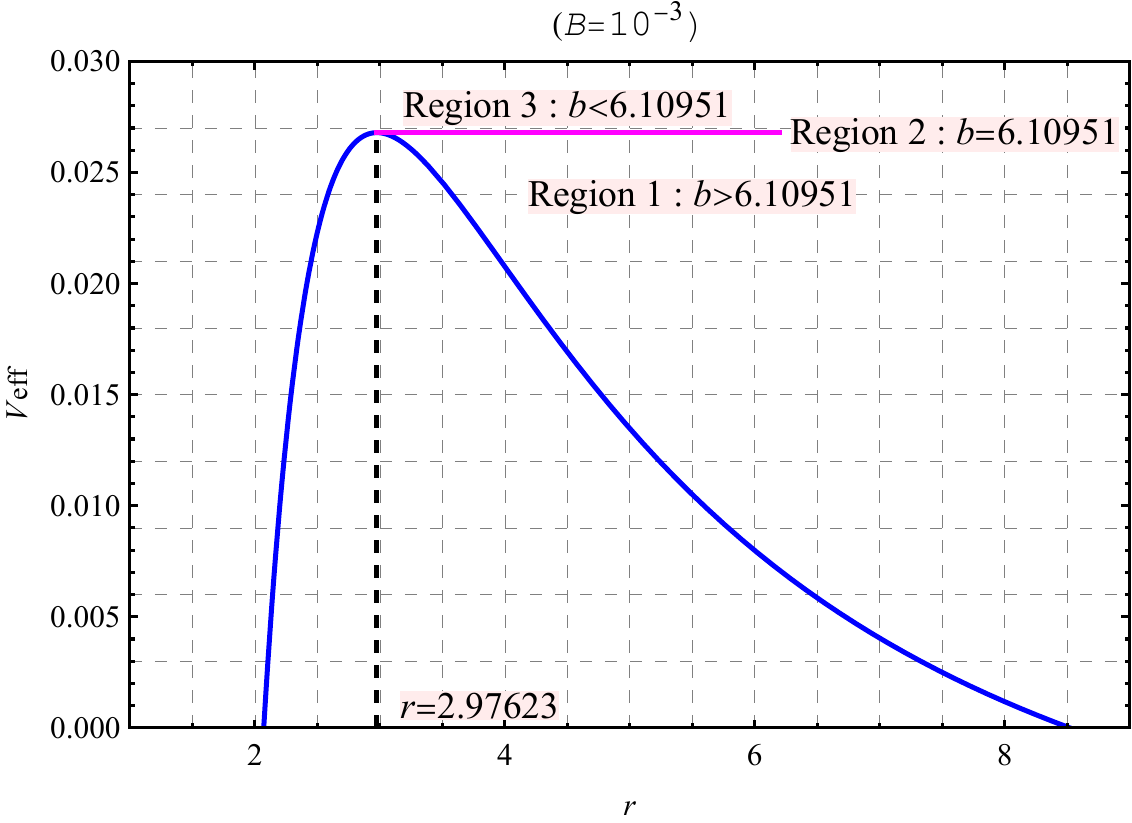}
\includegraphics[width=.32\textwidth]{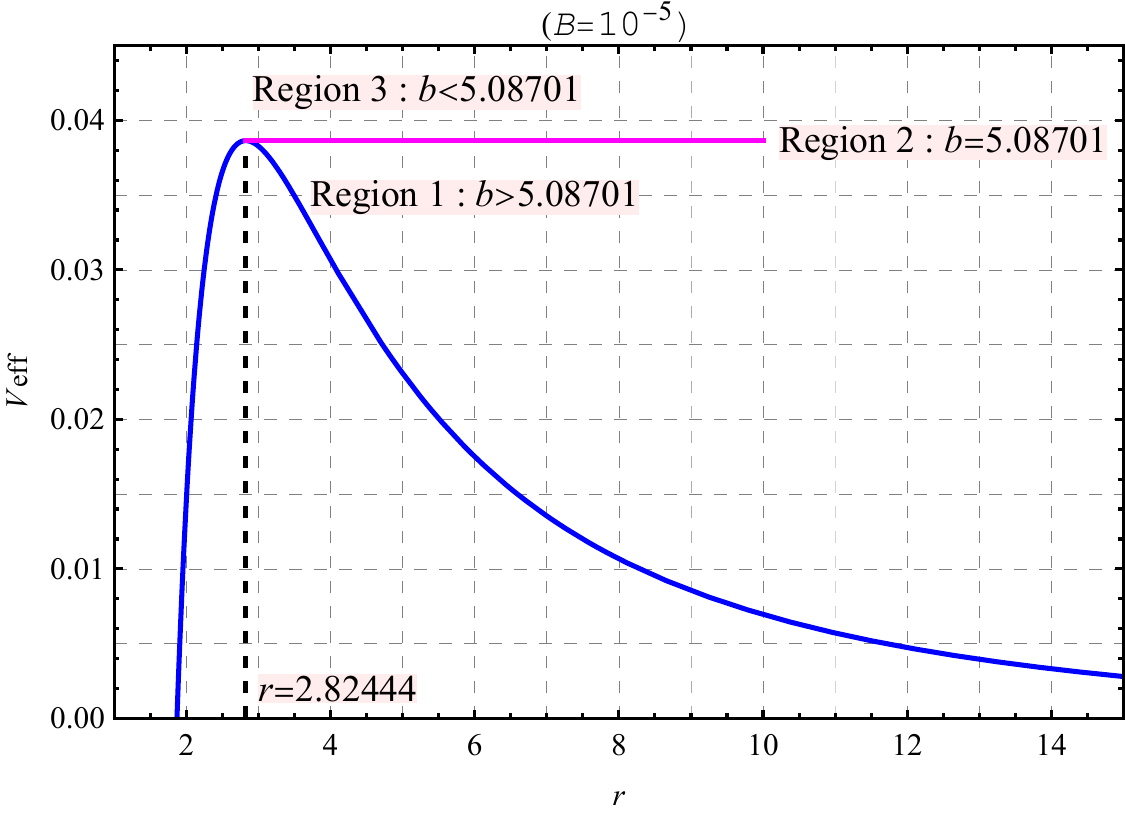}
\includegraphics[width=.32\textwidth]{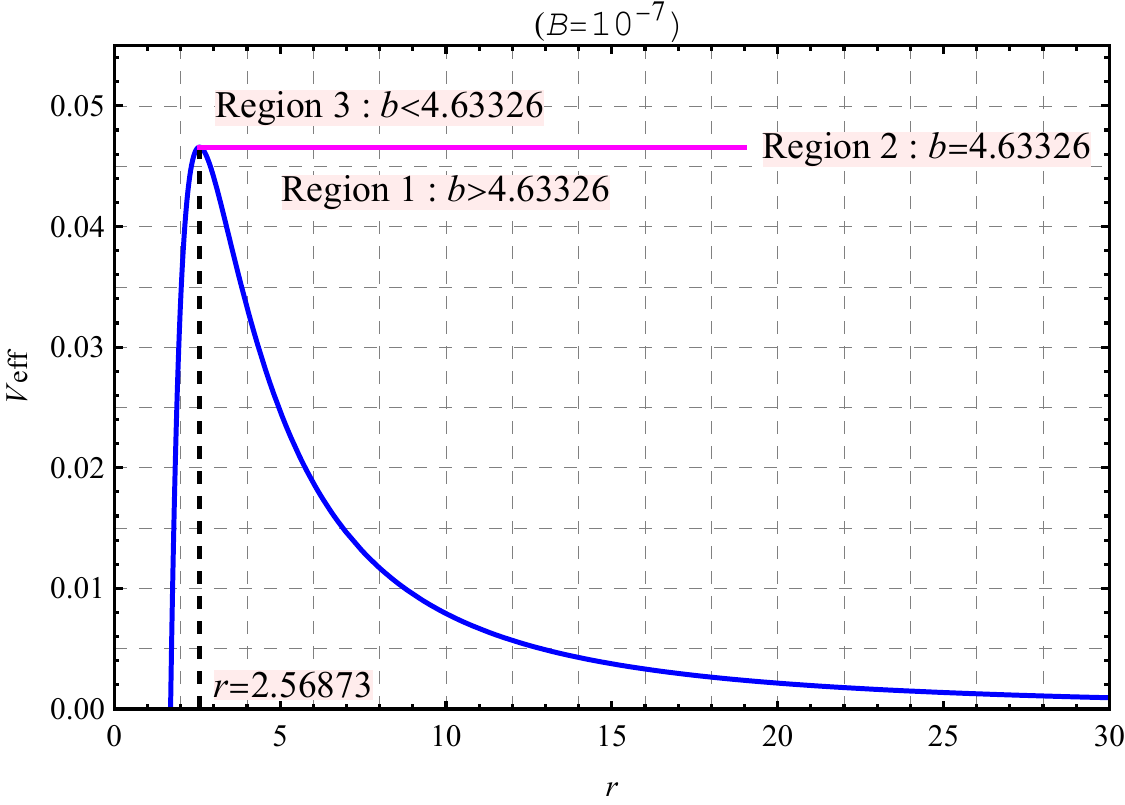}
\caption{\label{FigVeff}   The profile of the effective potential (blue lines) for $B=10^{-3}$ (left), $B=10^{-5}$ (middle) and $B=10^{-7}$ (right) with $M=1$, $q=0.2$. The dashed lines indicate the radii of the photon sphere $r_{\rm{ph}}$. Region 2 (pink lines) correspond to $V_{\rm{eff}}(r)=E_{\rm{ph}}^2$ ($b=b_{\rm{ph}}$), while  Region  1  and  Region 3  correspond to   $V_{\rm{eff}}(r)<E_{\rm{ph}}^2$ ($b>b_{\rm{ph}}$) and  $V_{\rm{eff}}(r)>E_{\rm{ph}}^2$  ($b<b_{\rm{ph}}$), respectively. }
\end{figure}

In Eq.~(\ref{vbr}), we observe that the particle's motion is contingent on the impact parameter and the effective potential. In Fig.~\ref{FigVeff}, the effective potential is depicted for $B=10^{-3}$, $B=10^{-5}$, and $B=10^{-7}$. It is evident that the effective potential becomes zero at the event horizon, subsequently increasing to reach a maximum at the photon sphere $r_{\rm{ph}}$, and then decreasing to zero at the pseudo-cosmological horizon. Notably, the structure of lightlike geodesics is simpler compared to timelike geodesics, with each effective potential curve exhibiting only one peak energy, corresponding to an unstable circular orbit. As the value of $B$ decreases, the maximum value of the effective potential gradually increases, and the radius of the unstable circular orbit gradually decreases, aligning with the findings from Table~\ref{tab:physical_quantities}. We consider a light ray moving radially inward. In Region 1, if the light ray initiates its motion at $r>r_{\rm{ph}}$, it will encounter a potential barrier and be reflected outward. If the photon starts its motion at $r<r_{\rm{ph}}$, it will fall into the singularity. In Region 2, specifically when $b=b_{\rm{ph}}$, as the light ray approaches the photon sphere, it will revolve around the BH indefinitely due to the non-zero angular velocity. In Region 3, the light ray will persist in moving inward without encountering a potential barrier. Eventually, it will enter the interior of the BH and plunge into the singularity.

The path of the light ray can be illustrated based on the equation of motion. Combining Eqs.~(\ref{psi}) and (\ref{radial}), we obtain
\begin{equation}
\frac{dr}{d\phi}=\pm r^2 \sqrt{\frac{1}{b^2}-\frac{1}{r^2}\left(1-\frac{2M}{r}-\frac{r^2}{3}\sqrt{B+\frac{q^2}{r^6}}+\frac{q}{3r}{\rm ArcSinh}\frac{q}{\sqrt{B}r^3}\right)}. \label{drp}
\end{equation}
To facilitate integration, we introduce the variable $u=1/r$. Thus, Eq.~(\ref{drp}) transforms into
\begin{equation}
\frac{du}{d\phi}=\mp \sqrt{\frac{1}{b^2}-\left(u^2-2 M u^3-\frac{1}{3}\sqrt{B+q^2u^6}+\frac{qu^3}{3}{\rm ArcSinh}\frac{qu^3}{\sqrt{B}}\right)}\equiv \Phi(u).\label{gu}
\end{equation}
The geometry of the geodesics is determined by the roots of the equation $ \Phi(u) =0$. Specifically, for $ b>b_{\rm{ph}}$, light will be deflected at the radial position $ u_i$ that satisfies $ \Phi(u_i) =0$. Therefore, finding the radial position $u_i$ is crucial for determining the trajectory of the light ray. Additionally, the observer's location is significant. While observers are typically situated at an infinite boundary for asymptotically flat spacetime, in our CDF-BH model, the pseudo-cosmological horizon is present. Physically, the observer should be positioned within the domain of outer communication, which lies between the event horizon and the pseudo-cosmological horizon, similar to de Sitter spacetime. Here, to study the trajectory of the light ray, we place the observer near the pseudo-cosmological horizon.
\begin{figure}[h]
\centering 
\includegraphics[width=.32\textwidth]{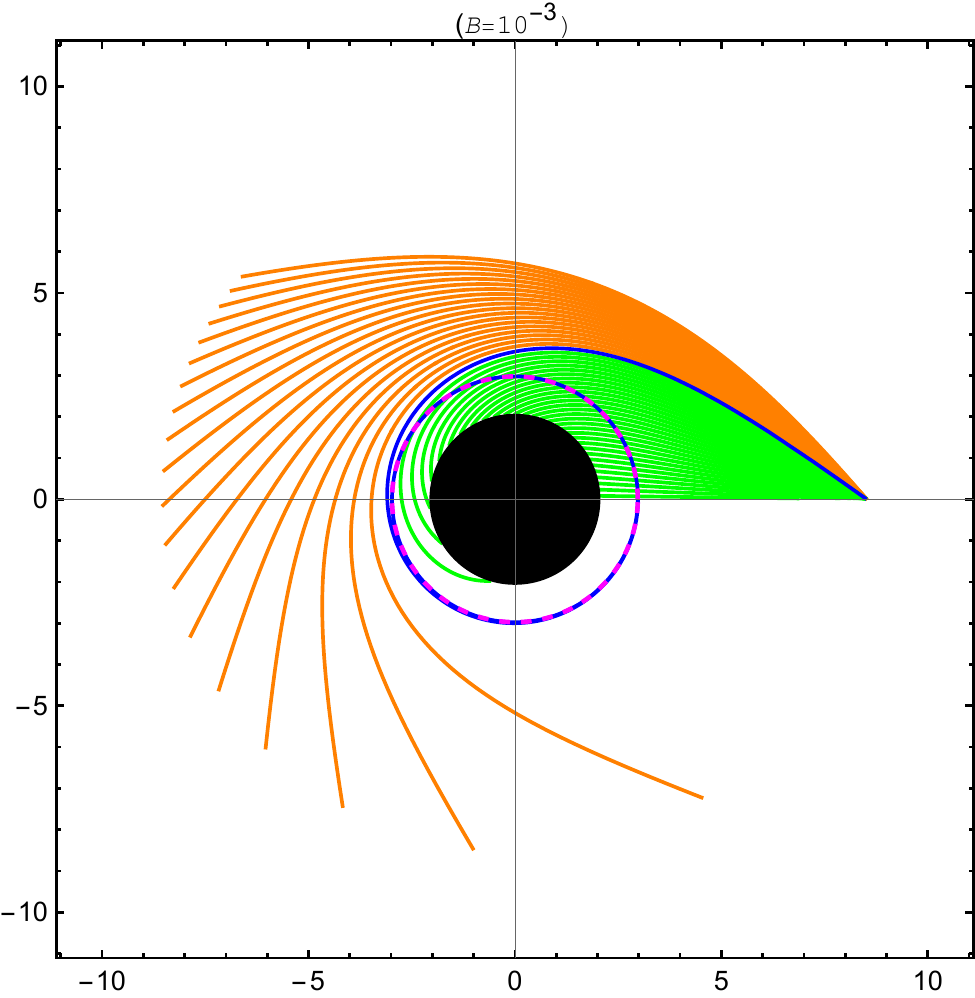}
\includegraphics[width=.32\textwidth]{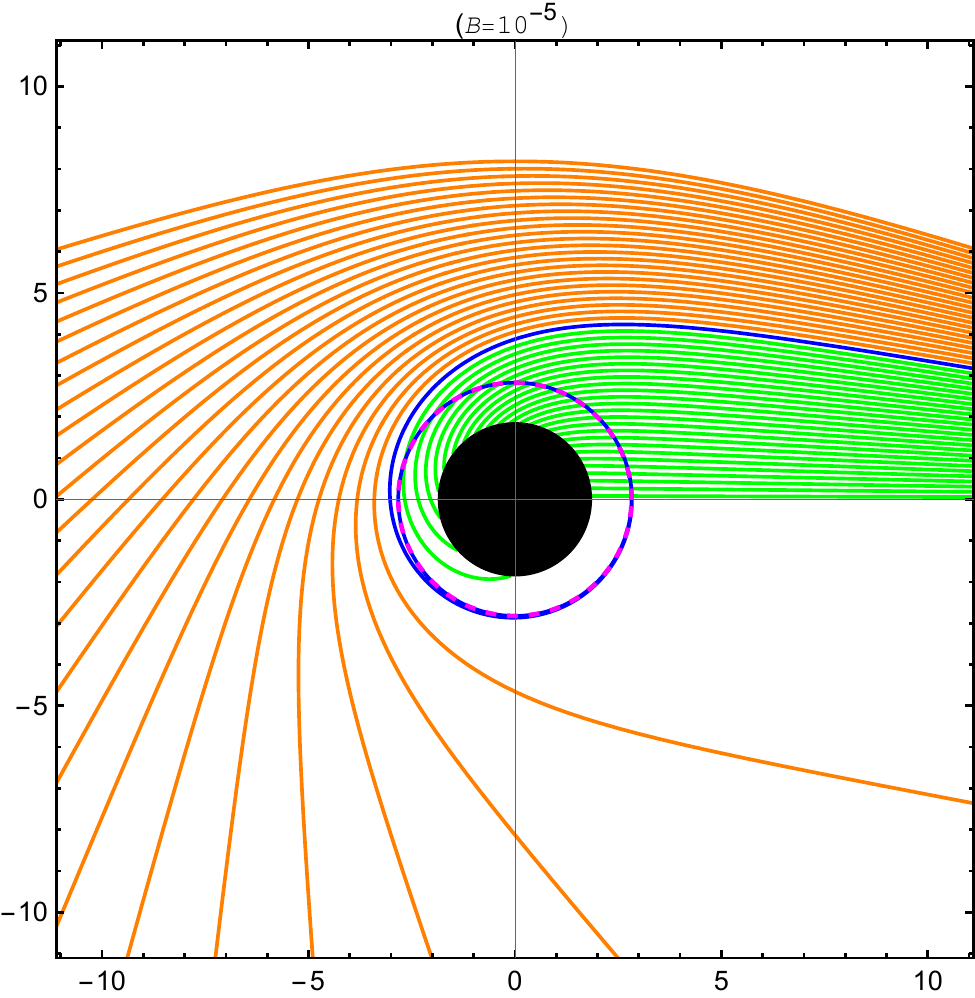}
\includegraphics[width=.32\textwidth]{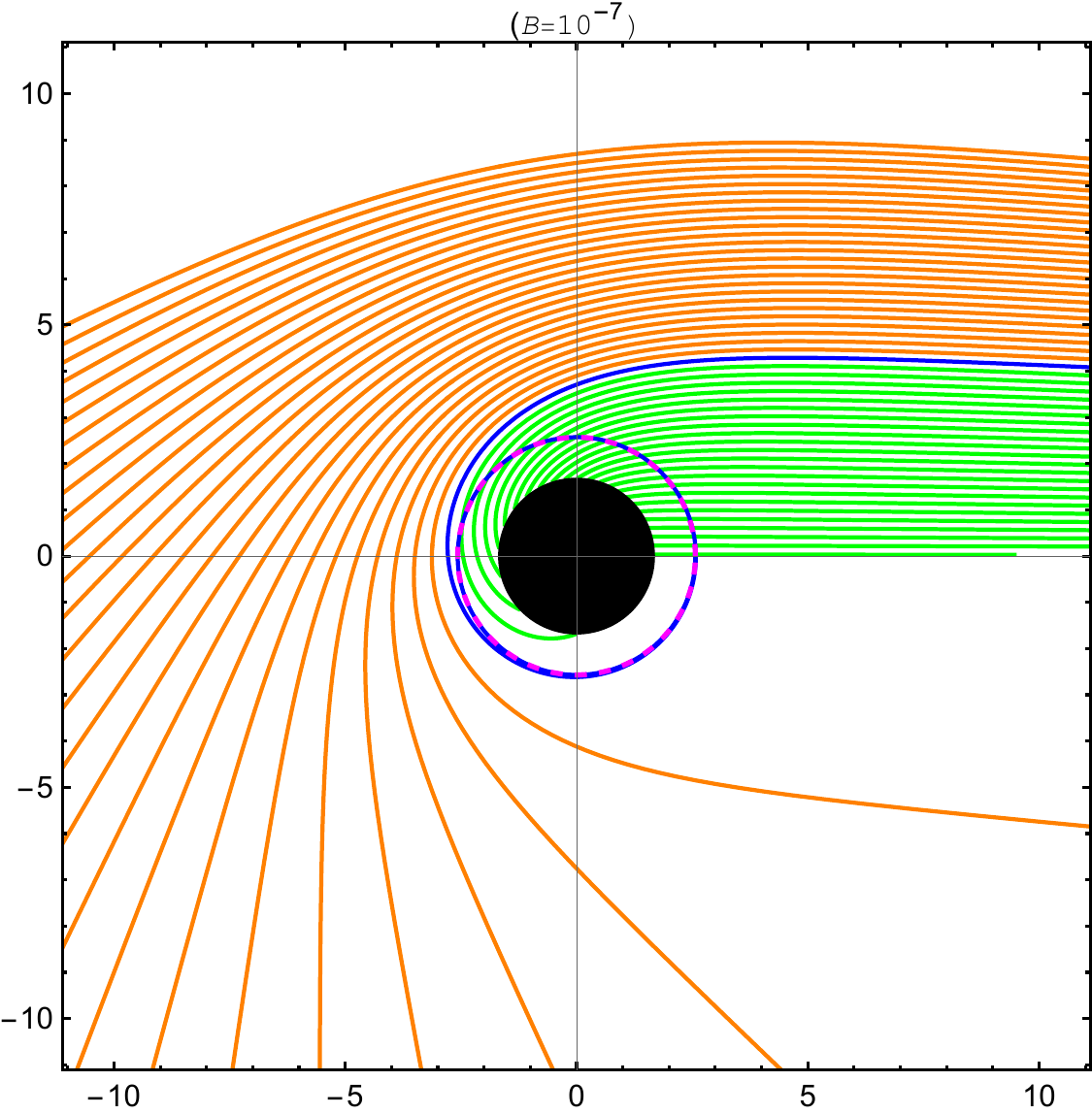}
\caption{\label{Figtrajectory1}  The trajectory of the light ray for $B=10^{-3}$ (left), $B=10^{-5}$ (middle) and $B=10^{-7}$ (right) with $M=1$, $q=0.2$. They are shown in the polar coordinates $(r, \phi)$.  The blue lines, green lines and orange lines correspond to $b=b_{\rm ph}$,  $b<b_{\rm ph}$ and $b>b_{\rm ph}$ respectively. The spacing of impact parameter is $\Delta b=0.2$ for each light ray. BH is shown as the solid black disk and the photon orbit as a dashed line.}
\end{figure}

Utilizing Eq.~\eqref{gu}, we can determine the trajectories of light rays, as illustrated in Fig.~\ref{Figtrajectory1}. All light rays approach the BH from the right side. The green, blue, and orange lines correspond to $b<b_{\rm{ph}}$, $b=b_{\rm{ph}}$, and $b>b_{\rm{ph}}$, respectively. Specifically, for the case of $b<b_{\rm{ph}}$, the light rays (green lines) descend into the BH entirely. For $b>b_{\rm{ph}}$, the light rays (orange lines) are deflected but never enter the BH. Notably, those light rays near the BH can even be reflected back to the right side, i.e., the side from which they approach the BH. For $b=b_{\rm{ph}}$, the light rays (blue lines) orbit around the BH. This conclusion aligns with the analysis of the effective potential in Fig. \ref{FigVeff}. Indeed, regions 1, 2, and 3 in Fig. \ref{FigVeff} correspond to the orange, blue, and green lines in Fig. \ref{Figtrajectory1}, respectively. As light undergoes deflection, it results in the formation of a BH shadow. The emitted light originates from accreting matter, making the profiles of the accretion matters crucial for determining the characteristics of BH shadows.

It is important to highlight that the presence of the pseudo-cosmological horizon leads to different geometries of light rays for varying values of $B$. Assuming the observer is in the domain of outer communications and positioned near the pseudo-cosmological horizon, relatively small values of $B$ (e.g., $B=10^{-7}$, right panel in Fig.\ref{Figtrajectory1}) result in the pseudo-cosmological horizon being farther from the event horizon, causing entering light rays to be approximately parallel. Conversely, for smaller values of $B$ (e.g., $B=10^{-3}$, left panel in Fig.\ref{Figtrajectory1}), the pseudo-cosmological horizon is closer to the event horizon, leading to noticeably non-parallel entering light rays.
\subsection{Shadow and observation constraints}
\begin{figure}[h]
\centering
\includegraphics[width=.49\textwidth]{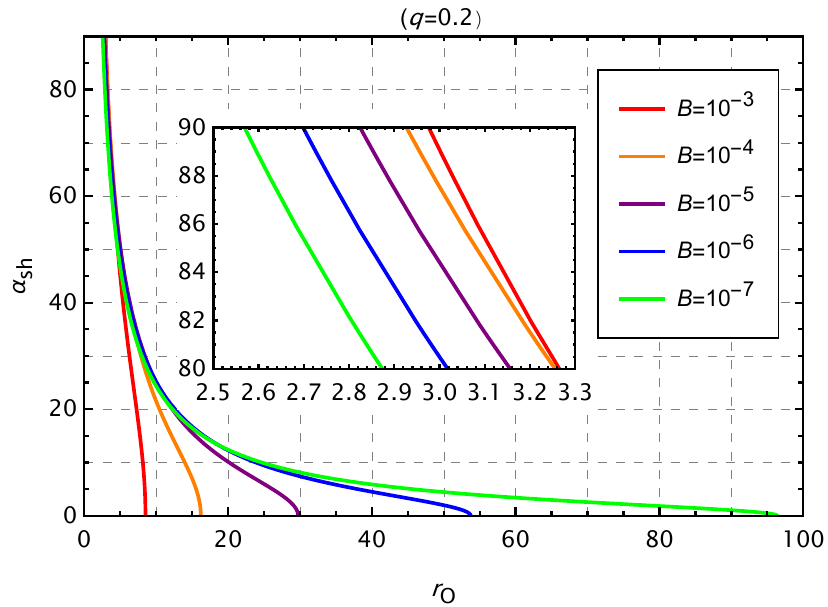}
\includegraphics[width=.49\textwidth]{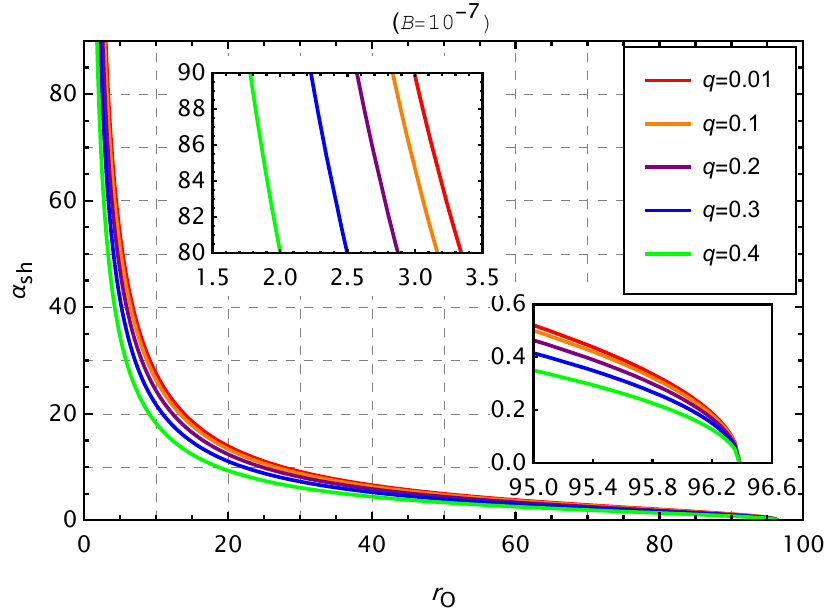}
\caption{\label{Figalpha}The behavior of the static angular radius of the shadow with $r_{\rm O}$ for different values of $B$ (fixing $q=0.2$, left panel), and for different values of $q$ (fixing $B=10^{-7}$, right panel). The point mass of the BH has been set as $M=1$.}
\end{figure}

For a BH spacetime with a (pseudo-)cosmological horizon, such as a Kottler BH (also known as Schwarzshild de Sitter BH), a quintessence-BH or a CDF-BH, the size of the BH shadow can explicitly depend on the radial coordinate of the observer, and on whether the observer is static or comoving. For a static observer located at a distance $r_{\rm O}$, the angular size of the BH shadow $\alpha_{\rm{sh}}$ is given by (see \cite{Perlick:2021aok})
\begin{equation}\label{Eqsinrsh}
{\rm sin}^2 \alpha_{{\rm sh}}=\frac{r^2_{{\rm ph}}}{f(r_{\rm ph})}\frac{f(r_{\rm O})}{r^2_{\rm O}}.
\end{equation}
The behavior of the static angular radius of the shadow $\alpha_{\rm sh}$ with $r_{\rm O}$ is depicted in Fig.~\ref{Figalpha}. For given values of $B$ and $q$, at $r_{\rm O}=r_{\rm ph}$, when the observer is at the photon sphere, $\alpha_{\rm sh}=90^{\circ}$, meaning that the shadow fills exactly one half of the observer's sky. When observed from a location in the vicinity of the pseudo-cosmological horizon, where the spacetime is highly curved, the angular radius of the BH shadow tends to approach zero, as shown in Fig.~\ref{Figalpha}. It is evident that the observed size of a BH shadow is dependent on the location of the observer. Therefore, specifying the size of a BH shadow within a metric that includes a pseudo-cosmological horizon is conditional.
\begin{figure}[h]
\centering 
\includegraphics[width=.485\textwidth]{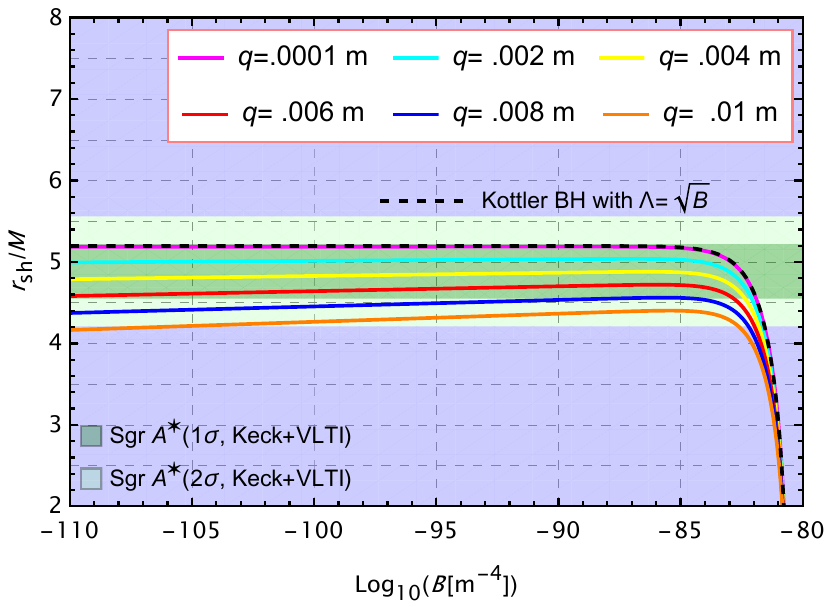}
\includegraphics[width=.49\textwidth]{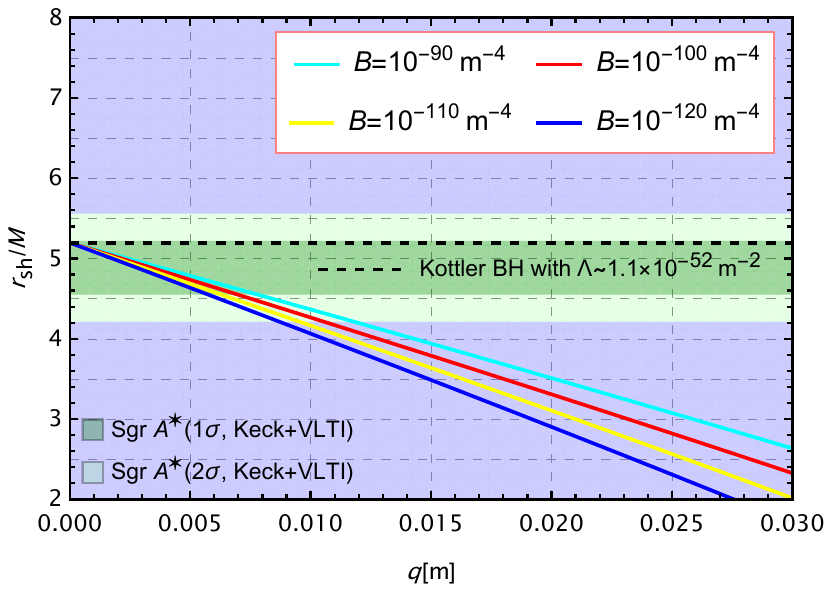}\\
\includegraphics[width=.485\textwidth]{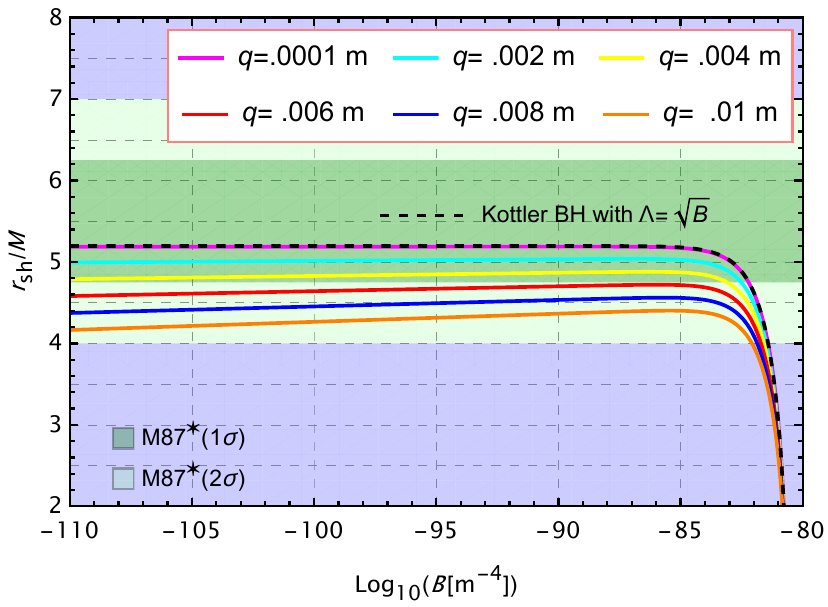}
\includegraphics[width=.49\textwidth]{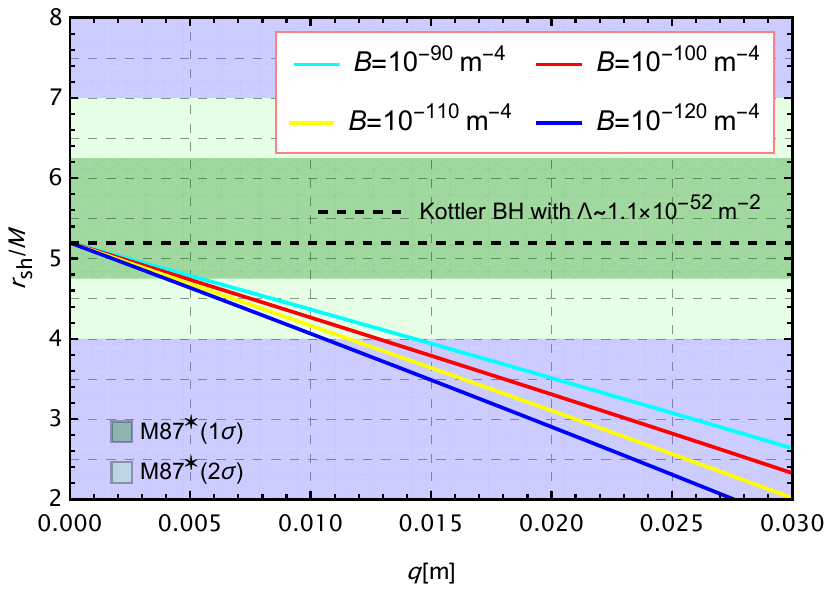}
\caption{\label{Figconstraints}  Shadow radius in $M$ unit as a function of $\mathrm{Log}_{10}B$ (left) and $q$ (right). The dark-green and light-green shaded regions represent the regions of $1\sigma$ and $2\sigma$ confidence intervals, respectively, with respect to the M$87^{\ast}$ and Sgr ${\rm A}^{\ast}$ observations. Here we set $r_{\rm O}\sim8\mathrm{kpc}\backsimeq2.5\times10^{20}\mathrm{m}$ to ensure $\sqrt{B}r_{\rm O}^2\ll1$.}
\end{figure}In the physically relevant small-angle approximation, it is easy to see that the shadow size is given by (see \cite{Perlick:2021aok} for more detailed discussions)
\begin{equation}\label{Eqrsh}
r_{{\rm sh}}=r_{{\rm ph}}\sqrt{\frac{f(r_{\rm O})}{f(r_{\rm ph})}}.
\end{equation}
Note that the CDF-BH shadow radius is affected by the parameters $B$ and $q$. The explicit dependence of the shadow size on the observer's position is clear in Eq.~(\ref{Eqrsh}). Considering an observer positioned far away from both the event horizon and the pseudo-cosmological horizon, the spacetime in the vicinity of the observer can be treated as approximately flat. This enables us to use astronomical observations on the BH shadow to constrain the CDF model. Considering the asymptotic behavior of $f(r)$ denoted in Eq.~(\ref{frasymp}), for $B<10^{-82}{\rm m}^{-4}$, the product $\sqrt{B}r^2_{\rm O}\ll1$ promote we considering $r_{\rm O}\thicksim 8{\rm kpc}\backsimeq2.5\times10^{20} {\rm m}$. In terms of observation, the BH shadow diameter $r_{\rm sh}$ can be measured with the EHT data, providing us constraints on parameters $B$ and $q$, as shown in Fig.~\ref{Figconstraints}. The bounds for the BH shadow radius in $M$ unit can be deduced as
\begin{align}
 \begin{array}{c}
                            \text{Sgr A$^{\ast}$\textsuperscript{\cite{Vagnozzi:2022moj}}} \\
                            \text{\footnotesize Keck+VLTI}
                          \end{array} &\left\{\begin{array}{c}
                            \text{$4.55\lesssim r_{\rm sh}/M \lesssim 5.22$~~($1\sigma$)} \\
                            \text{$4.21\lesssim r_{\rm sh}/M \lesssim 5.56$~~($2\sigma$)}
                          \end{array}\right., &
\text{M$87^{\ast}$\textsuperscript{\cite{Bambi:2019tjh,Allahyari:2019jqz}}} &\left\{\begin{array}{c}
                            \text{$4.75\lesssim r_{\rm sh}/M \lesssim 6.25$~~($1\sigma$)} \\
                            \text{$4\lesssim r_{\rm sh}/M \lesssim 7$~~($2\sigma$)}
                          \end{array}\right..
\end{align}
To comprehensively consider the impact of CDF, we also take into account the Kottler BH in Fig.~\ref{Figconstraints}, which has the following metric lapse function
\begin{equation}\label{Kottlermetric}
f_{\rm Kottler}(r)=1-\frac{2M}{r}-\frac{\Lambda}{3}r^2.
\end{equation}
Upon examination of the left panels of Fig.~\ref{Figconstraints}, it is evident that, for a fixed value of $q$, when $B$ is relatively small (i.e., $\sqrt{B}r^2_{\rm O}$ is significantly less than 1), the shadow radius $r_{\rm sh}$ of the CDF-BH experiences a gradual increase with the augmentation of $B$. Furthermore, with higher values of $q$, the influence of $B$ on the shadow size becomes more pronounced. When $q$ takes an extremely small value, the shadow radius approaches that of the Kottler BH with a cosmological constant of $\Lambda=\sqrt{B}$. In terms of fitting observational data, the CDF-BH offers greater degrees of freedom compared to the Kottler BH. Particularly for observational data related to Sgr A$^\ast$, adjusting the value of $q$ can yield theoretical results that better align with observations. When $B$ is relatively large, $\sqrt{B}r^2_{\rm O}$ approaches 1, the shadow radius rapidly contracts with increasing $B$. Additionally, the right panels of Fig.~\ref{Figconstraints} clearly illustrates that, for a fixed value of $B$, the observed shadow radius diminishes as the value of $q$ increases. Taking into account the asymptotic behavior of $f(r)$ at infinity, $\sqrt{B}$ can be regarded as the cosmological constant, which, according to the Planck 2018 results~\cite{Planck:2018vyg}, equals $1.1\times10^{-52}{\rm m}^{-2}$. As a result, using the $1\sigma$ and $2\sigma$ confidence intervals of the $r_{\rm sh}$ for Sgr A$^{\ast}$, the parameter $q$ can be constrained as $q<0.006696$~${\rm m}$ within $1\sigma$ and $q<0.010160$~${\rm m}$ within $2\sigma$; When considering the confidence intervals of $r_{\rm sh}$ for M$87^{\ast}$, it can be constrained as $q<0.004644$~${\rm m}$ within $1\sigma$ and $q<0.012285$~${\rm m}$ within $2\sigma$.

As we shall see later, not only is the position of observer important for determining the shadow radius, it is also crucial for the observation of the optical images of a BH. In a given BH spacetime, when the observer is close to the BH, light rays with a large impact parameter will fail to reach the observer. To avoid such a situation, in our following study on the optical images of the BH, we will adopt the convention employed in Fig.~\ref{Figtrajectory1}, where we place the observer as far away as possible from the BH, specifically inside the domain of outer communication and near the pseudo-cosmological horizon.
\section{Optical appearance with thin disk accretion }
\label{disk}
In this section, we examine a scenario involving an optically and geometrically thin disk accretion around the BH, observed face-on. As detailed in \cite{Gralla:2019xty}, a noteworthy characteristic of thin disk accretion is the presence of photon rings and lensing rings surrounding the BH shadow. The lensing ring is composed of light rays intersecting the plane of the disk twice outside the horizon, while the photon ring comprises light rays intersecting the plane of the disk three or more times. Therefore, understanding the trajectory of photons in our case is crucial for distinguishing between the photon ring and lensing ring.

\subsection{Number of orbits of the deflected light trajectories}
\begin{figure}[h]
\centering 
\includegraphics[width=.328\textwidth]{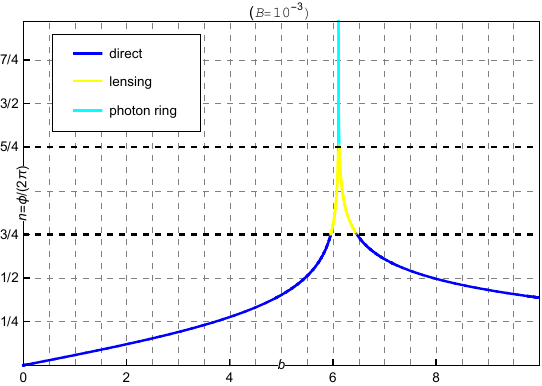}
\includegraphics[width=.328\textwidth]{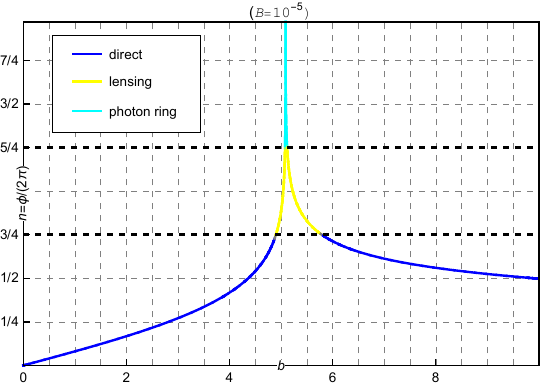}
\includegraphics[width=.328\textwidth]{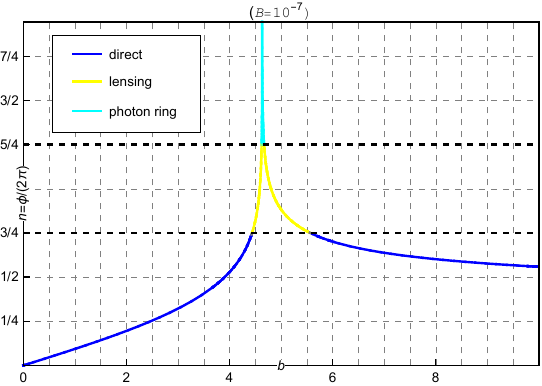}\\
\includegraphics[width=.328\textwidth]{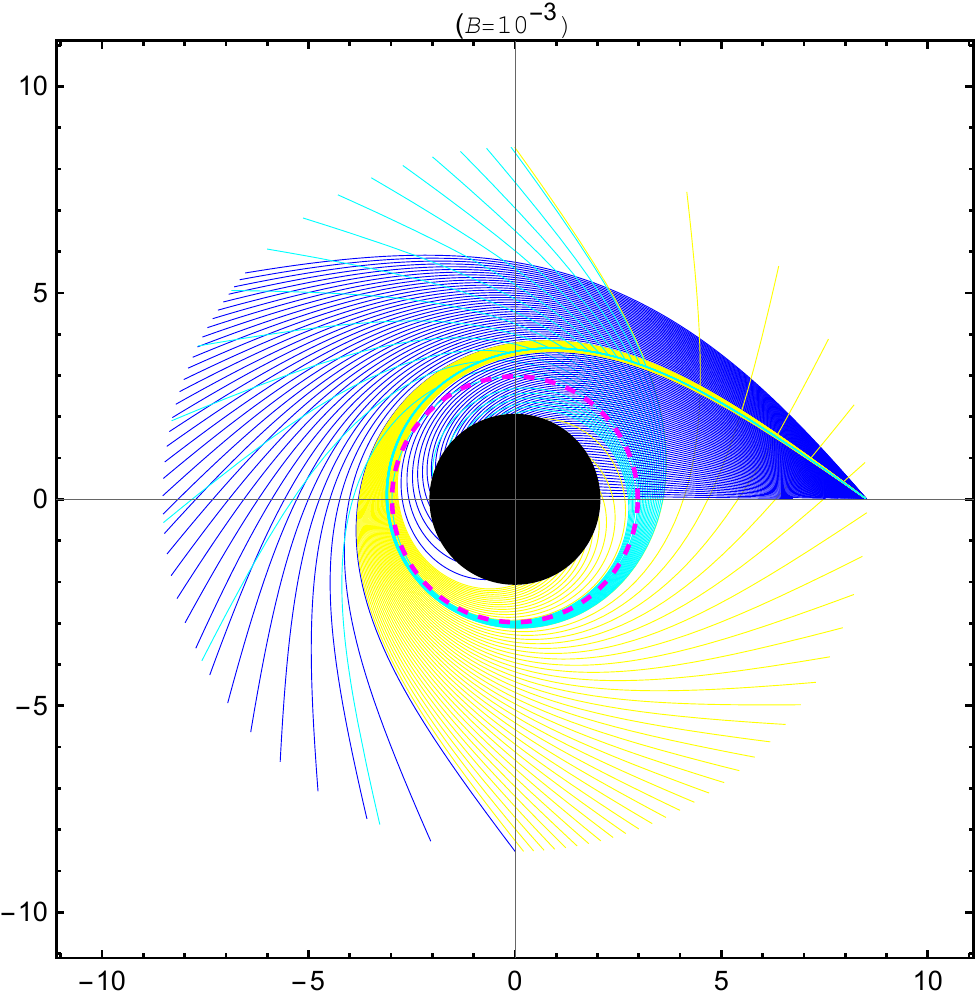}
\includegraphics[width=.328\textwidth]{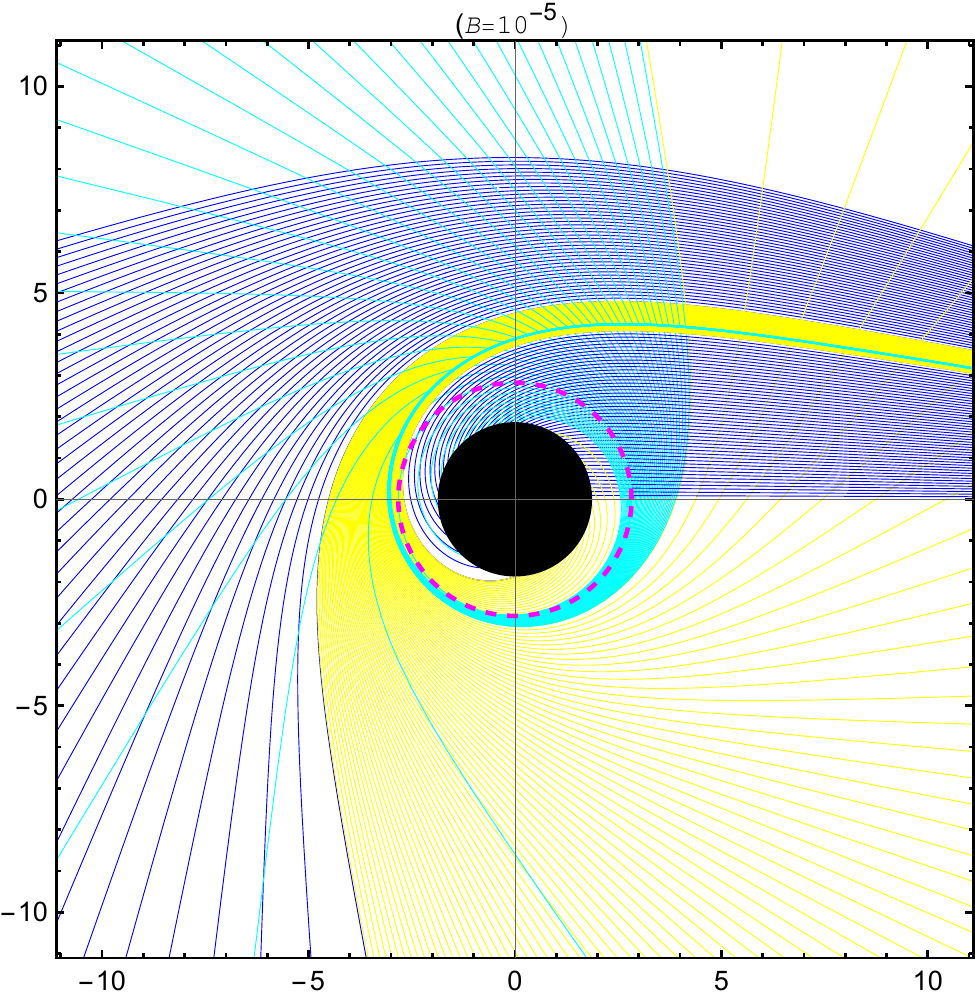}
\includegraphics[width=.328\textwidth]{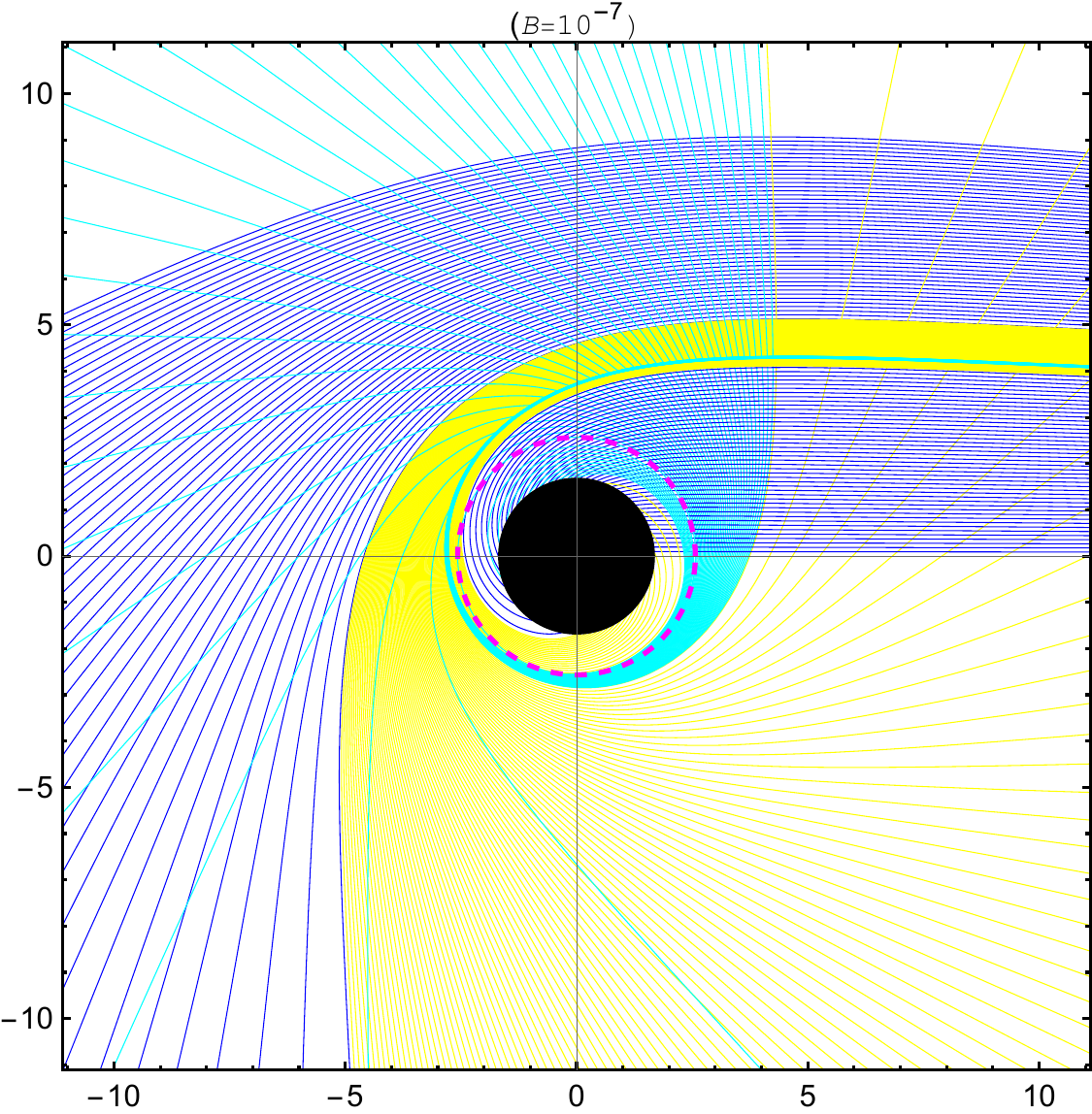}
\caption{\label{Figtrajectory2} Behavior of photons in the CDF-BH spacetime as a function of impact parameter $b$ with $M=1$,  $q=0.2$,  $B=10^{-3}$ (left), $B=10^{-5}$ (middle) and $B=10^{-7}$ (right). \textbf{Top row:} We show the fractional number of orbits, $n =\phi/2 \pi$, where $\phi$ is the total change in azimuthal angle outside the horizon. The direct, lensing and photon ring correspond to $ n<3/4$ (blue), $ 3/4<n<5/4$ (yellow), and $ n>5/4$ (cyan), respectively. \textbf{Bottom row:} We show a selection of associated photon trajectories in the polar coordinates $(r, \phi)$. The spacings in impact parameter are $1/10$, $1/100$ and $1/1000$, for the direct (blue), lensing (yellow), and photon ring (cyan) bands, respectively. The BHs are shown solid black disks.}
\end{figure}
The trajectories of light rays are depicted in the bottom row of Fig. \ref{Figtrajectory2} for $B=10^{-3}$ (left column), $B=10^{-5}$ (middle column), and $B=10^{-7}$ (right column), respectively. In these figures, the blue lines, yellow lines, and cyan lines represent direct emission, lensing rings, and photon rings, respectively. These designations are in accordance with the definitions in \cite{Gralla:2019xty}, where these light rays intersect the disk plane once, twice, and more than twice.

Another method to distinguish the trajectories of light rays is by using the total number of orbits, defined as $n=\phi/2 \pi$ \cite{Gralla:2019xty}. The total number of orbits is presented in the top row of Fig.~\ref{Figtrajectory2} for $B=10^{-3}$ (left column), $B=10^{-5}$ (middle column), and $B=10^{-7}$ (right column), respectively. The blue lines, yellow lines, and cyan line are still used to represent direct emission, lensing rings, and photon rings, respectively. Based on the definitions of these light rings, it is evident that direct emissions correspond to $n<3/4$, while lensing rings correspond to $3/4<n<5/4$, and photon rings correspond to $n>5/4$. The parameter intervals of $b$ for direct emission, photon rings, and lensing rings are listed in the last five columns of Table~\ref{tab:physical_quantities} for the cases of $B=10^{-3}$, $B=10^{-5}$, and $B=10^{-7}$.

\subsection{Observed specific intensities and transfer functions}
Subsequently, we explore the analysis of the observed specific intensity originating from the accretion of a thin disk. Our assumption involves isotropic emission in the rest frame of static worldlines for the thin disk, situated in the equatorial plane of the BH, with a static observer positioned at the North pole, aligning with the framework of \cite{Gralla:2019xty}. Let $I_{\mathrm{e}}(r)$ and $\nu_{\mathrm{e}}$ denote the emitted specific intensity and frequency, respectively, while $I_{\rm{O}}(r)$ and $\nu_{\rm O}$ represent the observed specific intensity and frequency. Liouville's theorem, which conserves $I_{\mathrm{e}}/\nu_{\mathrm{e}}^3$ along a light ray, enables us to express the observed specific intensity as
\begin{equation}
I_{\rm{O}}(r)=\left[\frac{f(r)}{f(r_{\rm O})}\right]^{3/2}I_{\mathrm{e}}(r).
\end{equation}
The total specific intensity is derived by integrating the specific intensity over different frequencies:
\begin{equation}
I_{\rm obs}(r)=\int I_{\rm{O}}(r) d\nu_{\rm{O}}=\int \left[\frac{f(r)}{f(r_{\rm O})}\right]^{2}I_{\mathrm{e}}(r)d\nu_{\mathrm{e}}=\left[\frac{f(r)}{f(r_{\rm O})}\right]^{2}I_{\rm{em}}(r),
\end{equation}
where $I_{\rm{em}}(r)=\int I_{\mathrm{e}}(r)d\nu_{\mathrm{e}}$ denotes the total emitted specific intensity near the accretion.

Should a light ray be retraced from the observer, intersecting the disk, it captures luminosity from the disk emission. For $3/4<n<5/4$, the light ray curves around the BH and strikes the opposite side of the disk from the back (refer to the yellow lines in Fig.~\ref{Figtrajectory2}). Consequently, it gains extra brightness from this second traversal through the disk. For $n>5/4$, the light ray bends around the BH more and subsequently strikes the front side of the disk once again (refer to the cyan lines in Fig.~\ref{Figtrajectory2}). This leads to additional brightness from the third traversal through the disk. Hence, the observed intensity is the sum of intensities from each intersection, expressed as
\begin{equation}
I_{\rm obs}(b)= \sum\limits_{m}\left[\frac{f(r)}{f(r_{\rm O})}\right]^{2}I_{\rm{em}}\mid_{r=r_m(b)}, \label{lir}
\end{equation}
where $r_{m(b)}$, denoted as the transfer function, signifies the radial position of the $m^{\rm{th}}$ intersection with the disk plane outside the horizon. For simplicity, we have overlooked light absorption in the thin accretion, potentially reducing the observed intensity from additional passages. From Eq.~(\ref{lir}), the observed intensity is dependent on the location of the observer. The observed intensity is relatively large when $r_{\rm O}$ is small or close to $r_{\rm c}$, while it decreases a lot when the observer is far from both the event and quasi-cosmological horizons.

\begin{figure}[h]
\centering 
\includegraphics[width=.32\textwidth]{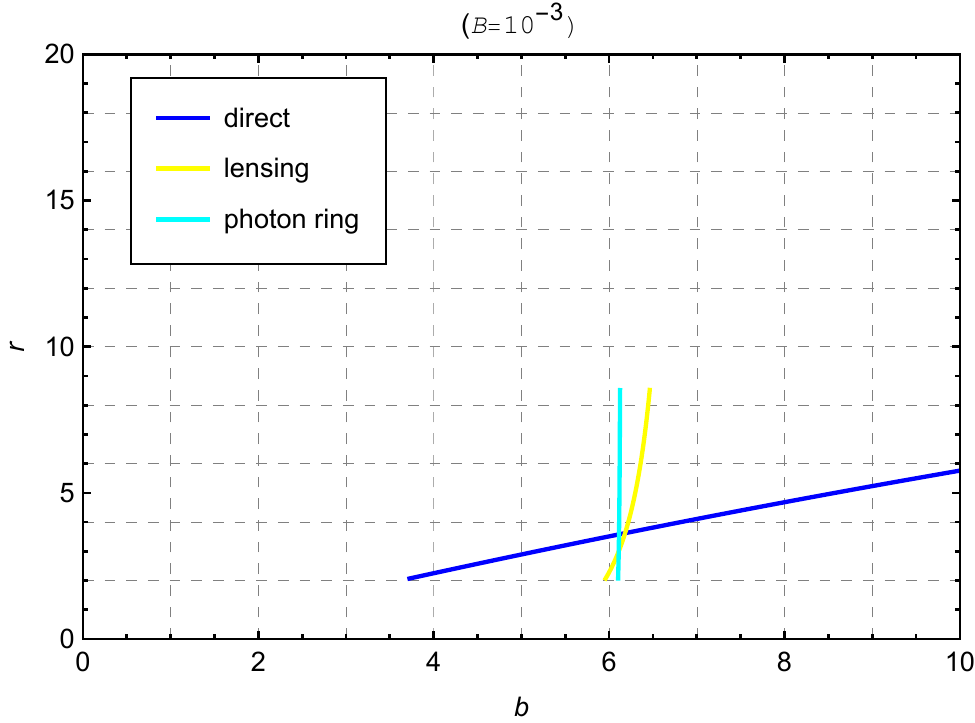}
\includegraphics[width=.32\textwidth]{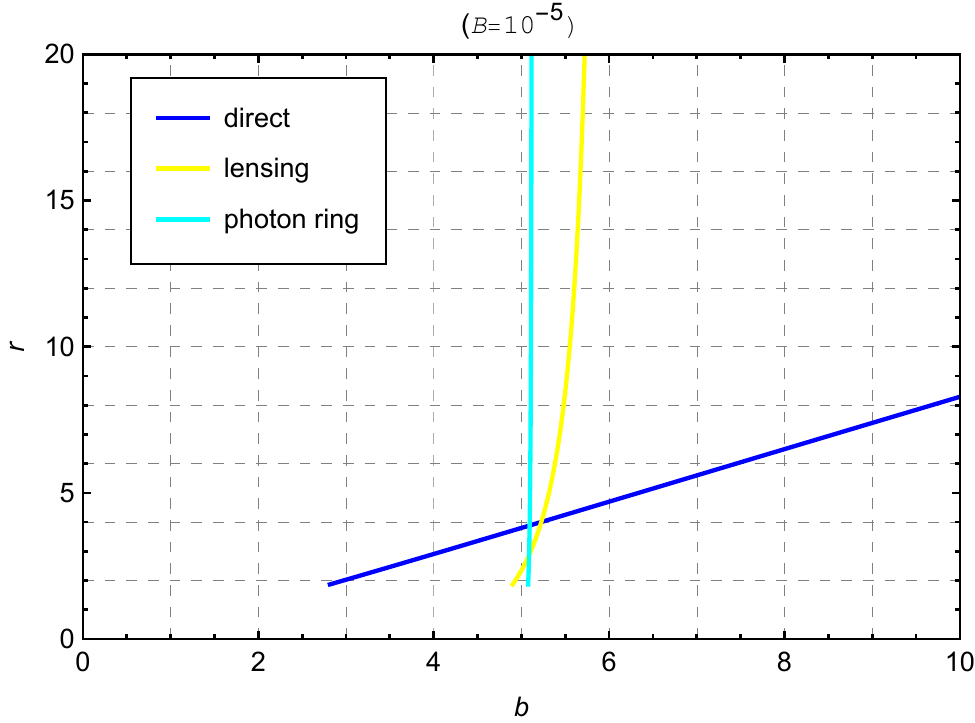}
\includegraphics[width=.32\textwidth]{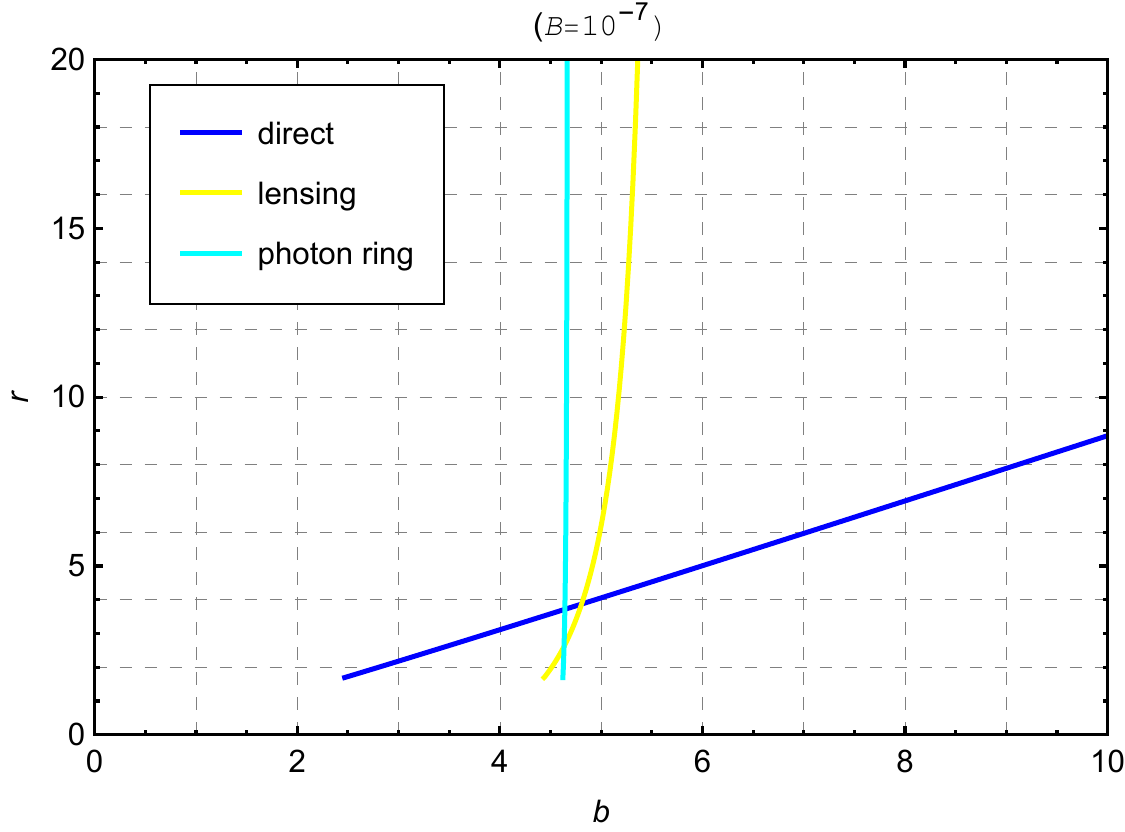}
\caption{\label{Figrmb} The first three transfer functions $r_m(b)$ for the thin disk in the CDF-BH with $M=1$,  $q=0.2$, $B=10^{-3}$ (left), $B=10^{-5}$ (middle) and $B=10^{-7}$ (right). The blue lines, yellow lines and cyan lines stand for  the radial coordinate of the first, second, and third intersections with a face-on thin disk outside the horizon.}
\end{figure}

The transfer function elucidates the connection between the radial coordinate $r$ and the impact parameter $b$. The slope of the transfer function, $dr/db$, functions as the demagnification factor. Fig.~\ref{Figrmb} illustrates the transfer functions concerning the impact parameter $b$ for distinct parameters $B$. The blue line, yellow line, and cyan line signify the first ($m=1$), second ($m=2$), and third ($m=3$) transfer functions, respectively. As such, the direct image profile essentially mirrors the redshifted source profile. The second transfer function corresponds to the lensing ring (encompassing the photon ring). In this context, the observer discerns a highly demagnified image of the back side of the disk. Finally, the third transfer function corresponds to the photon ring. Here, one observes an extremely demagnified image of the front side of the disk, given the slope is approximately infinite. Consequently, this contributes negligibly to the total brightness of the image.

\subsection{Direct image, lensing ring and photon ring}
Upon acquiring the transfer functions, we can proceed to deduce the specific intensity based on Eq.~(\ref{lir}), given the emitted specific intensity. In this study, we adopt three toy-model functions representing potential emissions from geometrically and optically thin matter. In Model 1, supposing the ring-shaped accretion disk is distributed between ISCO and OSCO, $I_{\rm{em}}(r)$ is presumed to be a second-order power decay function from the radial position of the ISCO $r_{\rm{ISCO}}$, as expressed in Eq.~(\ref{Eqfun1}). Model 2 posits that the emission follows a third-order power decay, detailed in Eq.~(\ref{Eqfun2}). Model 3 considers a moderate decay in emission, as depicted in Eq.~(\ref{Eqfun3}). The three models are shown below.
\begin{align}
\begin{split}
\text{Model 1:} \quad I_{\mathrm{em}}(r) &=
\begin{cases}
I_0\left(\frac{1}{r-(r_{\mathrm{ISCO}}-1)}\right)^2, \quad &r_{\mathrm{OSCO}}\geq r \geq r_{\mathrm{ISCO}} \\
0, & r < r_{\mathrm{ISCO}}\quad \text{or} \quad r > r_{\mathrm{OSCO}}
\end{cases}, \label{Eqfun1}
\end{split} \\
\begin{split}
\text{Model 2:} \quad I_{\mathrm{em}}(r) &=
\begin{cases}
I_0\left[\frac{1}{r-(r_{\mathrm{ph}}-1)}\right]^3, \quad & r \geq r_{\mathrm{ph}} \\
0, & r < r_{\mathrm{ph}}
\end{cases}, \label{Eqfun2}
\end{split} \\
\begin{split}
\text{Model 3:} \quad I_{\mathrm{em}}(r) &=
\begin{cases}
I_0\frac{\frac{\pi }{2}-\tan ^{-1}\left[r-(r_{\mathrm{ISCO}}-1)\right]}{\frac{\pi }{2}-\tan ^{-1}\left[r_{\mathrm{h}}-(r_{\mathrm{ISCO}}-1)\right]}
, \quad & r \geq r_{\mathrm{h}} \\
0, & r < r_{\mathrm{h}}
\end{cases}. \label{Eqfun3}
\end{split}
\end{align}
In order to compare the observational results arising from different disk accretion models and various CDF parameters, we neglect the influence of the observer's position on the observed intensity, by presenting plots depicting $f(r_{\rm O})^2I_{\rm obs}(b)/I_0$.

\begin{figure}
    \centering
\begin{minipage}[b]{0.02\linewidth}
\vspace{10mm}
\end{minipage}
\begin{minipage}[b]{0.4\linewidth}
    \includegraphics[width=\textwidth]{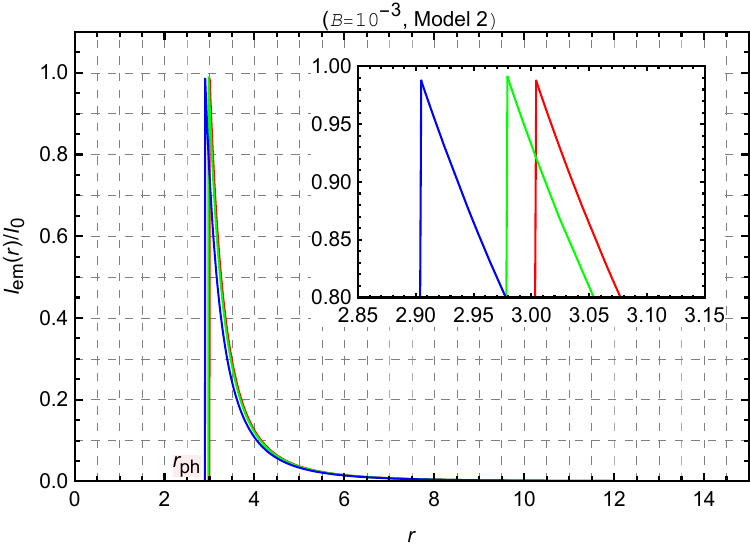}\\
    \includegraphics[width=\textwidth]{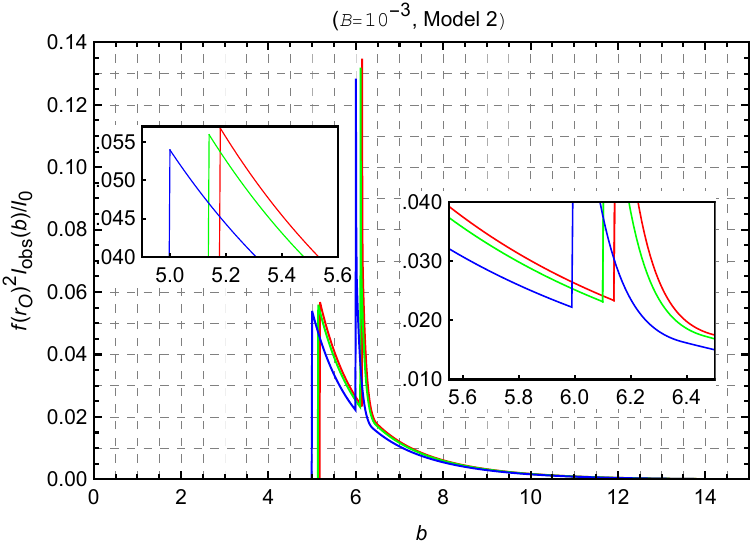}
\end{minipage}
\begin{minipage}[b]{0.02\linewidth}
\vspace{10mm}
\end{minipage}
\begin{minipage}[b]{0.56\linewidth}
    \includegraphics[width=\textwidth]{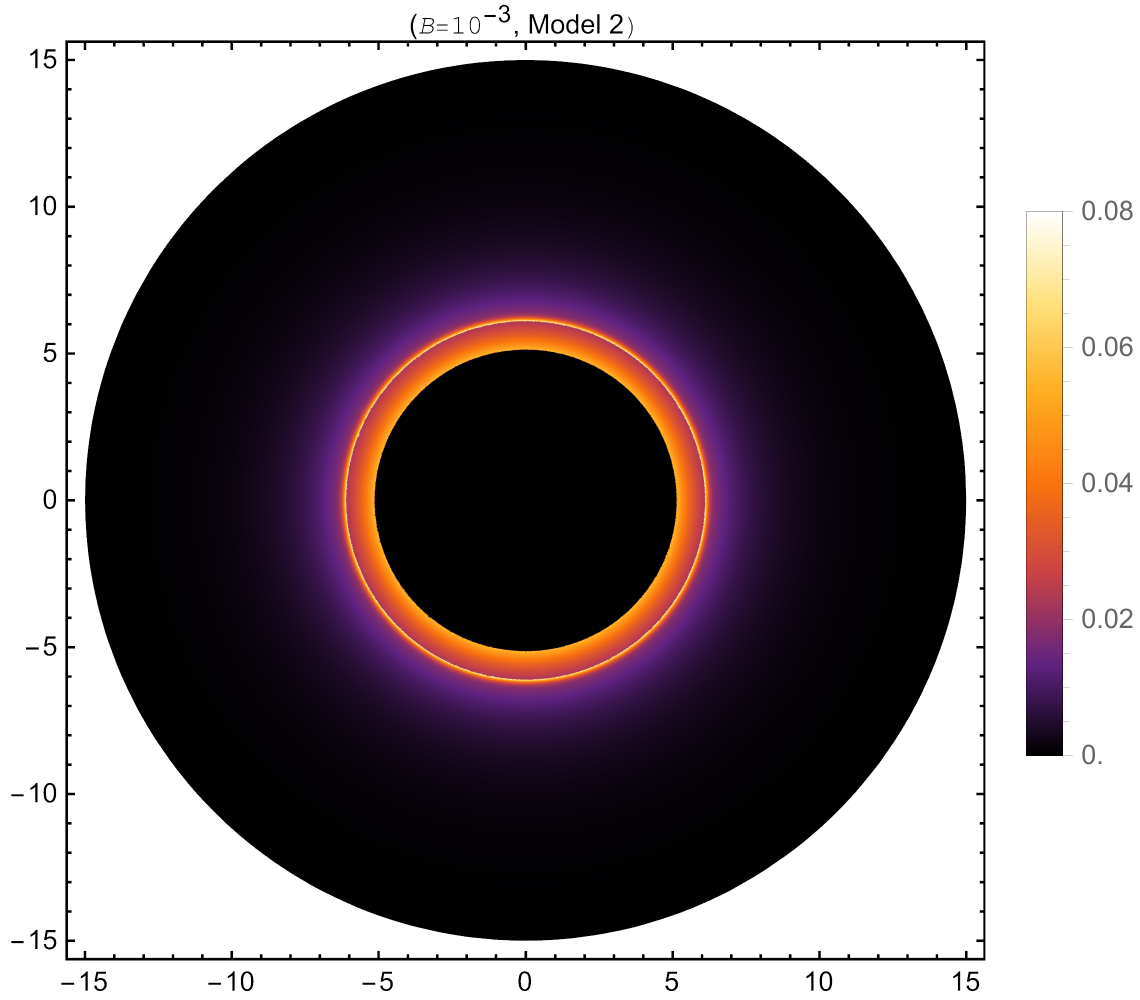}\vspace{8mm}
\end{minipage}
\caption{\label{FigBm3}  Observational appearances of a geometrically and optically thin disk with Model 2 emissivity profile near BH with $M=1$, $B=10^{-3}$, viewed from a face-on orientation. \textbf{Left column:} The profiles of $I_{\mathrm{em}}(r)$ (top) and the observed intensity $I_{\mathrm{obs}}(b)$ (bottom) as a function of the impact parameter $b$. The red, green and blue curves correspond to $q=0.01$, $q=0.2$ and $q=0.4$, respectively. \textbf{Right column:} The 2-dim density plots of the observed intensity $I_{\mathrm{obs}}(b)$ for $q=0.2$.}
\end{figure}

For $B=10^{-3}$, there is no ISCO and OSCO for timelike particles, regardless of the value of $q$. Consequently, we exclusively consider Model 2, and its profile is illustrated in the top-left panel of Fig.~\ref{FigBm3}. The observed specific intensities for $q=0.01$, $q=0.2$ and $q=0.4$ are shown in the bottom-left and right panels in Fig.~\ref{FigBm3}. The bottom-left panel is about the one dimensional functions of $I_{\rm{obs}}(b)$ with respect to $b$, while the right panel is about the two dimensional density plots of $I_{\rm{obs}}(b)$, viewed faced-on.
\begin{figure}[tbp]
\centering 
\includegraphics[width=.34\textwidth]{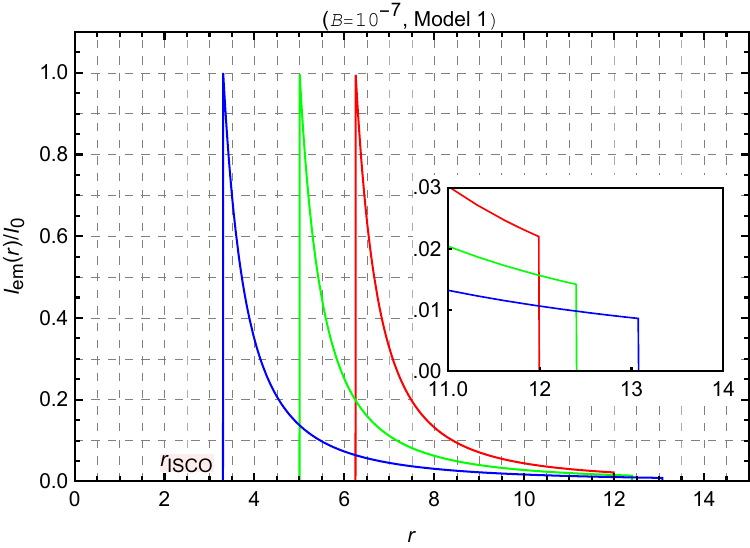}
\includegraphics[width=.34\textwidth]{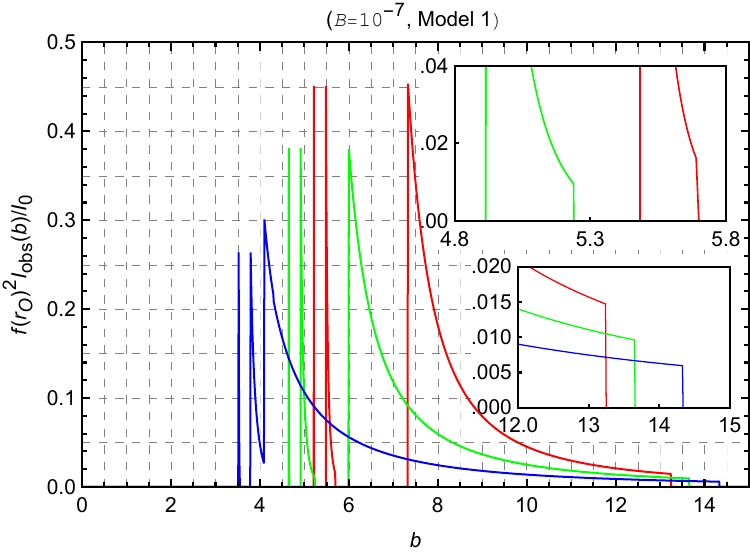}
\includegraphics[width=.28\textwidth]{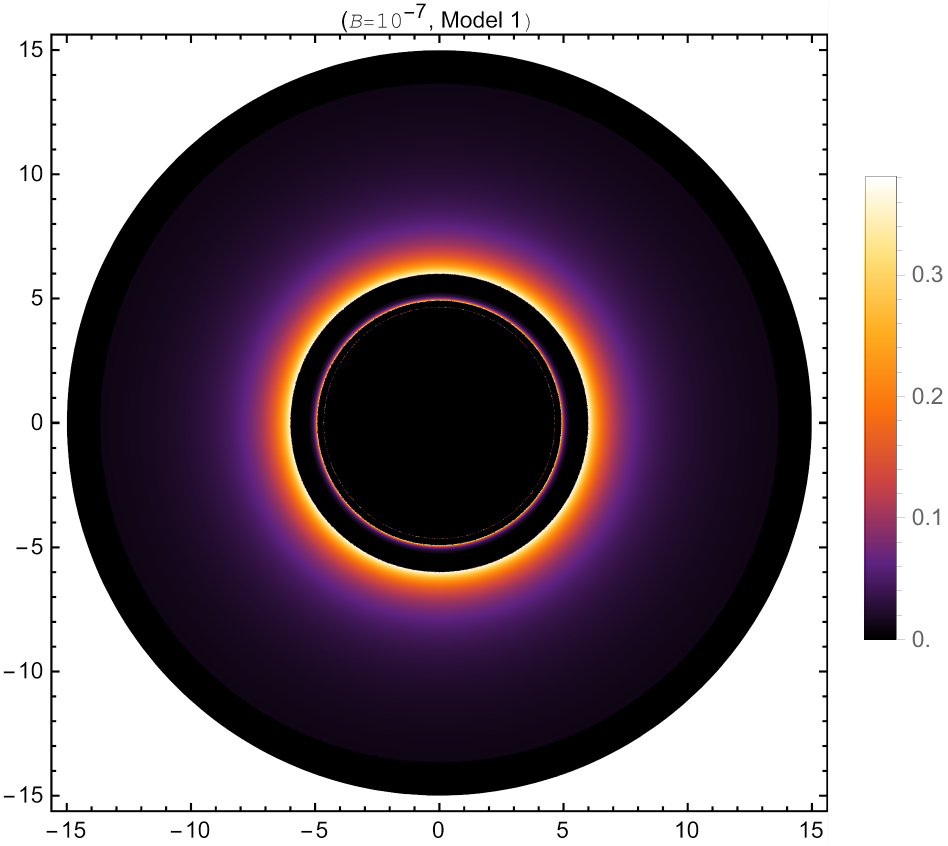}
\includegraphics[width=.34\textwidth]{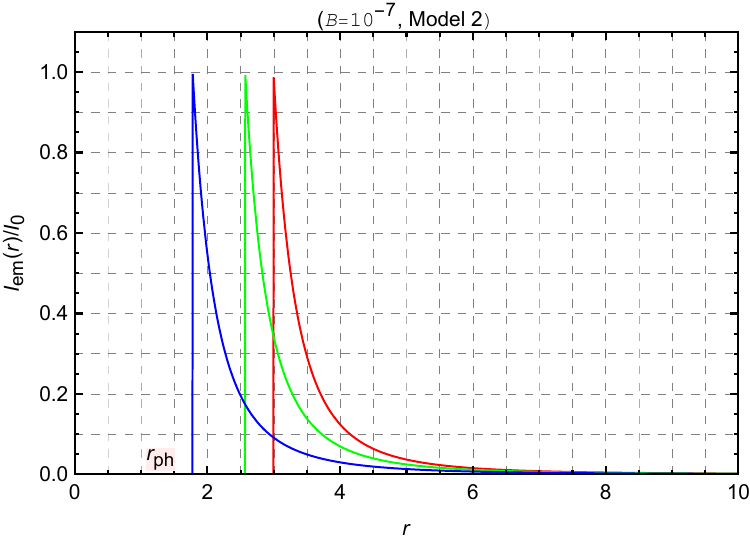}
\includegraphics[width=.34\textwidth]{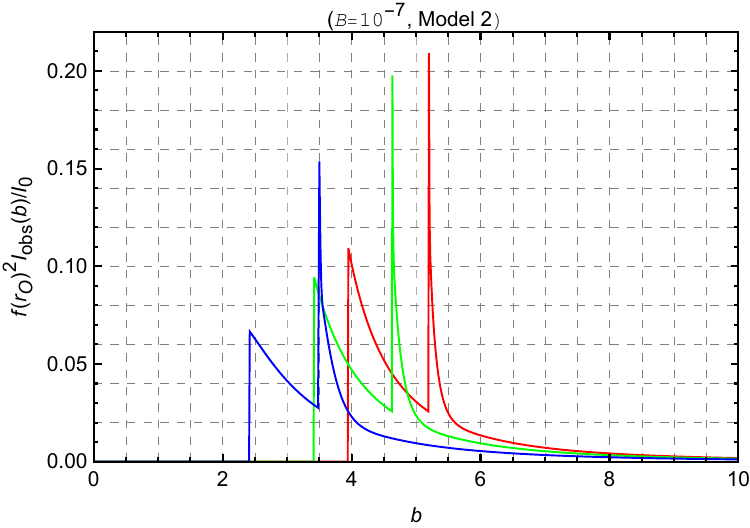}
\includegraphics[width=.28\textwidth]{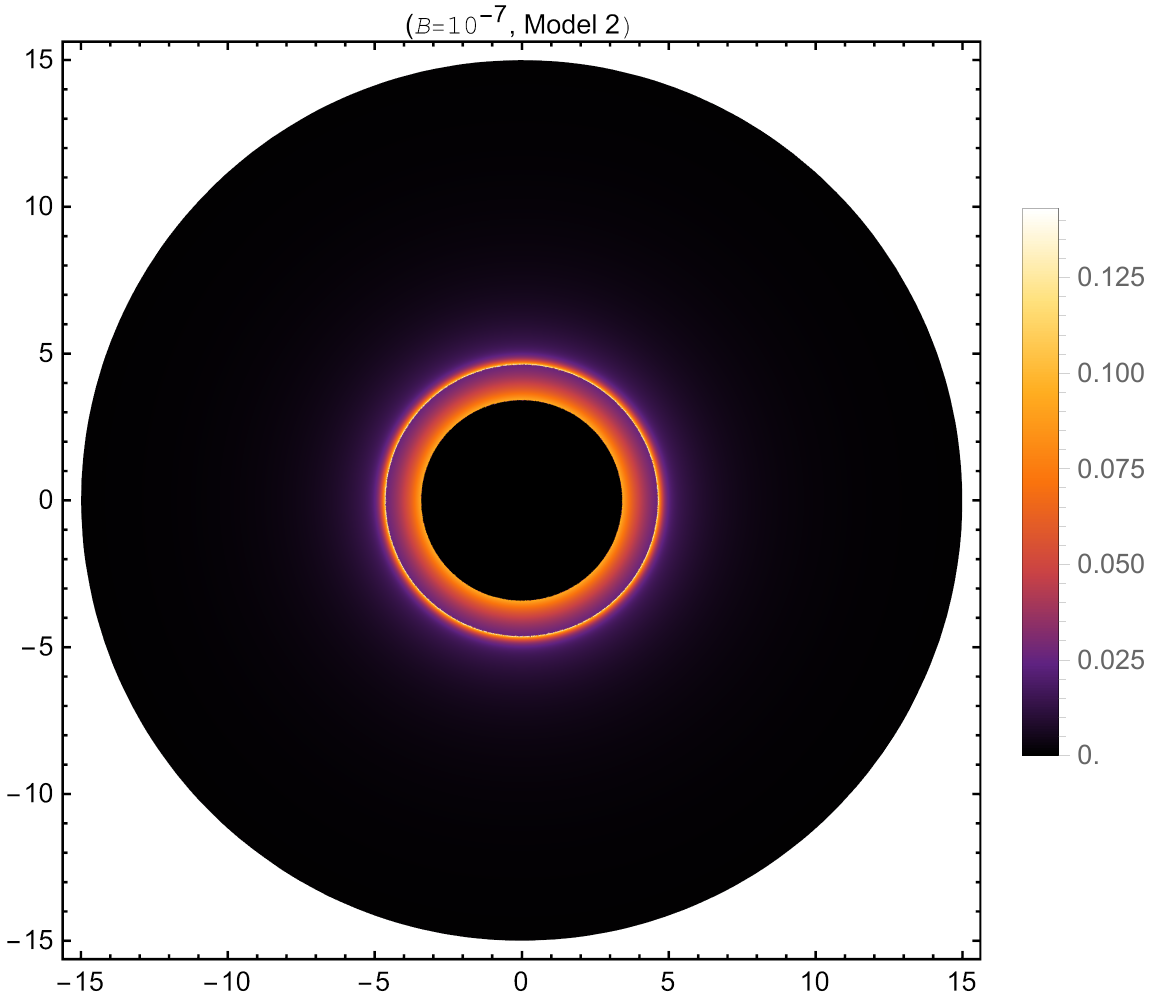}
\includegraphics[width=.34\textwidth]{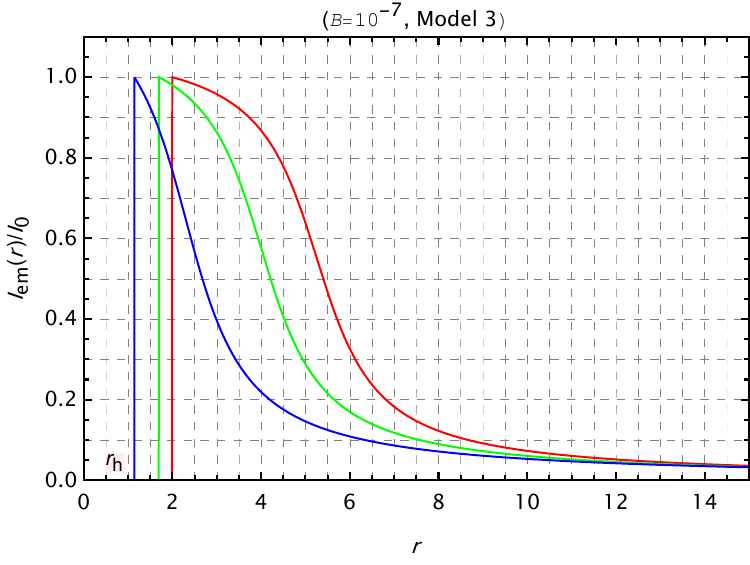}
\includegraphics[width=.34\textwidth]{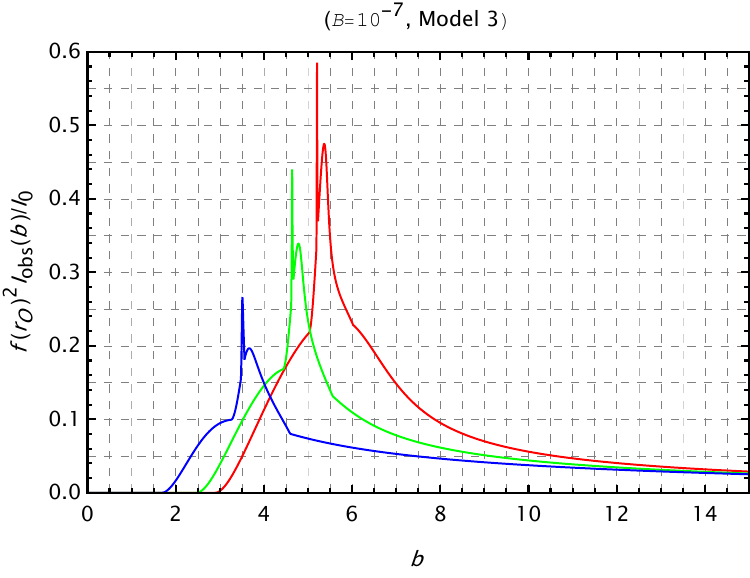}
\includegraphics[width=.28\textwidth]{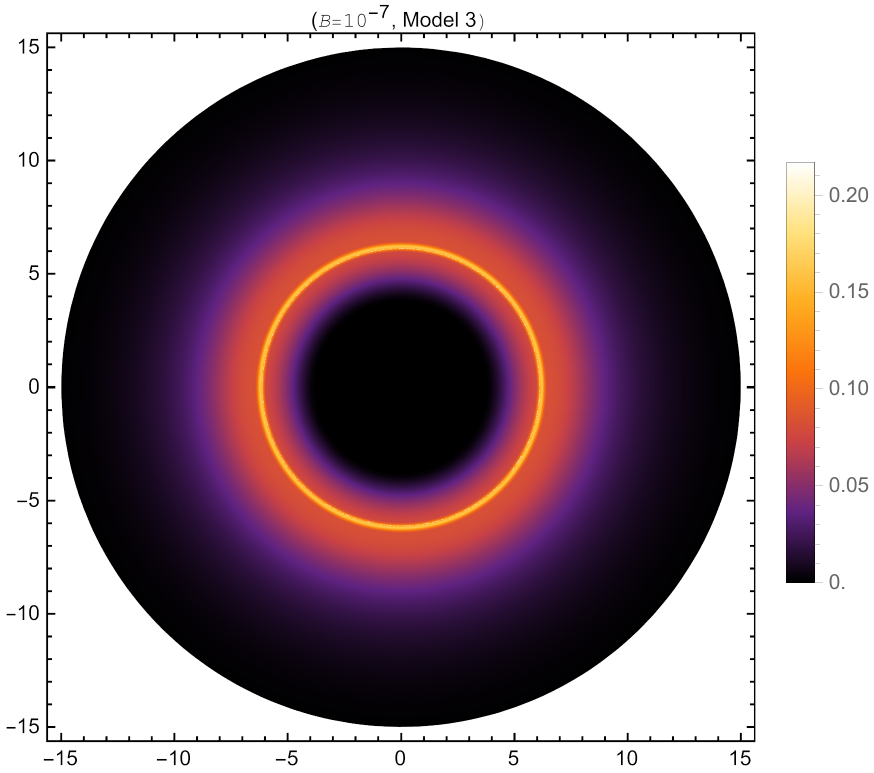}
\caption{\label{FigBm7}Observational appearances of a geometrically and optically thin disk with different emissivity profiles near BH with $M=1$, $B=10^{-7}$, viewed from a face-on orientation. \textbf{Left column:} The profiles of various emissions $I_{\mathrm{em}}(r)$. \textbf{Middle column:} The observed intensities $I_{\mathrm{obs}}(b)$ as a function of the impact parameter $b$. The red, green and blue curves correspond to $q=0.01$, $q=0.2$ and $q=0.4$, respectively. \textbf{Right column:} The 2-dim density plots of the observed intensities $I_{\mathrm{obs}}(b)$ for $q=0.2$.}
\end{figure}In Fig.~\ref{FigBm3}, the emission initiates from the photon sphere at $r_{\rm{ph}}=2.9762$ for $q=0.2$. The observed lensing ring and photon ring overlap with the direct image, which decays for $b>5.1334M$. The lensing ring exhibits a spike in brightness within the range of $6.1064M<b<6.4632M$, while the photon ring has an even narrower spike around $b\sim6.1095M$, which is challenging to distinguish from the lensing ring. It is evident that the lensing ring makes a minimal contribution to the total brightness, and the contribution of photon ring is negligible. The influence of the parameter $q$ on the emitted specific intensity and observed specific intensity is also apparent in Fig.~\ref{FigBm3}. With an increase in $q$, the $r_{\rm ph}$ decreases, causing a slight leftward shift in the $I_{\rm{em}}(r)$ curve. In the observed specific intensity plots, the peak of the direct image and the combined spike of the lensing ring and photon ring also move slightly to the left, accompanied by a modest decrease in these spikes as $q$ increases.

For the scenario with $B=10^{-7}$, the observed specific intensities are depicted in Fig.~\ref{FigBm7}. The middle column displays one-dimensional functions of $I_{\rm{obs}}(r)$ concerning $b$, while the right column presents two-dimensional density plots of $I_{\rm{obs}}(r)$, viewed face-on. In the first row, concerning the Model 1 emissivity profile, the emission from the accretion occurs within the domain between ISCO and OSCO (left panel). Here, we choose the case of $q=0.2$ to concretize our analysis.
\begin{itemize}
 \item The observed direct image diminishes for $b>5.9977M$ due to gravitational lensing, and the intensity experiences an abrupt drop to zero at the position $b\thicksim 13.6524M$, which is exactly the lensed position of $r_{\rm OSCO}$ .
 \item The observed lensing ring emission is confined to a narrow region $4.9150M < b <5.2381M$, where $b\thicksim 5.2381M$ is the position at which the observed lensing ring intensity experiences an abrupt drop to zero, also corresponding to $r_{\rm OSCO}$ in the emission profile.
 \item The photon ring emission appears as a spike at $b\thicksim4.6477M$, which is hardly visible in the right panel of the density plots. To discern the photon ring, one needs to zoom in on the plot.
\end{itemize}
Consequently, from the observer's perspective, the direct emission of the accretion contribute most in enhancing the brightness of the BH, the lensing ring contributes modestly to the total brightness, while the photon ring makes negligible contributions. We also observe the impact of the parameter $q$. With an increase in $q$, the spikes of direct emission, photon ring, and lensing ring in the observed intensity curve all shift leftward. Additionally, these spikes decrease. An interesting behavior is that as $q$ increases, the distance between the direct image and the lensing ring decreases. When $q=0.4$, the direct image and the lensing ring exhibit partial overlap.

In the second row of Fig.~\ref{FigBm7}, we illustrate the observed specific intensities related to the emissivity profile of Model 2, showcasing analogous patterns observed for $B=10^{-3}$. However, differences arise in the magnitudes of the intensities and the positions of the photon ring and lensing ring. The influence of the parameter $q$ on the observed outcomes mirrors that observed in the case of $B=10^{-3}$, albeit with a slightly greater impact.

Finally, in the third row of Fig.~\ref{FigBm7}, regarding the Model 3 emissivity profile, the emission extends to the event horizon at $r_{\mathrm{h}} =1.6962$ for $q=0.2$, and the decay of the emission is much more moderate compared to the cases of Model 1 and 2 emissivity profiles. The lensed position of the event horizon is identified as the inner edge of the observed intensity at $b\thicksim2.4683M$. The observed intensity increases outside the central dark area due to gravitational redshift. The very narrow spike observed at $b\thicksim4.6223M$ corresponds to the photon ring, while the broader bump within the range of $4.4375M < b <5.5704M$ is attributed to the lensing ring. In this instance, the lensing ring makes a substantial contribution to the observed intensity, whereas the photon ring remains entirely negligible. With an increase in the parameter $q$, both the inner edge and the peak of the observed intensity curve shift to the left, and the peak value decreases.
\section{Optical appearance with spherical accretions }
\label{spherical}
In this section, we will investigate the shadows and photon spheres related to spherical accretions under static and infalling accretion modes, respectively. We will examine the impacts of the CDF parameters $B$ and $q$ on the results.

\subsection{Static spherical accretion }
In this subsection, our focus is on investigating the shadow and photon sphere of a CDF-BH with a static spherical accretion. Specifically, we will concentrate on the specific intensity observed by the observer  $\rm (erg s^{-1} cm^{-2} str^{-1} Hz^{-1})$, which can be mathematically represented as described in \cite{Jaroszynski:1997bw, Bambi:2013nla}
\begin{equation} \label{intensity}
I_{\rm{obs}} = \int_\Gamma g^3 j (\nu_{\rm e}) dl_{\rm prop} ,
\end{equation}
where $g = \nu_{\rm o}/\nu_{\rm e}$ represents the redshift factor, $\nu_{\rm e}$ is the radiated
photon frequency, $\nu_{\rm o}$ is the observed photon frequency, $j (\nu_{ \rm e})$ denotes the
emissivity per unit volume measured in the static frame of the emitter, $dl_{\rm prop}$ is the
infinitesimal proper length, and $\Gamma$ stands for the trajectory of the light ray.
\begin{figure}[h]
\centering 
\includegraphics[width=.374\textwidth]{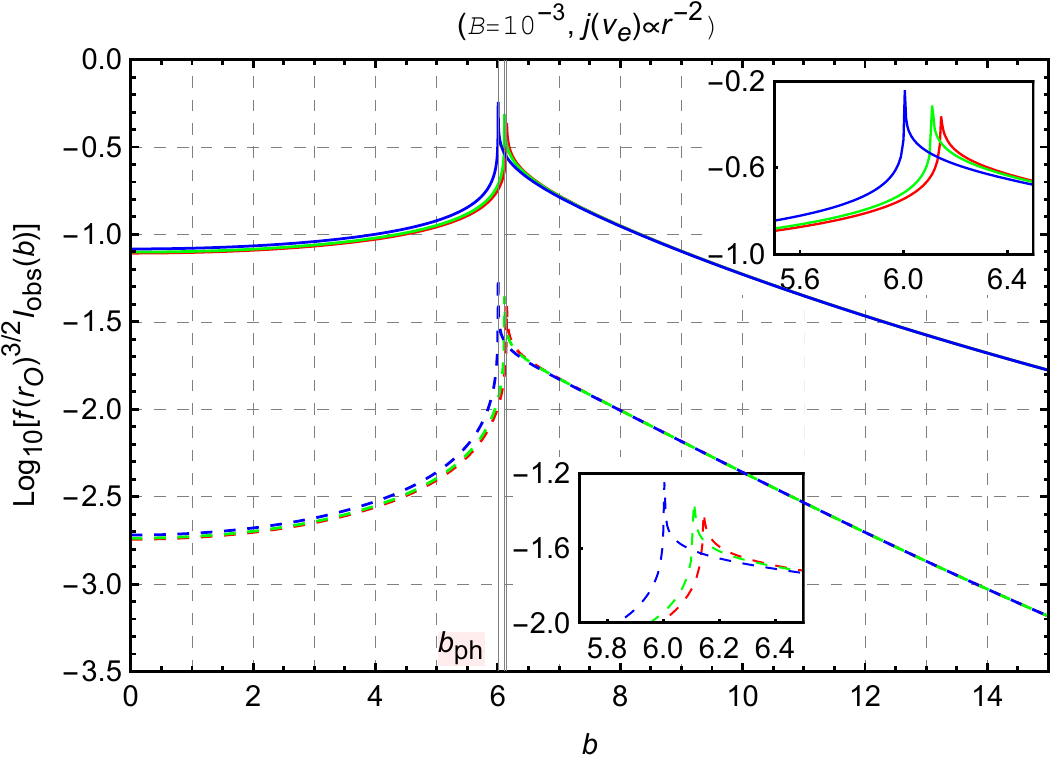}
\includegraphics[width=.305\textwidth]{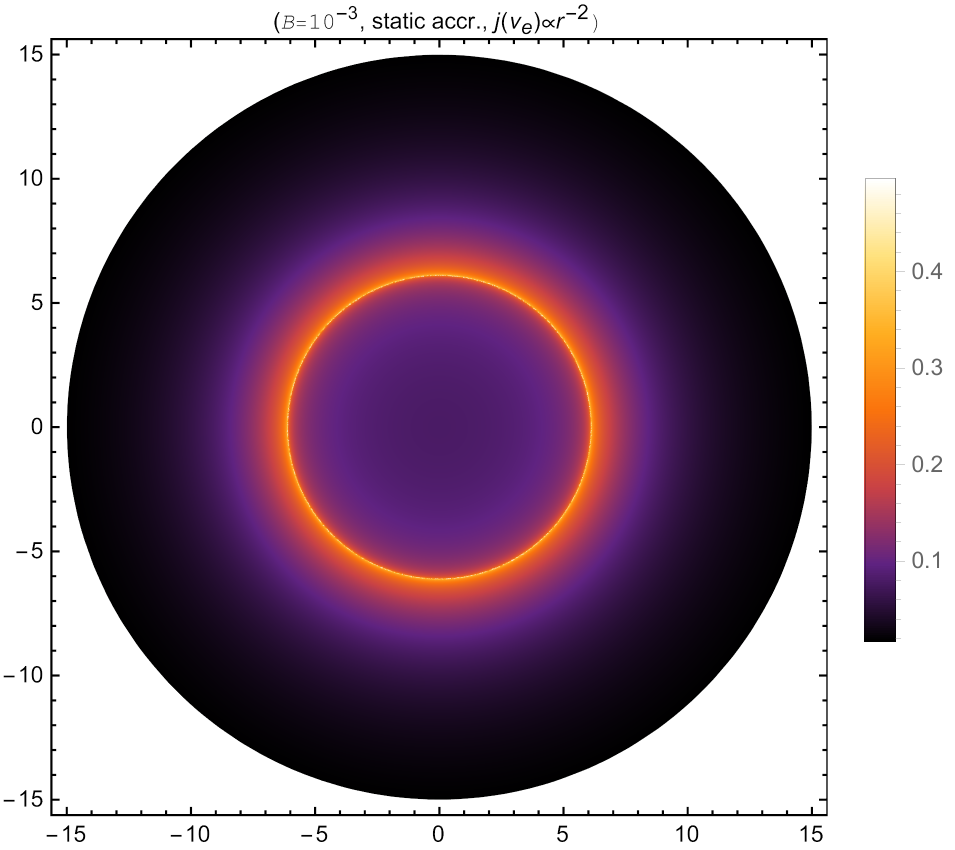}
\includegraphics[width=.305\textwidth]{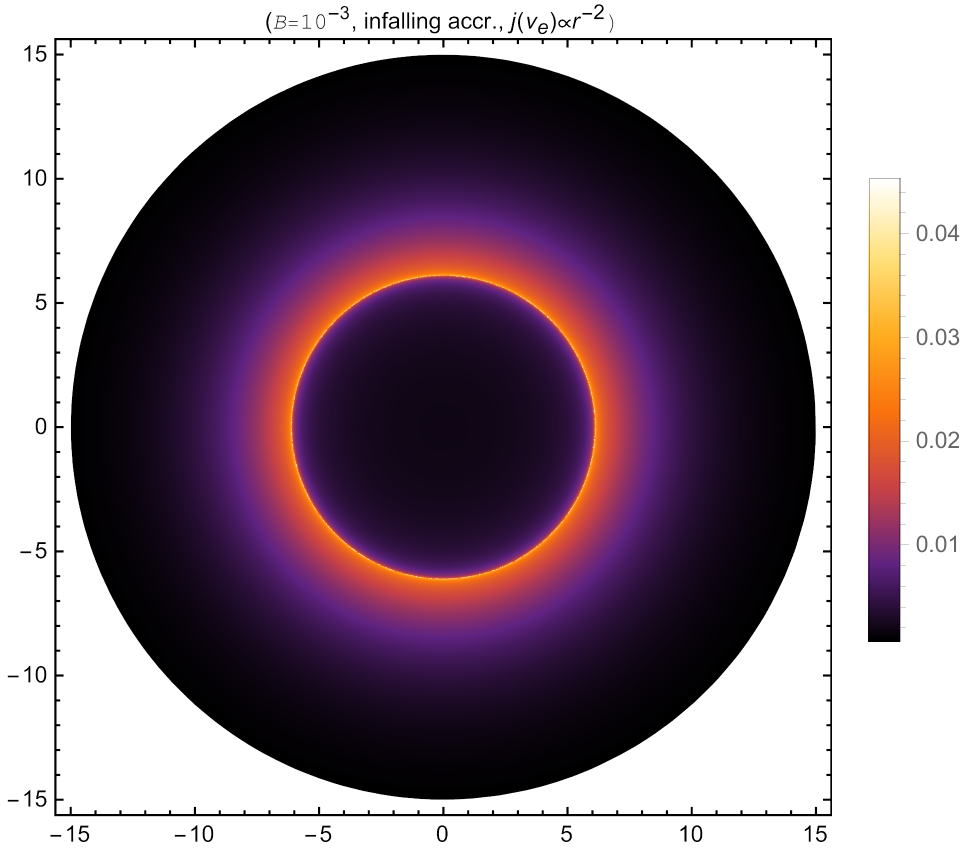}\\
\includegraphics[width=.374\textwidth]{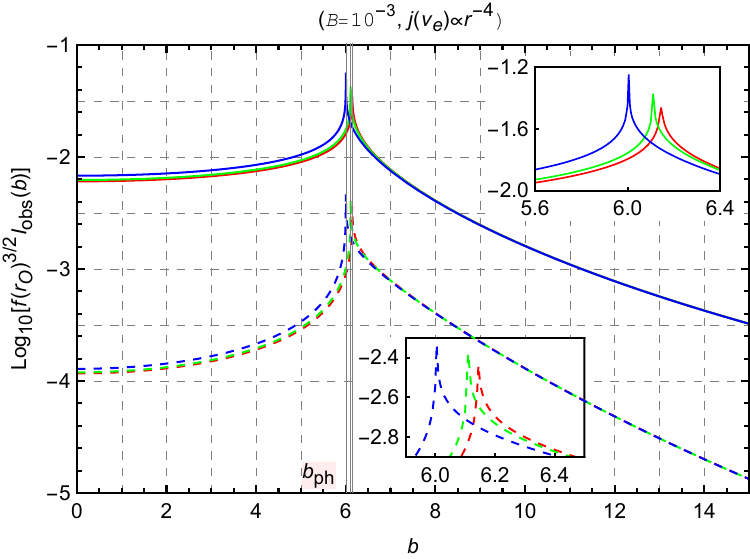}
\includegraphics[width=.305\textwidth]{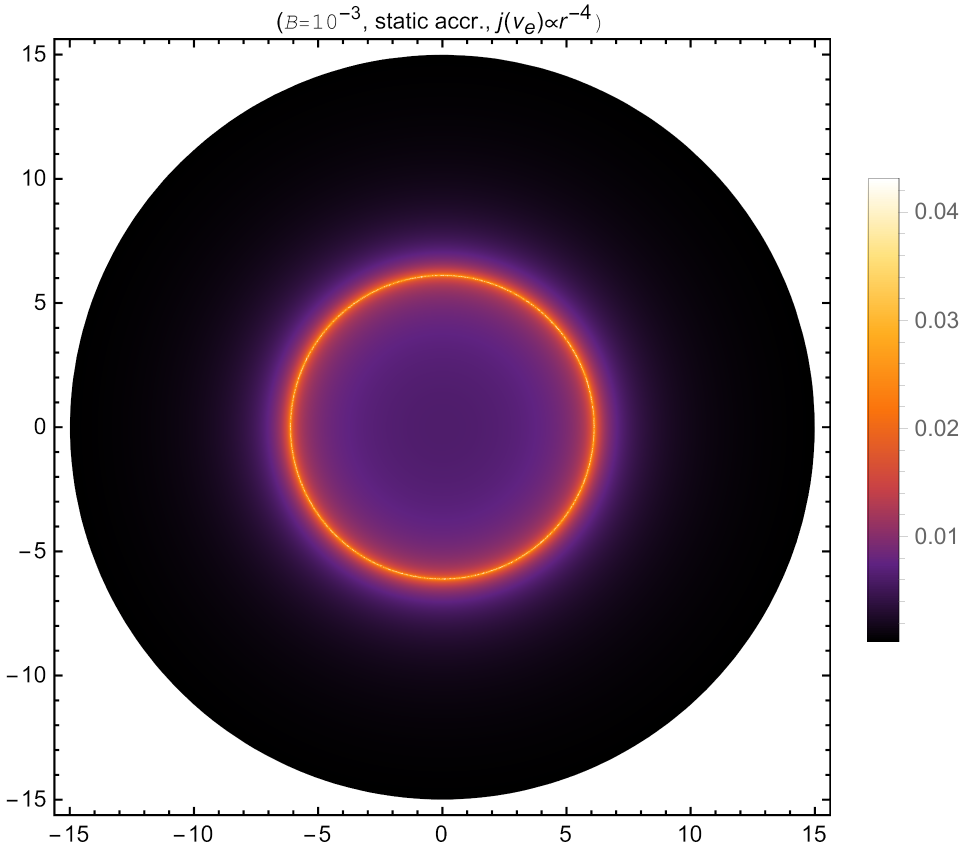}
\includegraphics[width=.305\textwidth]{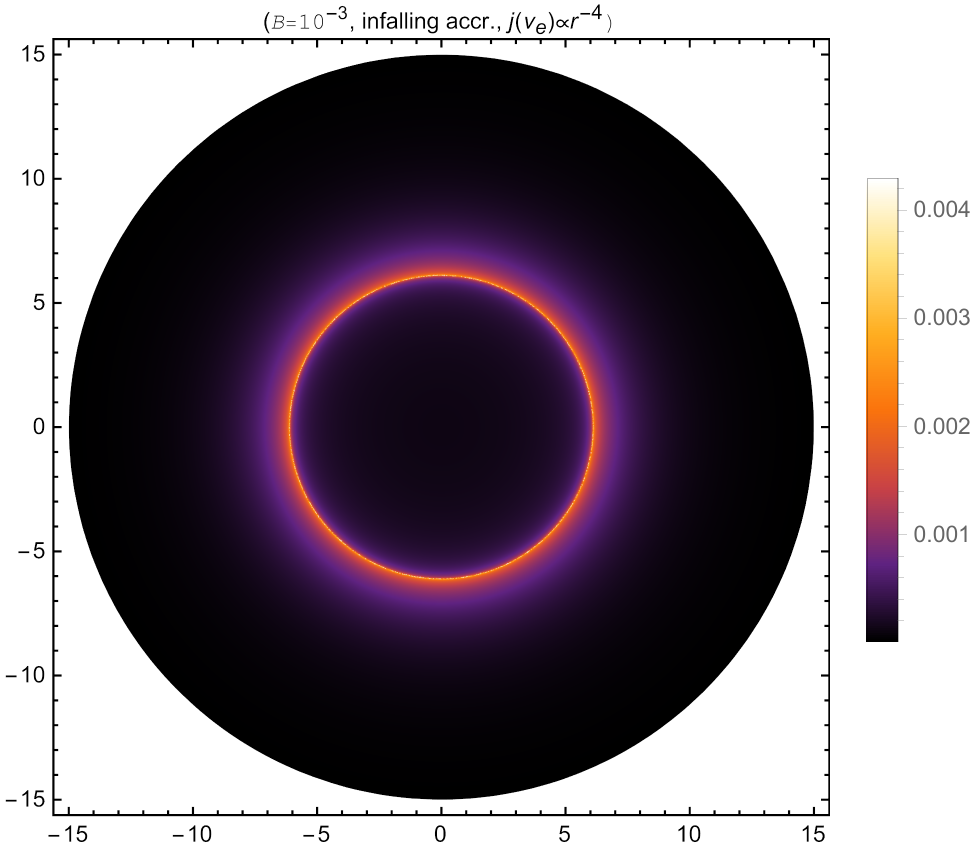}\\
\includegraphics[width=.374\textwidth]{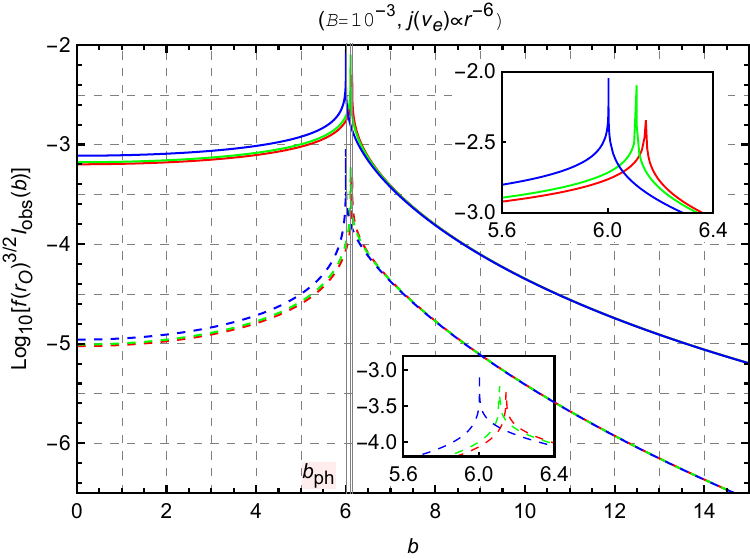}
\includegraphics[width=.305\textwidth]{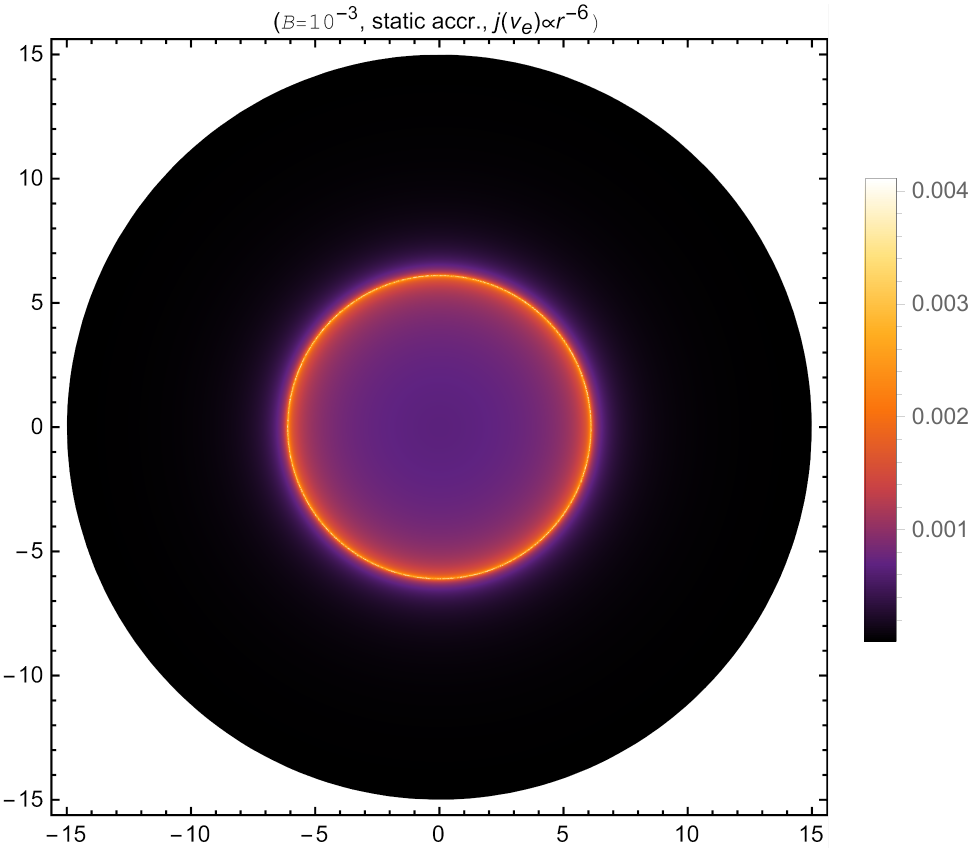}
\includegraphics[width=.305\textwidth]{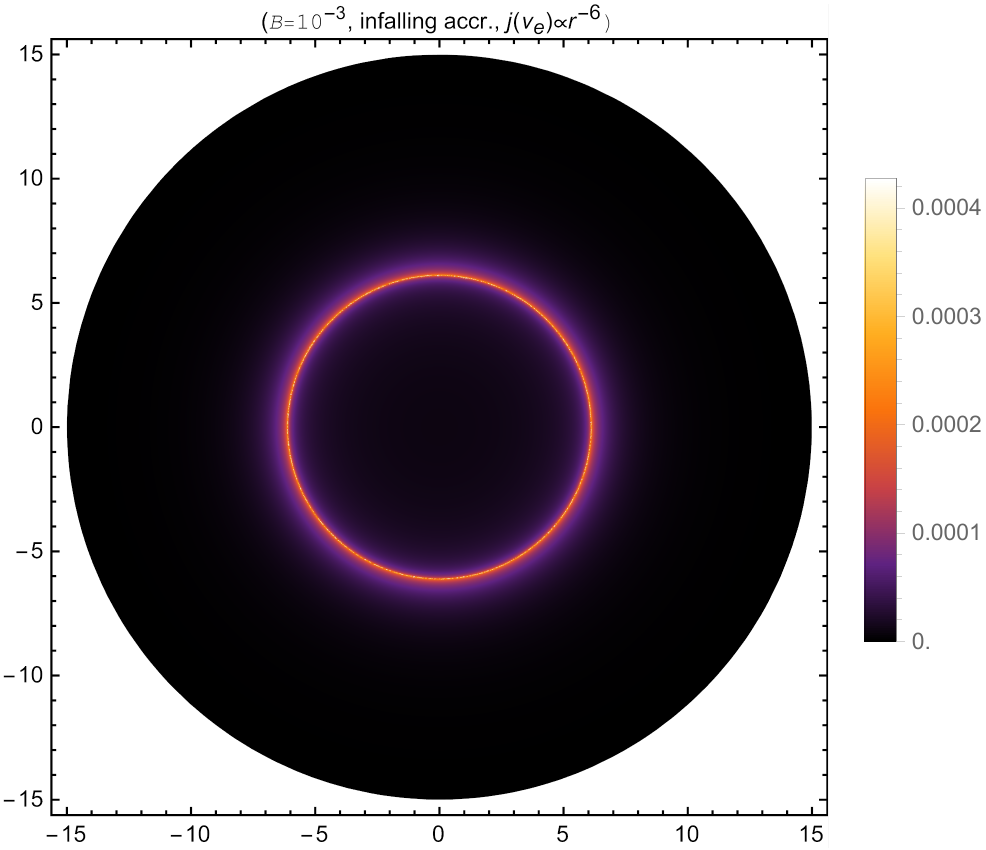}
\caption{\label{FigspheaccBm3} \textbf{Left column:} Profiles of the specific intensity $I_{\rm{obs}}(b)$ with spherical accretion cast by static (solid curves) and infalling (dashed curves) spherical accretion for different emissivity profiles, viewed face-on by an observer near the pseudo-cosmological horizon. The red, green and blue curves corresponds to $q=0.01$, $q=0.2$ and $q=0.4$, respectively. \textbf{Middle column:} Images of the BH shadows with static spherical accretion for $q=0.2$. \textbf{Right column:} Images of the BH shadows with infalling spherical accretion for $q=0.2$. The radial profile of emission from top to bottom is
$j_2 = 1/r^2$, $j_4 = 1/r^4$ and $j_6 = 1/r^6$, respectively. Here we set $M=1$ and $B=10^{-3}$.}
\end{figure}

In the spacetime of the CDF-BH, the redshift factor is given by $g=\left[f(r)/f(r_{\rm O})\right]^{1/2}$. Assuming monochromatic radiation with a fixed frequency $\nu_s$, the specific emissivity is expressed as
\begin{equation}
j (\nu_{\rm e})\propto \delta(\nu_e-\nu_s)j_l(r), \label{profile}
\end{equation}
where the radial profile is represented as $j_l(r)=1/r^l$, with $l$ assuming values of 2~\cite{Bambi:2013nla}, 4~\cite{Kocherlakota:2022jnz}, and 6~\cite{Bauer:2021atk}. The proper length measured in the rest frame of the emitter, according to Eq.~(\ref{ffr}), is
\begin{eqnarray}
dl_{\rm prop}&=&\sqrt{f(r)^{-1}d r^2+r^2 d \phi^2} \nonumber \\
  &=&\sqrt{f(r)^{-1}+r^2 \left(\frac {d \phi}{d r}\right)^2}~ dr, \label{dl}
\end{eqnarray}
where $ d \phi/dr$ is given by the inverse of Eq.~(\ref{drp}). Consequently, the specific intensity observed by the static observer becomes
\begin{equation}\label{finalintensity}
I_{\rm{obs}} = \int_\Gamma \left[\frac{f(r)}{f(r_{\rm O})}\right]^{3/2}\frac{1}{r^l} \sqrt{f(r)^{-1}+r^2 \left(\frac {d \phi}{dr}\right)^2}~ dr.
\end{equation}

Now, we will utilize Eq.~(\ref{finalintensity}) to explore the shadow of the CDF-BH with static spherical accretion. Since the intensity depends on the trajectory of the light ray, determined by the impact parameter $b$, we will investigate the variation of intensity with respect to the impact parameter $b$. The observed specific intensities are depicted in the left and middle panels of Fig.~\ref{FigspheaccBm3} for $B=10^{-3}$ and in Fig.~\ref{FigspheaccBm7} for $B=10^{-7}$. We present plots depicting $f(r_{\rm O})^{3/2}I_{\rm obs}(b)$ to conceal the influence of the observer's position on the observed intensity. Analyzing Figs.~\ref{FigspheaccBm3} and \ref{FigspheaccBm7}, it is evident that as $b$ increases, the intensity initially rises, reaches a peak at $b_{\rm{ph}}$ (e.g., when $q=0.2$, $b_{\rm{ph}}=6.1095M$ for $B=10^{-3}$ and $b_{\rm{ph}}=4.6333M$ for $B=10^{-7}$), and then rapidly decreases. For $b<b_{\rm{ph}}$, the intensity originating from the accretion matter is mostly absorbed by the BH, resulting in minimal observed intensity. At $b=b_{\rm{ph}}$, where the light ray revolves around the BH several times, the observed intensity peaks. Beyond $b_{\rm{ph}}$, only refracted light contributes to the observed intensity, and as $b$ increases, the refracted light diminishes, causing the observed intensity to vanish for sufficiently large $b$. For fixed values of $M$ and $B$, an increase in $q$ leads to a decrease in the radius of the photon sphere (consistent with Table~\ref{tab:physical_quantities}), accompanied by an increase in the observed intensity at $r_{\rm{ph}}$.

Upon comparing Figs.~\ref{FigspheaccBm3} and \ref{FigspheaccBm7}, we can examine the influence of the parameter $B$ on the observed intensity. For fixed values of $M$ and $q$, it is observed that a larger parameter $B$ results in a weaker intensity; specifically, the intensity of $B=10^{-7}$ is stronger than that of $B=10^{-3}$. The 2-dimensional plot of the shadows is presented in the middle columns of Figs.~\ref{FigspheaccBm3} and \ref{FigspheaccBm7}, with the impact parameter $b$ represented on the radius axis. Notably, the shadow exhibits circular symmetry for the CDF-BH, and beyond the BH, a bright ring, corresponding to the photon sphere, is evident. The radii of the photon sphere for different $B$ values are listed in Table~\ref{tab:physical_quantities}. Clearly, the outcomes depicted in Figs.~\ref{FigspheaccBm3} and \ref{FigspheaccBm7} align with the information presented in Table~\ref{tab:physical_quantities}; that is, an increased radius of the photon sphere $r_{\rm ph}$ corresponds to a larger value of the parameter $B$.
\subsection{Infalling  spherical accretion }
\begin{figure}[h]
\centering 
\includegraphics[width=.374\textwidth]{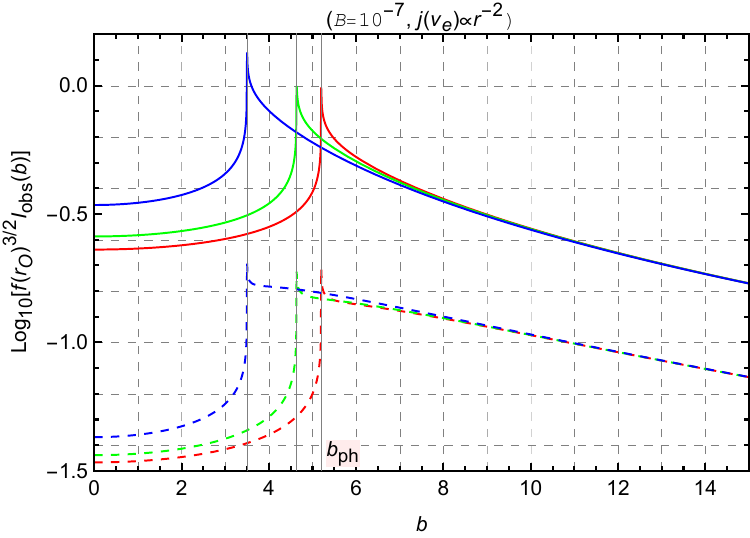}
\includegraphics[width=.305\textwidth]{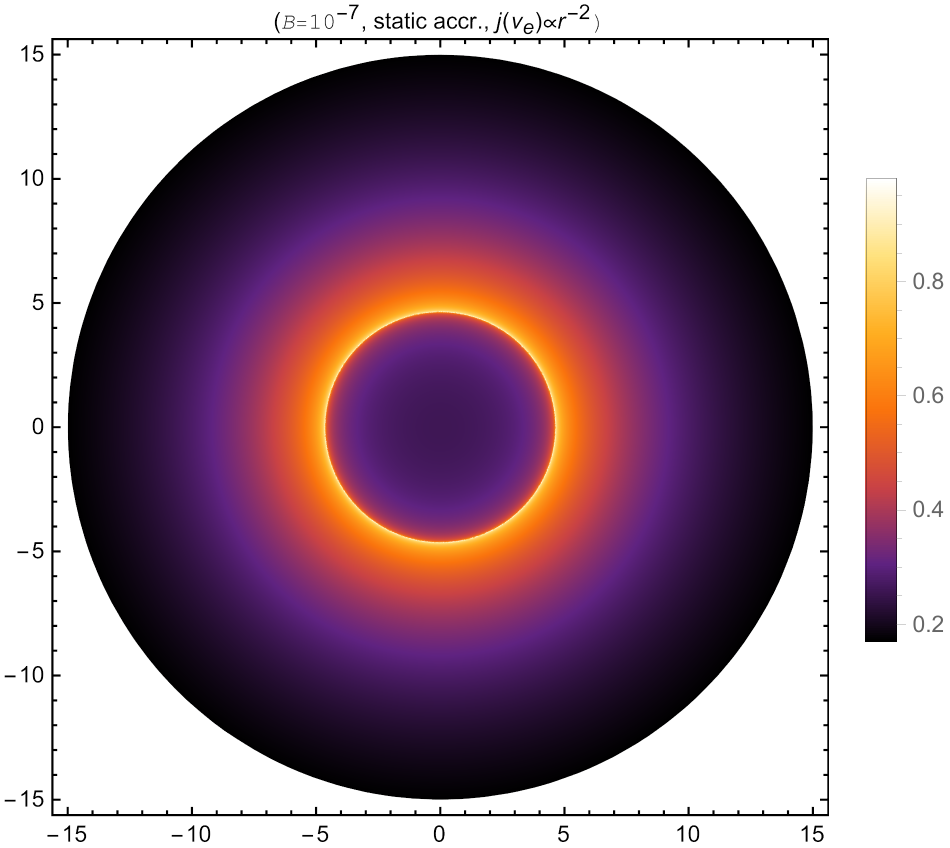}
\includegraphics[width=.305\textwidth]{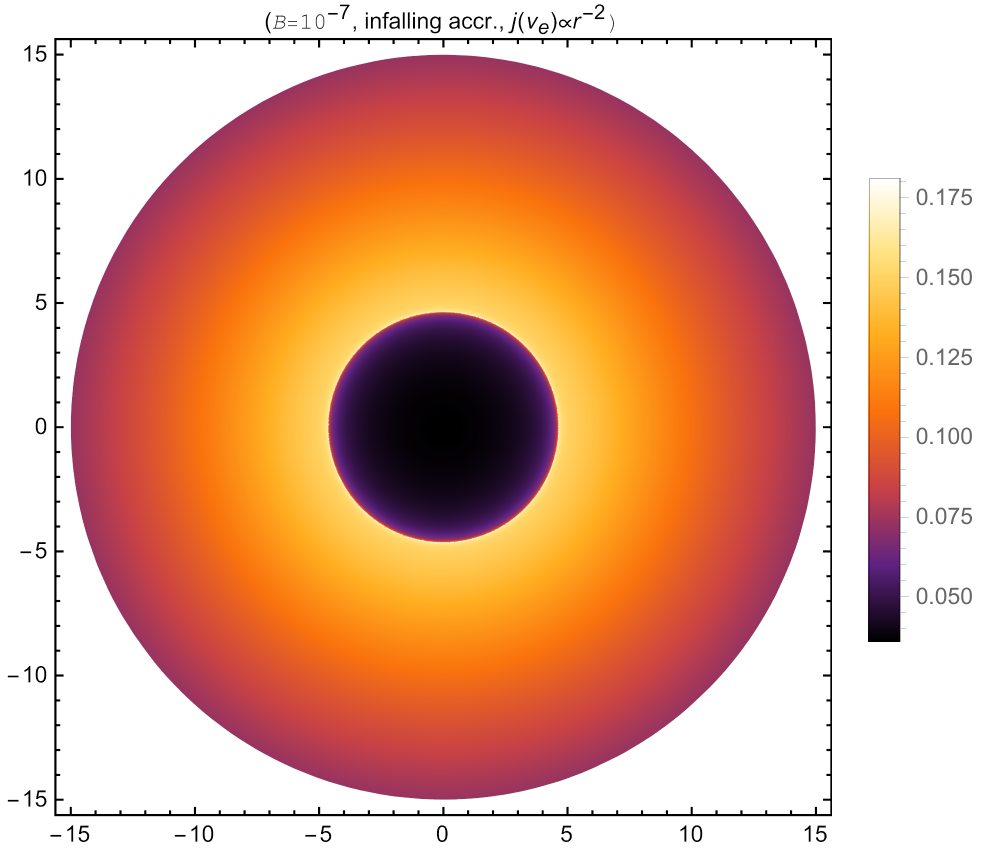}\\
\includegraphics[width=.374\textwidth]{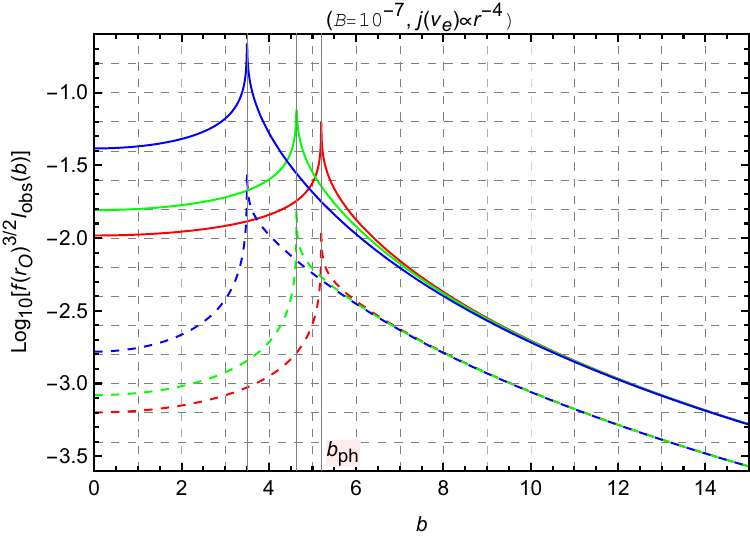}
\includegraphics[width=.305\textwidth]{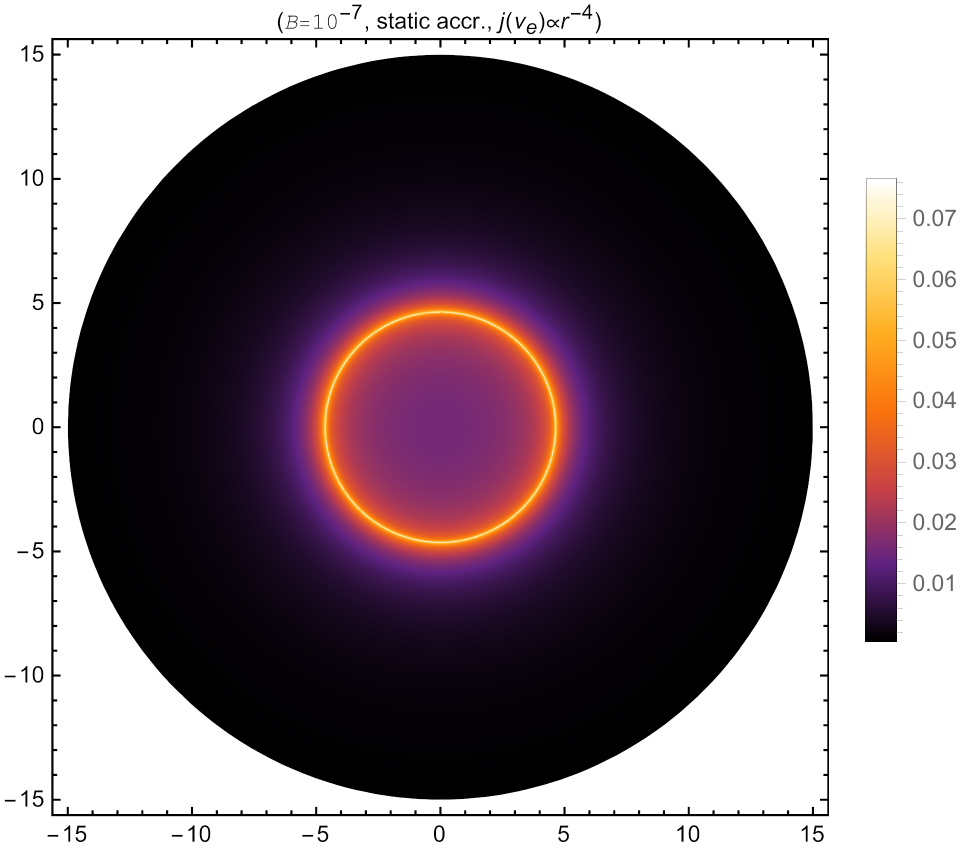}
\includegraphics[width=.305\textwidth]{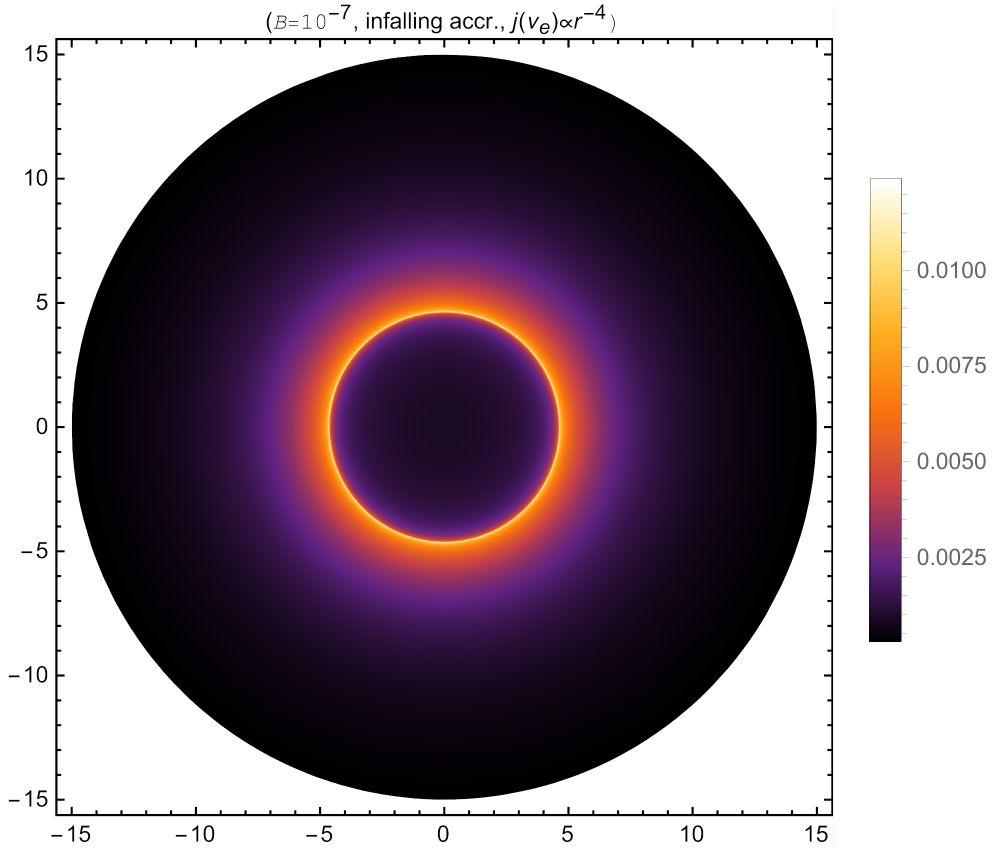}\\
\includegraphics[width=.37\textwidth]{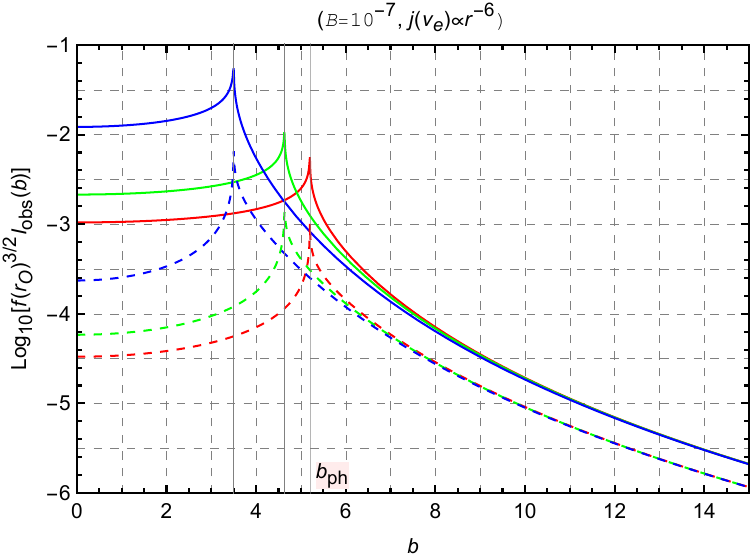}
\includegraphics[width=.305\textwidth]{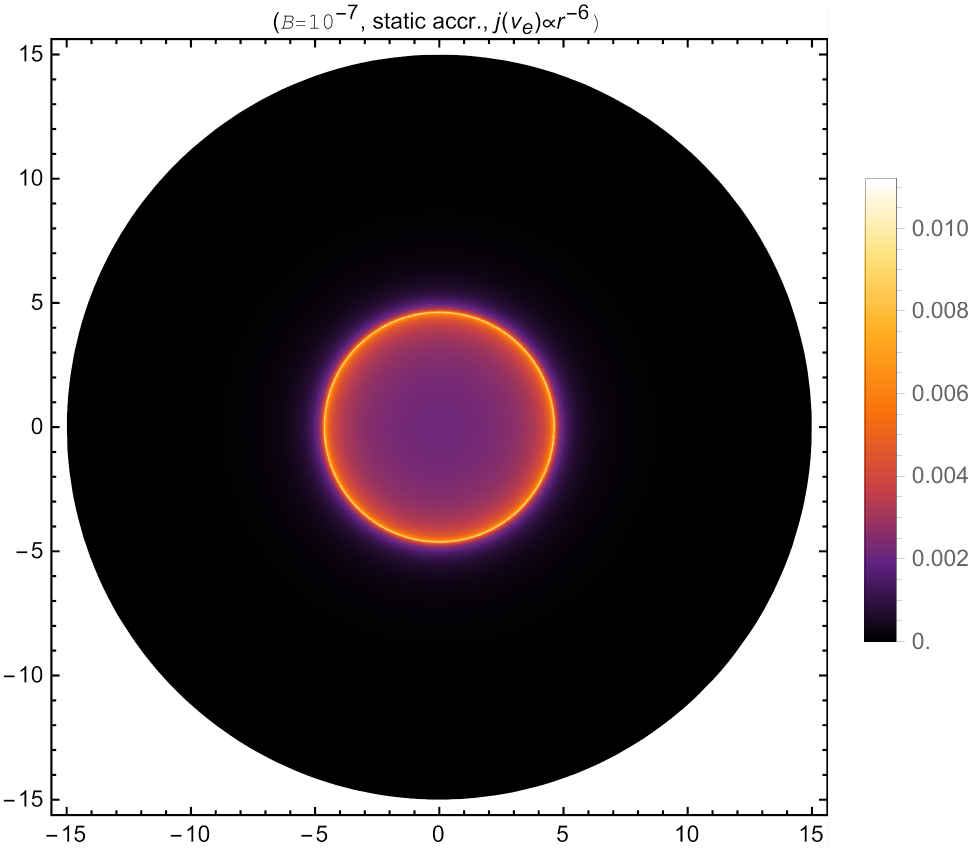}
\includegraphics[width=.305\textwidth]{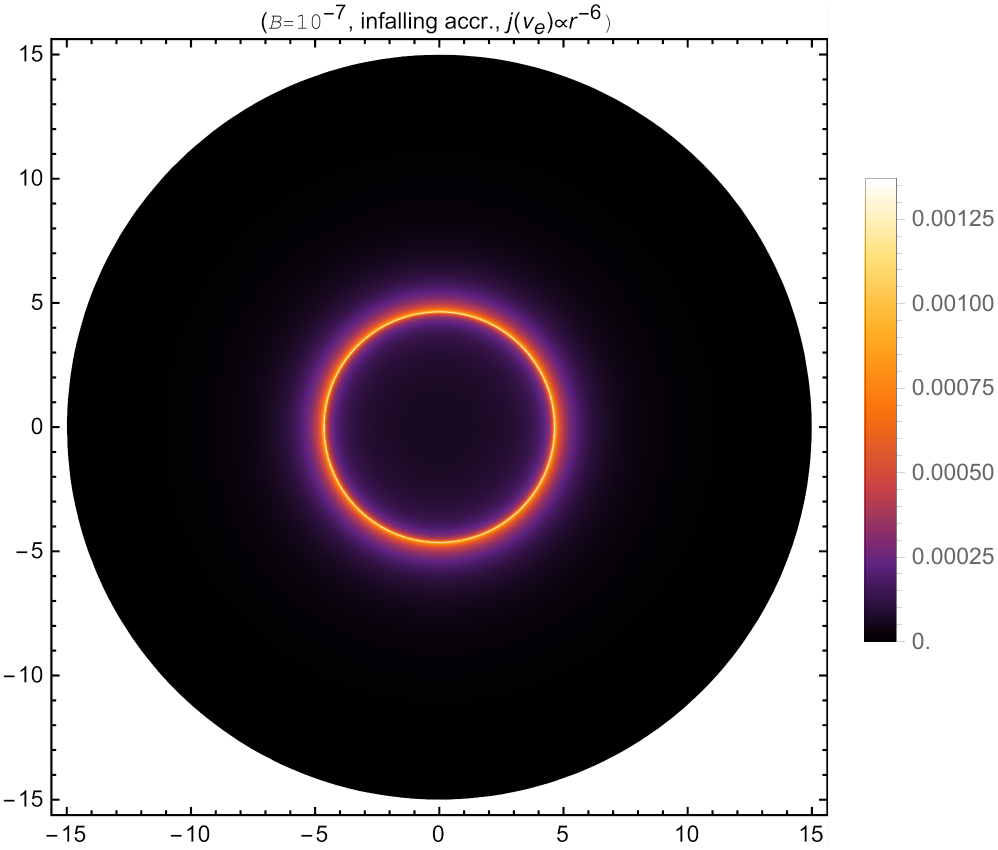}
\caption{\label{FigspheaccBm7} \textbf{Left column:} Profiles of the specific intensity $I_{\rm{obs}}(b)$ with spherical accretion cast by static (solid curves) and infalling (dashed curves) spherical accretion for different emissivity profiles, viewed face-on by an observer near the pseudo-cosmological horizon. The red, green and blue curves corresponds to $q=0.01$, $q=0.2$ and $q=0.4$, respectively. \textbf{Middle column:} Images of the BH shadows with static spherical accretion for $q=0.2$. \textbf{Right column:} Images of the BH shadows with infalling spherical accretion for $q=0.2$. The radial profile of emission from top to bottom is
$j_2 = 1/r^2$, $j_4 = 1/r^4$ and $j_6 = 1/r^6$, respectively. Here we set $M=1$ and $B=10^{-7}$.}
\end{figure}
In this section, we consider optically thin accretion involving infalling matter. This infalling model is deemed more realistic than the static accretion model, given the dynamic nature of most accretion matters in the universe. We continue to employ Eq.~(\ref{intensity}) to explore the shadow cast by infalling accretion. Unlike static accretion, the redshift factor for infalling accretion is linked to the velocity of the accretion, expressed as
\begin{equation}
g = \frac{k_\beta u^\beta_{\rm O}}{k_\gamma u^\gamma_{\rm e}} , \label{redf}
\end{equation}
where $k^\mu=\dot{x_\mu}$ is the four-velocity of the photon, $u^\mu_{\rm O} = (\frac{1}{\sqrt{f(r_{\rm O})}},0,0,0)$ is the four-velocity of the static observer, and $u^\mu_{\rm e}$ is the four-velocity of the accretion under consideration, defined as
\begin{eqnarray}
u^t_{\rm e} = \frac{1}{f(r)},~~
u^r_{\rm e} = - \sqrt{ 1 - f(r) },~~
u^\theta_{\rm e} = u^\phi_{\rm e} = 0.
\end{eqnarray}
The four-velocity of the photon can be obtained from Eq.~(\ref{time}) to Eq.~(\ref{radial}) by setting $\delta=0$. Redefining the affine parameter $s$ as $s/|L|$, $k_t$ can be regarded as a constant $k_t=1/b$, and $k_r$ can be determined from the equation $k_\alpha k^\alpha = 0$, yielding
\begin{equation}
\frac{k_r}{k_t} = \pm \frac{1}{f(r)}\sqrt{ 1 - \frac{b^2 f(r)}{r^2} } , \label{krkt}
\end{equation}
with the $\pm$ corresponding to the photon approaching/moving away from the BH. With Eq.~(\ref{krkt}), the redshift factor in Eq.~(\ref{redf}) can be simplified as
\begin{equation}
g = \frac{u^t_{\rm O}}{u^{t}_e+k_r/k_t u^r_e}, \label{sredf}
\end{equation}
which differs from the static accretion case.

Additionally, the proper distance is defined as
\begin{equation}
dl_{\rm prop} = k_\mu u^\mu_{\rm e}ds=\frac{k_t}{g | k_r |} dr.
\end{equation}
Regarding the specific emissivity, we maintain the assumption of monochromatic emission, allowing Eq.~(\ref{profile}) to remain valid. Integrating Eq.~(\ref{intensity}) over the observed frequencies, we obtain
\begin{equation}
I_{\rm{obs}} \propto \int_\Gamma \frac{g^3 k_t dr}{r^l | k_r |}. \label{finaltensity}
\end{equation}
Notably, there is a factor of $\frac{1}{f(r_{\rm O})^{3/2}}$ in the right hand side of Eq.~(\ref{finaltensity}), thus we also employ $f(r_{\rm O})^{3/2}I_{\rm obs}(b)$ to numerically investigate the shadow of the CDF-BH with infalling accretion. Also note that, there is an absolute sign for $ k_r$ in the denominator. Therefore, when the photon changes its motion direction, the sign before $ k_r$ should also change. For different values of the parameter $q$, the observed intensity with respect to $b$ is depicted in the left and right panels in Fig.~\ref{FigspheaccBm3} for $B=10^{-3}$ and Fig.~\ref{FigspheaccBm7} for $B=10^{-7}$. From these figures, it is observed that as $b$ increases, the intensity also increases up to a peak at $b=b_{\rm{ph}}$, after which it decreases to smaller values. This behavior is similar to that in static accretion. However, the relative observed intensity inside the brightest ring for infalling emissions is lower compared to that for static emissions. For fixed values of $M$ and $B$, as $q$ increases, the observed intensity at $r_{\rm{ph}}$ increases. The effect of $B$ on the intensity is also evident, with the intensity decreasing as the value of $B$ increases; that is, the intensity for $B=10^{-7}$ is stronger than $B=10^{-3}$. The two-dimensional image of the intensity is presented in the right columns in Figs.~\ref{FigspheaccBm3} and~\ref{FigspheaccBm7}. It is apparent that the radii of the shadows and the locations of the photon spheres remain consistent with the static case. This implies that the motion of the accretion does not affect the radii of the shadows and the locations of the photon spheres. However, in contrast to static accretion, the central region of the intensity for infalling accretion is darker, a phenomenon attributed to the Doppler effect.

\section{Conclusions and discussions}
\label{conclusion}
We investigated the geodesic structure, shadow, and optical appearance of BH surrounded by the CDF. Based on the analysis of the effective potential for timelike particles in the CDF-BH spacetime, we found that the existence of the ISCO and OSCO depends on the parameters $B$ and $q$. The analysis of epicyclic frequencies and orbital stability for different CDF parameters corroborates the analysis of the existence of ISCO/OSCO. We also analyzed the dependence of conserved quantities and Keplerian frequencies on CDF parameters. Based on the analysis of the effective potential for null particles, we found that the photon orbits are all unstable. As CDF-BH possesses a cosmological event horizon, the observation of the shadow radius depends on the observer's position and whether the observer is at rest. Assuming the asymptotic behavior of CDF-BH spacetime at infinity to be identical to the de Sitter spacetime governed by the cosmological constant responsible for our expanding universe, and considering observers located radially far from both the event horizon and the cosmological horizon, we used EHT observations of the BH shadow radius to constrain the parameters of the CDF model.

Assuming the observer is positioned near the cosmological horizon within the domain of outer communications, we explored the optical representations of the CDF-BH surrounded by both a geometrically and optically thin disk accretion and a spherically symmetric accretion. Our findings indicate that, in terms of the observed specific intensities, a dark interior (shadow), lensing ring, and photon ring emerge as a result of the thin disk accretion. However, their specific positions are dependent on the emissivity profile of the accretion near the BH. We have a particular interest in the Model 1 emission profile because it is associated with the ISCO and OSCO of timelike particles. The presence of OSCO results in sudden transitions to zero in the observed direct image intensity and lensing ring intensity. This is distinct from the situation in asymptotically flat spacetime, such as Schwarzschild black hole and RN BH, where OSCO does not exist. Despite these distinctions, the direct emission of the accretion plays a significant role in enhancing the brightness of the BH, while the contribution from the lensing ring is minor, and that from the photon ring is negligible. The CDF parameters $B$ and $q$ also play a crucial role in the formation of optical images of the BH, by affecting features such as the event horizon radius, photon sphere radius and the radius of the ISCO/OSCO, as well as the distance between the observer and the emitting matter. We conducted an investigation into static and infalling spherically symmetric accretions, revealing that the images of the BH exhibit a dark interior and a photon sphere. Notably, in the case of infalling accretion, the interior appears darker compared to the static scenario, a phenomenon attributed to the Doppler effect induced by infalling matter. The positions of the photon spheres in the images undergo alterations with varying CDF parameters.

It is important to acknowledge the limitation of this study, namely the lack of study on
the impact of expanding universe on the shadow and optical appearance of CDF-BH. Indeed,
simultaneously constraining dark energy models using both observations of expanding universe
and BH shadow is a non-trivial task. On one hand, observations of BH shadow depend
on the propagation of light through cosmological space, and hence on the evolution of the
cosmological background. Therefore, studying the shadow and optical appearance of BH in an
expanding universe requires simultaneous investigation into the universe expanding processes
governed by dark energy and other cosmological components, as well as the propagation of
light in the strong gravitational region of BH and in the expanding universe. On the other
hand, the relationship between the equations of state of dark energy in FLRW spacetime
and black hole spacetime is not straightforward. We have the following considerations: the
field properties of dark energy are unknown, and if we divide its energy into kinetic and
potential components, in homogeneous and isotropic FLRW spacetime, the kinetic component
manifests as variations in the field with respect to time, whereas in black hole spacetime, the
kinetic component manifests as variations in the field with respect to space coordinates. Thus,
describing the equations of state of dark energy in an expanding universe background and in
a curved BH spacetime may have a non-trivial relationship. In the literatures studying BHs
immersed in cosmological dark fluids, researchers directly introduce linear~\cite{Kiselev:2002dx} or nonlinear~\cite{Li:2019lhr,Li:2019ndh,Li:2022csn,Li:2023zfl} equations of state in the curved BH spacetime, rather than in the spatially flat universe
spacetime. This is why, in our study, we refer to the dark fluid involved as Chaplygin-like dark
fluid, rather than Chaplygin gas, which have already been introduced in FLRW spacetime to
illustrate the evolution of our universe. In summary, our discussions in this paragraph suggest potential areas for future investigation, such as the description of a cosmological fluid in
different spacetimes, the shadow of a BH in an expanding universe governed by various dark
energy models, as well as the effects of expanding universe on the ring structure in BH image.

\acknowledgments

This work is supported  by the National Natural Science Foundation of China under Grant No. 12305070 and the Basic Research Program of Shanxi Province under Grant No. 20210302123152, 202303021222018, 202303021221033. The authors would like to express their gratitude to Minyong Guo from Beijing Normal University for his assistance with the Mathematica program for computing black hole optical images.




\begin{thebibliography}{99}

\bibitem{Narayan:2019imo}
R.~Narayan, M.~D.~Johnson and C.~F.~Gammie,
``The Shadow of a Spherically Accreting Black Hole,''
Astrophys. J. Lett. \textbf{885} (2019) no.2, L33 [arXiv:1910.02957 [astro-ph.HE]].

\bibitem{EventHorizonTelescope:2019dse}
K.~Akiyama \textit{et al.} [Event Horizon Telescope],
``First M87 Event Horizon Telescope Results. I. The Shadow of the Supermassive Black Hole,''
Astrophys. J. Lett. \textbf{875} (2019), L1 [arXiv:1906.11238 [astro-ph.GA]].

\bibitem{EventHorizonTelescope:2019uob}
K.~Akiyama \textit{et al.} [Event Horizon Telescope],
`First M87 Event Horizon Telescope Results. II. Array and Instrumentation,''
Astrophys. J. Lett. \textbf{875} (2019) no.1, L2 [arXiv:1906.11239 [astro-ph.IM]].


\bibitem{EventHorizonTelescope:2019jan}
K.~Akiyama \textit{et al.} [Event Horizon Telescope],
``First M87 Event Horizon Telescope Results. III. Data Processing and Calibration,''
Astrophys. J. Lett. \textbf{875} (2019) no.1, L3 [arXiv:1906.11240 [astro-ph.GA]].

\bibitem{EventHorizonTelescope:2019ths}
K.~Akiyama \textit{et al.} [Event Horizon Telescope],
``First M87 Event Horizon Telescope Results. IV. Imaging the Central Supermassive Black Hole,''
Astrophys. J. Lett. \textbf{875} (2019) no.1, L4 [arXiv:1906.11241 [astro-ph.GA]].

\bibitem{EventHorizonTelescope:2019pgp}
K.~Akiyama \textit{et al.} [Event Horizon Telescope],
``First M87 Event Horizon Telescope Results. V. Physical Origin of the Asymmetric Ring,''
Astrophys. J. Lett. \textbf{875} (2019) no.1, L5 [arXiv:1906.11242 [astro-ph.GA]].

\bibitem{EventHorizonTelescope:2019ggy}
K.~Akiyama \textit{et al.} [Event Horizon Telescope],
``First M87 Event Horizon Telescope Results. VI. The Shadow and Mass of the Central Black Hole,''
Astrophys. J. Lett. \textbf{875} (2019) no.1, L6 [arXiv:1906.11243 [astro-ph.GA]].

\bibitem{EventHorizonTelescope:2022wkp}
K.~Akiyama \textit{et al.} [Event Horizon Telescope],
``First Sagittarius A* Event Horizon Telescope Results. I. The Shadow of the Supermassive Black Hole in the Center of the Milky Way,''
Astrophys. J. Lett. \textbf{930} (2022) no.2, L12
[arXiv:2311.08680 [astro-ph.HE]].

\bibitem{Takahashi:2004xh}
R.~Takahashi,
``Shapes and positions of black hole shadows in accretion disks and spin parameters of black holes,''
J. Korean Phys. Soc. \textbf{45} (2004), S1808-S1812 [arXiv:astro-ph/0405099 [astro-ph]].

\bibitem{Synge:1966okc}
J.~L.~Synge,
``The Escape of Photons from Gravitationally Intense Stars,''
Mon. Not. Roy. Astron. Soc. \textbf{131} (1966) no.3, 463-466.


\bibitem{Bardeen:1972fi}
J.~M.~Bardeen, W.~H.~Press and S.~A.~Teukolsky,
``Rotating black holes: Locally nonrotating frames, energy extraction, and scalar synchrotron radiation,''
Astrophys. J. \textbf{178} (1972), 347.

\bibitem{Luminet:1979nyg}
J.~P.~Luminet,
``Image of a spherical black hole with thin accretion disk,''
Astron. Astrophys. \textbf{75} (1979), 228-235.

\bibitem{Falcke:1999pj}
H.~Falcke, F.~Melia and E.~Agol,
``Viewing the shadow of the black hole at the galactic center,''
Astrophys. J. Lett. \textbf{528} (2000), L13 [arXiv:astro-ph/9912263 [astro-ph]].

\bibitem{Perlick:2018iye}
V.~Perlick, O.~Y.~Tsupko and G.~S.~Bisnovatyi-Kogan,
``Black hole shadow in an expanding universe with a cosmological constant,''
Phys. Rev. D \textbf{97} (2018) no.10, 104062
[arXiv:1804.04898 [gr-qc]].

\bibitem{Gibbons:2008rj}
G.~W.~Gibbons and M.~C.~Werner,
``Applications of the Gauss-Bonnet theorem to gravitational lensing,''
Class. Quant. Grav. \textbf{25} (2008), 235009
[arXiv:0807.0854 [gr-qc]].

\bibitem{Werner:2012rc}
M.~C.~Werner,
``Gravitational lensing in the Kerr-Randers optical geometry,''
Gen. Rel. Grav. \textbf{44} (2012), 3047-3057
[arXiv:1205.3876 [gr-qc]].

\bibitem{Jusufi:2018kry}
K.~Jusufi,
``Gravitational deflection of relativistic massive particles by Kerr black holes and Teo wormholes viewed as a topological effect,''
Phys. Rev. D \textbf{98} (2018) no.6, 064017
[arXiv:1806.01256 [gr-qc]].

\bibitem{Jusufi:2017mav}
K.~Jusufi and A.~\"Ovg\"un,
``Gravitational Lensing by Rotating Wormholes,''
Phys. Rev. D \textbf{97} (2018) no.2, 024042
[arXiv:1708.06725 [gr-qc]].

\bibitem{Crisnejo:2018uyn}
G.~Crisnejo and E.~Gallo,
``Weak lensing in a plasma medium and gravitational deflection of massive particles using the Gauss-Bonnet theorem. A unified treatment,''
Phys. Rev. D \textbf{97} (2018) no.12, 124016
[arXiv:1804.05473 [gr-qc]].

\bibitem{Kumar:2018ple}
R.~Kumar and S.~G.~Ghosh,
``Black Hole Parameter Estimation from Its Shadow,''
Astrophys. J. \textbf{892} (2020), 78
[arXiv:1811.01260 [gr-qc]].

\bibitem{Guo:2019lur}
M.~Guo, S.~Song and H.~Yan,
``Observational signature of a near-extremal Kerr-Sen black hole in the heterotic string theory,''
Phys. Rev. D \textbf{101} (2020) no.2, 024055
[arXiv:1911.04796 [gr-qc]].


\bibitem{Gralla:2019drh}
S.~E.~Gralla and A.~Lupsasca,
``Lensing by Kerr Black Holes,''
Phys. Rev. D \textbf{101} (2020) no.4, 044031
[arXiv:1910.12873 [gr-qc]].

\bibitem{Jusufi:2020cpn}
K.~Jusufi, M.~Jamil and T.~Zhu,
``Shadows of Sgr A$^{*}$ black hole surrounded by superfluid dark matter halo,''
Eur. Phys. J. C \textbf{80} (2020) no.5, 354
[arXiv:2005.05299 [gr-qc]].

\bibitem{Kumar:2020owy}
R.~Kumar and S.~G.~Ghosh,
``Rotating black holes in $4D$ Einstein-Gauss-Bonnet gravity and its shadow,''
JCAP \textbf{07} (2020), 053
[arXiv:2003.08927 [gr-qc]].

\bibitem{Zeng:2020dco}
X.~X.~Zeng, H.~Q.~Zhang and H.~Zhang,
``Shadows and photon spheres with spherical accretions in the four-dimensional Gauss\textendash{}Bonnet black hole,''
Eur. Phys. J. C \textbf{80} (2020) no.9, 872
[arXiv:2004.12074 [gr-qc]].

\bibitem{Gan:2021pwu}
Q.~Gan, P.~Wang, H.~Wu and H.~Yang,
``Photon spheres and spherical accretion image of a hairy black hole,''
Phys. Rev. D \textbf{104} (2021) no.2, 024003
[arXiv:2104.08703 [gr-qc]].

\bibitem{Guo:2021bwr}
S.~Guo, K.~J.~He, G.~R.~Li and G.~P.~Li,
``The shadow and photon sphere of the charged black hole in Rastall gravity,''
Class. Quant. Grav. \textbf{38} (2021) no.16, 165013
[arXiv:2205.07242 [gr-qc]].

\bibitem{Jusufi:2020zln}
K.~Jusufi and Saurabh,
``Black hole shadows in Verlinde\textquoteright{}s emergent gravity,''
Mon. Not. Roy. Astron. Soc. \textbf{503} (2021) no.1, 1310-1318
[arXiv:2010.15870 [gr-qc]].

\bibitem{Saurabh:2020zqg}
K.~Saurabh and K.~Jusufi,
``Imprints of dark matter on black hole shadows using spherical accretions,''
Eur. Phys. J. C \textbf{81} (2021) no.6, 490
[arXiv:2009.10599 [gr-qc]].

\bibitem{Gralla:2019xty}
S.~E.~Gralla, D.~E.~Holz and R.~M.~Wald,
``Black Hole Shadows, Photon Rings, and Lensing Rings,''
Phys. Rev. D \textbf{100} (2019) no.2, 024018
[arXiv:1906.00873 [astro-ph.HE]].

\bibitem{Peng:2020wun}
J.~Peng, M.~Guo and X.~H.~Feng,
``Influence of quantum correction on black hole shadows, photon rings, and lensing rings,''
Chin. Phys. C \textbf{45} (2021) no.8, 085103
[arXiv:2008.00657 [gr-qc]].

\bibitem{Chakhchi:2022fls}
L.~Chakhchi, H.~El Moumni and K.~Masmar,
``Shadows and optical appearance of a power-Yang-Mills black hole surrounded by different accretion disk profiles,''
Phys. Rev. D \textbf{105} (2022) no.6, 064031.

\bibitem{Guo:2021bhr}
S.~Guo, G.~R.~Li and E.~W.~Liang,
``Influence of accretion flow and magnetic charge on the observed shadows and rings of the Hayward black hole,''
Phys. Rev. D \textbf{105} (2022) no.2, 023024
doi:10.1103/PhysRevD.105.023024
[arXiv:2112.11227 [astro-ph.HE]].

\bibitem{He:2022yse}
K.~J.~He, S.~C.~Tan and G.~P.~Li,
``Influence of torsion charge on shadow and observation signature of black hole surrounded by various profiles of accretions,''
Eur. Phys. J. C \textbf{82} (2022) no.1, 81.

\bibitem{Li:2021riw}
G.~P.~Li and K.~J.~He,
``Shadows and rings of the Kehagias-Sfetsos black hole surrounded by thin disk accretion,''
JCAP \textbf{06} (2021), 037
[arXiv:2105.08521 [gr-qc]].

\bibitem{Zeng:2021dlj}
X.~X.~Zeng, G.~P.~Li and K.~J.~He,
``The shadows and observational appearance of a noncommutative black hole surrounded by various profiles of accretions,''
Nucl. Phys. B \textbf{974} (2022), 115639
[arXiv:2106.14478 [hep-th]].

\bibitem{Zeng:2021mok}
X.~X.~Zeng, K.~J.~He and G.~P.~Li,
``Effects of dark matter on shadows and rings of Brane-World black holes illuminated by various accretions,''
Sci. China Phys. Mech. Astron. \textbf{65} (2022) no.9, 290411
[arXiv:2111.05090 [gr-qc]].

\bibitem{He:2021htq}
K.~J.~He, S.~Guo, S.~C.~Tan and G.~P.~Li,
``Shadow images and observed luminosity of the Bardeen black hole surrounded by different accretions *,''
Chin. Phys. C \textbf{46} (2022) no.8, 085106
[arXiv:2103.13664 [hep-th]].

\bibitem{Guo:2022rql}
S.~Guo, G.~R.~Li and E.~W.~Liang,
``Observable characteristics of the charged black hole surrounded by thin disk accretion in Rastall gravity,''
Class. Quant. Grav. \textbf{39} (2022) no.13, 135004
[arXiv:2205.11241 [astro-ph.HE]].

\bibitem{Guerrero:2021ues}
M.~Guerrero, G.~J.~Olmo, D.~Rubiera-Garcia and D.~S.~C.~G\'omez,
``Shadows and optical appearance of black bounces illuminated by a thin accretion disk,''
JCAP \textbf{08} (2021), 036
[arXiv:2105.15073 [gr-qc]].

\bibitem{Yan:2021ygy}
H.~Yan, Z.~Hu, M.~Guo and B.~Chen,
``Photon emissions from near-horizon extremal and near-extremal Kerr equatorial emitters,''
Phys. Rev. D \textbf{104} (2021) no.12, 124005
[arXiv:2108.09051 [gr-qc]].


\bibitem{Guerrero:2022qkh}
M.~Guerrero, G.~J.~Olmo, D.~Rubiera-Garcia and D.~G\'omez S\'aez-Chill\'on,
``Light ring images of double photon spheres in black hole and wormhole spacetimes,''
Phys. Rev. D \textbf{105} (2022) no.8, 084057
[arXiv:2202.03809 [gr-qc]].

\bibitem{Rosa:2022tfv}
J.~L.~Rosa and D.~Rubiera-Garcia,
``Shadows of boson and Proca stars with thin accretion disks,''
Phys. Rev. D \textbf{106} (2022) no.8, 084004
[arXiv:2204.12949 [gr-qc]].

\bibitem{Atamurotov:2021hck}
F.~Atamurotov, U.~Papnoi and K.~Jusufi,
``Shadow and deflection angle of charged rotating black hole surrounded by perfect fluid dark matter,''
Class. Quant. Grav. \textbf{39} (2022) no.2, 025014
[arXiv:2104.14898 [gr-qc]].

\bibitem{Kumar:2019ohr}
R.~Kumar, B.~P.~Singh and S.~G.~Ghosh,
``Shadow and deflection angle of rotating black hole in asymptotically safe gravity,''
Annals Phys. \textbf{420} (2020), 168252
[arXiv:1904.07652 [gr-qc]].

\bibitem{Kumar:2017tdw}
R.~Kumar, B.~P.~Singh, M.~S.~Ali and S.~G.~Ghosh,
``Shadows of black hole surrounded by anisotropic fluid in Rastall theory,''
Phys. Dark Univ. \textbf{34} (2021), 100881
[arXiv:1712.09793 [gr-qc]].

\bibitem{Hu:2022lek}
S.~Hu, C.~Deng, D.~Li, X.~Wu and E.~Liang,
``Observational signatures of Schwarzschild-MOG black holes in scalar-tensor-vector gravity: shadows and rings with different accretions,''
Eur. Phys. J. C \textbf{82} (2022) no.10, 885.

\bibitem{Heydari-Fard:2023ent}
M.~Heydari-Fard, M.~Heydari-Fard and N.~Riazi,
``Shadows and photon rings of a spherically accreting Kehagias\textendash{}Sfetsos black hole,''
Int. J. Mod. Phys. D \textbf{32} (2023) no.13, 2350088
[arXiv:2307.01529 [gr-qc]].

\bibitem{Pulice:2023dqw}
B.~Puli\c{c}e, R.~C.~Pantig, A.~\"Ovg\"un and D.~Demir,
``Constraints on charged symmergent black hole from shadow and lensing,''
Class. Quant. Grav. \textbf{40} (2023) no.19, 195003
[arXiv:2308.08415 [gr-qc]].

\bibitem{Yang:2022btw}
J.~Yang, C.~Zhang and Y.~Ma,
``Shadow and stability of quantum-corrected black holes,''
Eur. Phys. J. C \textbf{83} (2023) no.7, 619
[arXiv:2211.04263 [gr-qc]].

\bibitem{Ma:2022jsy}
S.~J.~Ma, T.~C.~Ma, J.~B.~Deng and X.~R.~Hu,
``Shadow of Schwarzschild black hole in the cold dark matter halo,''
Mod. Phys. Lett. A \textbf{38} (2023) no.24n25, 2350104
[arXiv:2206.12820 [gr-qc]].

\bibitem{Wang:2023rjl}
Z.~L.~Wang,
``Shadows and rings of a de Sitter\textendash{}Schwarzschild black hole,''
Eur. Phys. J. Plus \textbf{138} (2023) no.12, 1131
[arXiv:2307.12361 [gr-qc]].

\bibitem{Kumaran:2023brp}
Y.~Kumaran and A.~\"Ovg\"un,
``Shadow and deflection angle of asymptotic, magnetically-charged, non-singular black hole,''
Eur. Phys. J. C \textbf{83} (2023) no.9, 812
[arXiv:2306.04705 [gr-qc]].

\bibitem{Huang:2023ilm}
Y.~X.~Huang, S.~Guo, Y.~H.~Cui, Q.~Q.~Jiang and K.~Lin,
``Influence of accretion disk on the optical appearance of the Kazakov-Solodukhin black hole,''
Phys. Rev. D \textbf{107} (2023) no.12, 123009
[arXiv:2311.00302 [gr-qc]].

\bibitem{Zhang:2022osx}
Z.~Zhang, H.~Yan, M.~Guo and B.~Chen,
``Shadows of Kerr black holes with a Gaussian-distributed plasma in the polar direction,''
Phys. Rev. D \textbf{107} (2023) no.2, 024027
[arXiv:2206.04430 [gr-qc]].


\bibitem{Hu:2023bzy}
S.~Hu, C.~Deng, S.~Guo, X.~Wu and E.~Liang,
``Observational signatures of Schwarzschild-MOG black holes in scalar\textendash{}tensor\textendash{}vector gravity: images of the accretion disk,''
Eur. Phys. J. C \textbf{83} (2023) no.3, 264.

\bibitem{SupernovaCosmologyProject:1998vns}
S.~Perlmutter \textit{et al.} [Supernova Cosmology Project],
``Measurements of $\Omega$ and $\Lambda$ from 42 high redshift supernovae,''
Astrophys. J. \textbf{517} (1999), 565-586
[arXiv:astro-ph/9812133 [astro-ph]].

\bibitem{SupernovaSearchTeam:1998fmf}
A.~G.~Riess \textit{et al.} [Supernova Search Team],
``Observational evidence from supernovae for an accelerating universe and a cosmological constant,''
Astron. J. \textbf{116} (1998), 1009-1038
[arXiv:astro-ph/9805201 [astro-ph]].

\bibitem{SupernovaSearchTeam:1998cav}
P.~M.~Garnavich \textit{et al.} [Supernova Search Team],
``Supernova limits on the cosmic equation of state,''
Astrophys. J. \textbf{509} (1998), 74-79
[arXiv:astro-ph/9806396 [astro-ph]].

\bibitem{Wang:1999fa}
L.~M.~Wang, R.~R.~Caldwell, J.~P.~Ostriker and P.~J.~Steinhardt,
``Cosmic concordance and quintessence,''
Astrophys. J. \textbf{530} (2000), 17-35
[arXiv:astro-ph/9901388 [astro-ph]].


\bibitem{Bahcall:1999xn}
N.~A.~Bahcall, J.~P.~Ostriker, S.~Perlmutter and P.~J.~Steinhardt,
``The Cosmic triangle: Assessing the state of the universe,''
Science \textbf{284} (1999), 1481-1488
[arXiv:astro-ph/9906463 [astro-ph]].

\bibitem{Kiselev:2002dx}
V.~V.~Kiselev,
``Quintessence and black holes,''
Class. Quant. Grav. \textbf{20} (2003), 1187-1198
[arXiv:gr-qc/0210040 [gr-qc]].

\bibitem{Lacroix:2012nz}
T.~Lacroix and J.~Silk,
``Constraining the distribution of dark matter at the Galactic Centre using the high-resolution Event Horizon Telescope,''
Astron. Astrophys. \textbf{554} (2013), A36
[arXiv:1211.4861 [astro-ph.GA]].

\bibitem{Haroon:2018ryd}
S.~Haroon, M.~Jamil, K.~Jusufi, K.~Lin and R.~B.~Mann,
``Shadow and Deflection Angle of Rotating Black Holes in Perfect Fluid Dark Matter with a Cosmological Constant,''
Phys. Rev. D \textbf{99} (2019) no.4, 044015
[arXiv:1810.04103 [gr-qc]].

\bibitem{Khan:2020ngg}
S.~U.~Khan and J.~Ren,
``Shadow cast by a rotating charged black hole in quintessential dark energy,''
Phys. Dark Univ. \textbf{30} (2020), 100644
[arXiv:2006.11289 [gr-qc]].

\bibitem{Zeng:2020vsj}
X.~X.~Zeng and H.~Q.~Zhang,
``Influence of quintessence dark energy on the shadow of black hole,''
Eur. Phys. J. C \textbf{80} (2020) no.11, 1058
[arXiv:2007.06333 [gr-qc]].

\bibitem{He:2021aeo}
A.~He, J.~Tao, Y.~Xue and L.~Zhang,
``Shadow and photon sphere of black hole in clouds of strings and quintessence *,''
Chin. Phys. C \textbf{46} (2022) no.6, 065102
[arXiv:2109.13807 [gr-qc]].

\bibitem{Heydari-Fard:2022jdu}
M.~Heydari-Fard,
``Effect of quintessence dark energy on the shadow of Hayward black holes with spherical accretion,''
Indian J. Phys. \textbf{1} (2023), 14
[arXiv:2209.09103 [gr-qc]].

\bibitem{Kamenshchik:2001cp}
A.~Y.~Kamenshchik, U.~Moschella and V.~Pasquier,
``An Alternative to quintessence,''
Phys. Lett. B \textbf{511} (2001), 265-268
[arXiv:gr-qc/0103004 [gr-qc]].

\bibitem{Bilic:2001cg}
N.~Bilic, G.~B.~Tupper and R.~D.~Viollier,
``Unification of dark matter and dark energy: The Inhomogeneous Chaplygin gas,''
Phys. Lett. B \textbf{535} (2002), 17-21
[arXiv:astro-ph/0111325 [astro-ph]].

\bibitem{Bento:2002ps}
M.~C.~Bento, O.~Bertolami and A.~A.~Sen,
``Generalized Chaplygin gas, accelerated expansion and dark energy matter unification,''
Phys. Rev. D \textbf{66} (2002), 043507
[arXiv:gr-qc/0202064 [gr-qc]].

\bibitem{Sengupta:2023yxh}
R.~Sengupta, P.~Paul, B.~C.~Paul and M.~Kalam,
``Can extended Chaplygin gas source a Hubble tension resolved emergent universe ?,''
[arXiv:2307.02602 [gr-qc]].

\bibitem{Abdullah:2021tee}
A.~Abdullah, A.~A.~El-Zant and A.~Ellithi,
``Growth of fluctuations in Chaplygin gas cosmologies: A nonlinear Jeans scale for unified dark matter,''
Phys. Rev. D \textbf{106} (2022) no.8, 083524
[arXiv:2108.03260 [astro-ph.CO]].

\bibitem{Ogawa:2000gj}
N.~Ogawa,
``A Note on classical solution of Chaplygin gas as d-branes,''
Phys. Rev. D \textbf{62} (2000), 085023 [arXiv:hep-th/0003288 [hep-th]].

\bibitem{Bordemann:1993ep}
M.~Bordemann and J.~Hoppe,
``The Dynamics of relativistic membranes. 1. Reduction to two-dimensional fluid dynamics,''
Phys. Lett. B \textbf{317} (1993), 315-320 [arXiv:hep-th/9307036 [hep-th]].

\bibitem{Jackiw:2000cc}
R.~Jackiw and A.~P.~Polychronakos,
``Supersymmetric fluid mechanics,''
Phys. Rev. D \textbf{62} (2000), 085019 [arXiv:hep-th/0004083 [hep-th]].

\bibitem{Li:2019lhr}
X.~Q.~Li, B.~Chen and L.~l.~Xing,
``Charged Lovelock black holes in the presence of dark fluid with a nonlinear equation of state,''
Eur. Phys. J. Plus \textbf{135} (2020) no.2, 175
[arXiv:1905.08156 [gr-qc]].

\bibitem{Li:2019ndh}
X.~Q.~Li, B.~Chen and L.~l.~Xing,
``Black holes in Einstein\textendash{}Gauss\textendash{}Bonnet gravity with a background of modified Chaplygin gas,''
Eur. Phys. J. Plus \textbf{137} (2022) no.10, 1167
[arXiv:1908.09827 [gr-qc]].

\bibitem{Li:2022csn}
X.~Q.~Li, B.~Chen and L.~L.~Xing,
``Black holes surrounded by modified Chaplygin gas in Lovelock theory of gravity,''
Annals Phys. \textbf{446} (2022), 169125.

\bibitem{Li:2023zfl}
X.~Q.~Li, H.~P.~Yan, L.~L.~Xing and S.~W.~Zhou,
``Critical behavior of AdS black holes surrounded by dark fluid with Chaplygin-like equation of state,''
Phys. Rev. D \textbf{107} (2023) no.10, 104055
[arXiv:2305.03028 [gr-qc]].

\bibitem{Ali:2020omz}
A.~Ali and K.~Saifullah,
``Magnetized black holes surrounded by dark fluid in Lovelock-power-Yang-Mills gravity,''
JCAP \textbf{10} (2021), 058
[arXiv:2006.15610 [gr-qc]].

\bibitem{Ali:2024rrm}
A.~Ali and K.~Saifullah,
``Dimensionally continued black holes sourced by a conformally coupled scalar field and Chaplygin-like dark fluid,''
Eur. Phys. J. C \textbf{84} (2024) no.1, 41.

\bibitem{Arora:2023mve}
D.~Arora, M.~Yasir, H.~Chaudhary, F.~Javed, G.~Mustafa, X.~Tiecheng and F.~Atamurotov,
``Joule-Thomson expansion and tidal force effects of AdS black holes surrounded by Chaplygin dark fluid,''
[arXiv:2312.16224 [gr-qc]].

\bibitem{Sekhmani:2023plr}
Y.~Sekhmani, J.~Rayimbaev, G.~G.~Luciano, R.~Myrzakulov and D.~J.~Gogoi,
``Phase structure of charged AdS black holes surrounded by exotic fluid with modified Chaplygin equation of state,''
Eur. Phys. J. C \textbf{84} (2024) no.3, 227 [arXiv:2311.02448 [gr-qc]].

\bibitem{Zhang:2024fxj}
M.~Y.~Zhang, H.~Chen, H.~Hassanabadi, Z.~W.~Long and H.~Yang,
``Critical behavior and Joule-Thomson expansion of charged AdS black holes surrounded by exotic fluid with modified Chaplygin equation of state,''
[arXiv:2401.17589 [gr-qc]].

\bibitem{Semiz:2008ny}
I.~Semiz,
``All 'static' spherically symmetric perfect fluid solutions of Einstein's equations with constant equation of state parameter and finite-polynomial 'mass function',''
Rev. Math. Phys. \textbf{23} (2011), 865-882
[arXiv:0810.0634 [gr-qc]].

\bibitem{Raposo:2018rjn}
G.~Raposo, P.~Pani, M.~Bezares, C.~Palenzuela and V.~Cardoso,
``Anisotropic stars as ultracompact objects in General Relativity,''
Phys. Rev. D \textbf{99} (2019) no.10, 104072
[arXiv:1811.07917 [gr-qc]].

\bibitem{Friedman:1993ty}
J.~L.~Friedman, K.~Schleich and D.~M.~Witt,
``Topological censorship,''
Phys. Rev. Lett. \textbf{71} (1993), 1486-1489
[erratum: Phys. Rev. Lett. \textbf{75} (1995), 1872]
[arXiv:gr-qc/9305017 [gr-qc]].

\bibitem{Chrusciel:1994tr}
P.~T.~Chrusciel and R.~M.~Wald,
``On the topology of stationary black holes,''
Class. Quant. Grav. \textbf{11} (1994), L147-L152
[arXiv:gr-qc/9410004 [gr-qc]].

\bibitem{Perlick:2021aok}
V.~Perlick and O.~Y.~Tsupko,
``Calculating black hole shadows: Review of analytical studies,''
Phys. Rept. \textbf{947} (2022), 1-39
[arXiv:2105.07101 [gr-qc]].


\bibitem{Vagnozzi:2022moj}
S.~Vagnozzi, R.~Roy, Y.~D.~Tsai, L.~Visinelli, M.~Afrin, A.~Allahyari, P.~Bambhaniya, D.~Dey, S.~G.~Ghosh and P.~S.~Joshi, \textit{et al.}
``Horizon-scale tests of gravity theories and fundamental physics from the Event Horizon Telescope image of Sagittarius A,''
Class. Quant. Grav. \textbf{40} (2023) no.16, 165007 [arXiv:2205.07787 [gr-qc]].


\bibitem{Bambi:2019tjh}
C.~Bambi, K.~Freese, S.~Vagnozzi and L.~Visinelli,
``Testing the rotational nature of the supermassive object M87* from the circularity and size of its first image,''
Phys. Rev. D \textbf{100} (2019) no.4, 044057
[arXiv:1904.12983 [gr-qc]].


\bibitem{Allahyari:2019jqz}
A.~Allahyari, M.~Khodadi, S.~Vagnozzi and D.~F.~Mota,
``Magnetically charged black holes from non-linear electrodynamics and the Event Horizon Telescope,''
JCAP \textbf{02} (2020), 003
[arXiv:1912.08231 [gr-qc]].

\bibitem{Planck:2018vyg}
N.~Aghanim \textit{et al.} [Planck],
``Planck 2018 results. VI. Cosmological parameters,''
Astron. Astrophys. \textbf{641} (2020), A6
[erratum: Astron. Astrophys. \textbf{652} (2021), C4]
[arXiv:1807.06209 [astro-ph.CO]].

\bibitem{Jaroszynski:1997bw}
M.~Jaroszynski and A.~Kurpiewski,
``Optics near kerr black holes: spectra of advection dominated accretion flows,''
Astron. Astrophys. \textbf{326} (1997), 419
[arXiv:astro-ph/9705044 [astro-ph]].

\bibitem{Bambi:2013nla}
C.~Bambi,
``Can the supermassive objects at the centers of galaxies be traversable wormholes? The first test of strong gravity for mm/sub-mm very long baseline interferometry facilities,''
Phys. Rev. D \textbf{87} (2013), 107501
[arXiv:1304.5691 [gr-qc]].

\bibitem{Kocherlakota:2022jnz}
P.~Kocherlakota and L.~Rezzolla,
``Distinguishing gravitational and emission physics in black hole imaging: spherical symmetry,''
Mon. Not. Roy. Astron. Soc. \textbf{513} (2022) no.1, 1229-1243
[arXiv:2201.05641 [gr-qc]].


\bibitem{Bauer:2021atk}
A.~M.~Bauer, A.~C\'ardenas-Avenda\~no, C.~F.~Gammie and N.~Yunes,
``Spherical Accretion in Alternative Theories of Gravity,''
Astrophys. J. \textbf{925} (2022) no.2, 119
[arXiv:2111.02178 [gr-qc]].




\end{thebibliography}
\end{document}